\documentclass[10pt,journal,compsoc]{IEEEtran}
\makeatletter \def\mdseries@tt{m} \makeatother 

\usepackage{booktabs}
\usepackage{fancyvrb}
\usepackage{graphicx}
\usepackage[singlespacing]{setspace}
\usepackage{threeparttable}
\usepackage{multirow}
\usepackage{caption}
\usepackage{subcaption}
\usepackage[utf8]{inputenc}
\usepackage[english]{babel}
\usepackage{amsthm}
\usepackage{changepage} 
\usepackage{mathtools}
\usepackage{textgreek}
\usepackage{amsmath}
\usepackage{tikz}
\usetikzlibrary{shapes,positioning,arrows,calc}
\DeclareFontFamily{OT1}{pzc}{}
\DeclareFontShape{OT1}{pzc}{m}{it}{<-> s * [1.1] pzcmi7t}{}
\DeclareMathAlphabet{\mathpzc}{OT1}{pzc}{m}{it}
\theoremstyle{definition}

\DeclareMathAlphabet\mathbfcal{OMS}{cmsy}{b}{n}
\usepackage{centernot}
\usepackage{enumitem}
\makeatletter \def\algbackskip{\hskip-\ALG@thistlm} \makeatother
\usepackage{xcolor}
\usepackage[ruled, vlined, linesnumbered]{algorithm2e}
\SetAlFnt{\small\sffamily}

\SetCommentSty{mycommfont}
\usepackage{amsmath, listings, amsthm, amssymb, proof, xspace}
\usepackage{hyperref}
\hypersetup{colorlinks,allcolors=black}
\theoremstyle{remark}

\usepackage{forest}
\usetikzlibrary{angles}
\usepackage{array}
\newcolumntype{?}{!{\vrule width 2pt}}
\usepackage{xurl}
\usepackage[backgroundcolor=white,bordercolor=blue,linecolor=blue]{todonotes}
\usepackage[most]{tcolorbox}
\usepackage[T1]{fontenc}
\usepackage{beramono}
\usepackage[]{mdframed}
\usepackage{float}

\input{code-style}
\newcommand{\specialToken}[1]{\text{\scriptsize[<{\MakeUppercase{#1}}>]}}
\newcommand{\specialTokenL}[1]{\text{\footnotesize[<{#1}>]}}

\newcommand\approachName{\textsc{TaRGET}}
\newcommand\approachNameAbbr{\textsc{\textbf{T}est \textbf{R}epair \textbf{GE}nera\textbf{T}or}}
\newcommand\benchmarkName{\textsc{TaRBench}}
\newcommand\ceprot{\textsc{Ceprot}}
\newcommand\nocontext{\textsc{NoContext}}
\newcommand\sutcopy{\textsc{SUTCopy}}

\newcommand{\TR}[2]{\textcolor{black}{#2}}
\newcommand{\TRC}[1]{\textcolor{black}{#1}}

\newcommand{\TRR}[2]{\textcolor{black}{#2}}
\newcommand{\TRRC}[1]{\textcolor{black}{#1}}

\begin{document}

\bstctlcite{IEEEexample:BSTcontrol}

\title{Automated Test Case Repair\\ Using Language Models}

\author{
    Ahmadreza Saboor Yaraghi,
    Darren Holden,
    Nafiseh Kahani,
    and Lionel Briand, \IEEEmembership{Fellow, IEEE} %
\IEEEcompsocitemizethanks{\IEEEcompsocthanksitem A. Saboor Yaraghi and L. Briand are with the School of Electrical Engineering and Computer Science, University of Ottawa, Ottawa, Ontario, Canada, K1N 5N6. L. Briand also holds an appointment with the Lero, the Research Ireland Centre for Software, University of Limerick, Ireland. \protect\\
E-mail: \{a.saboor, lbriand\}@uottawa.ca
\IEEEcompsocthanksitem D. Holden and N. Kahani are with the Department of Systems and Computer Engineering, Carleton University, Ottawa, Ontario, Canada, K1S 5B6. \protect\\
E-mail: darren.holden@carleton.ca , kahani@sce.carleton.ca}
}

\IEEEtitleabstractindextext{
\begin{abstract}

Ensuring the quality of software systems through testing is essential, yet maintaining test cases poses significant challenges and costs. The need for frequent updates to align with the evolving system under test often entails high complexity and cost for maintaining these test cases. Further, unrepaired broken test cases can degrade test suite quality and disrupt the software development process, wasting developers' time.
To address this challenge, we present \approachName{} (\approachNameAbbr{}), a novel approach leveraging pre-trained code language models for automated test case repair. \approachName{} treats test repair as a language translation task, employing a two-step process to fine-tune a language model based on essential context data characterizing the test breakage. 
To evaluate our approach, we introduce \benchmarkName{}, a comprehensive benchmark we developed covering 45,373 broken test repairs across 59 open-source projects. Our results demonstrate \approachName{}'s effectiveness, achieving a 66.1\% exact match accuracy. Furthermore, our study examines the effectiveness of \approachName{} across different test repair scenarios. We provide a practical guide to predict situations where the generated test repairs might be less reliable. We also explore whether project-specific data is always necessary for fine-tuning and if our approach can be effective on new projects.

\end{abstract}

}

\maketitle

\section{Introduction}
\label{sec:introdution}
Testing is crucial for ensuring the quality of software systems~\cite{duvall2007continuous,herzig2015art}. However, maintaining test cases can be costly as developers often need to update them when the source code of the System Under Test (SUT) evolves, posing a maintainability challenge~\cite{kasurinen2010software,labuschagne2017measuring,daniel2010test,painpointsCI}. Failing to do so can result in broken test cases that are no longer valid  and, without proper maintenance, these broken test cases might be discarded, ultimately reducing the overall quality of the test suites. Two recent analyses of test case failures of 211 Apache software foundation projects~\cite{vahabzadeh2015empirical} and 61 open-source projects~\cite{labuschagne2017measuring} reported that broken test cases account for 14\% and 22\% of failures, respectively. Additionally, false alarms resulting from broken test cases, which are failures caused by the test code rather than the SUT, disrupt software build processes and waste valuable developer time~\cite{daniel2009reassert,daniel2010test}.

\begin{figure*}[tbp]
    \centering
    \begin{subfigure}{\linewidth}
        \centering
        \begin{tcolorbox}[codebox]
        \begin{lstlisting}[language=Java,gobble=10,numbers=left,style=codeHighlighting,showstringspaces=false,breaklines=true]
            <@\textcolor{javagreen}{// Updates in the system under test (change of exception type)}@>
            &- return new &^FailedPreconditionRuntimeException^& ( t ) ;&
            `+ return new `@@InternalRuntimeException@@` ( t ) ;`
            
            <@\textcolor{javagreen}{// Repairing the broken test case}@>
            @Test
            public void close() throws Exception {
            <@\quad@>ByteBuffer buf = BufferUtils.getIncreasingByteBuffer(TEST_BLOCK_SIZE);
            <@\quad@>Assert.assertEquals(TEST_BLOCK_SIZE, mWriter.append(buf));
            <@\quad@>mWriter.close();
            <@\quad@>&- &^mThrown^&.&^expect^&(&^FailedPreconditionRuntimeException^&.class);&
            <@\quad@>^- mWriter.append(buf);^
            <@\quad@>`+ `@@Assert@@`.`@@assertThrows@@`(`@@InternalRuntimeException@@`.class`@@, (@@`)`@@ -> mWriter.append(buf))@@`;`
            }
        \end{lstlisting}
        \end{tcolorbox}
        \subcaption{A broken test example causing a runtime error, from the \textit{alluxio/alluxio} project at commit \href{https://github.com/Alluxio/alluxio/commit/6982d6c759}{\textcolor{blue}{\textit{\underline{6982d6c759}}}}}
        \label{fig:runtime_example}
    \end{subfigure}
    \begin{subfigure}{\linewidth}
        \centering
        \begin{tcolorbox}[codebox]
        \begin{lstlisting}[language=Java,gobble=10,numbers=left,style=codeHighlighting,showstringspaces=false,breaklines=true]
            <@\textcolor{javagreen}{// Updates in the system under test (Columns.add is not public anymore)}@>
            &- &^public^& Columns add ( Column < ? > column ) {&
            `+ Columns add ( Column < ? > column ) {`
            
            <@\textcolor{javagreen}{// Repairing the broken test case}@>
            @Test
            public void testValidateColumns() {
            <@\quad@>Schema schema = SchemaBuilders.schema().mapper("field1", stringMapper()).build();
            <@\quad@>&- Columns columns = new Columns ( ) .&^add^& ( &^Column.builder ( ^&"field1" &^)^& &^.buildWithComposed ( ^&"value" ,& &UTF8Type.instance ) &^) ^&;&
            <@\quad@>`+ Columns columns = new Columns ( ) .`@@addComposed@@` ( "field1" `@@,@@` "value" , UTF8Type.instance ) ;`
            <@\quad@>schema.validate(columns);
            <@\quad@>schema.close();
            }
        \end{lstlisting}
        \end{tcolorbox}
        \subcaption{A broken test case example causing a compilation error, from the \textit{stratio/cassandra-lucene-index} project at commit \href{https://github.com/Stratio/cassandra-lucene-index/commit/fdd34e53}{\textcolor{blue}{\textit{\underline{fdd34e53}}}}}
        \label{fig:compile_example}
    \end{subfigure}

    \caption{\TRC{Two real-world broken test examples causing compilation and runtime errors.}}
    \label{fig:intro_examples}
\end{figure*}

\TR{R3.2}{
To illustrate the motivation and scope of the problem, we present two representative examples of broken test cases from real-world software projects in Figure~\ref{fig:intro_examples}. The first example (Figure~\ref{fig:runtime_example}) demonstrates a scenario where the test fails at runtime due to a change in the type of exception thrown by the SUT. In this case, the test originally expected a \textit{FailedPreconditionRuntimeException}, but after the code change, the SUT threw an \textit{InternalRuntimeException} instead. The second example (Figure~\ref{fig:compile_example}) showcases a compilation error caused by a change in method visibility. In this case, the method \textit{Columns.add} was modified from public to non-public, causing a test that relied on this method to fail compilation. These examples cover two main types of test breakages: runtime errors that arise during execution and compilation errors that prevent tests from running in the first place.}

Overall, the failure of broken test cases accounts for a significant portion of test case failures, negatively impacting the quality of testing. To address this challenge, researchers have developed two complementary groups of techniques~\cite{imtiaz2019survey}: detection and repair. The main focus of our study is the latter. It aims at automatically repairing broken tests and often uses source code analysis and search-based techniques~\cite{daniel2009reassert,daniel2010test,mirzaaghaei2014automatic,xu2014using,li2019intent}. On the other hand, detection focuses on determining whether a test failure results from a broken test case and pinpoints the specific location of the breakage within the test code~\cite{pinto2012myth,hao2013bugorobsolete,gao2015ref,rapos2016simulink,nguyen2017interaction}. Apart from the aforementioned studies on broken tests, which necessitate access to and analysis of the source code of the SUT, there exist methods to repair broken black-box GUI tests solely through GUI analysis~\cite{pan2020gui,geneticrepairing}.

\TR{R1.18 R3.1}{Existing automated test repair studies show limitations from both methodological and evaluation perspectives, restricting their applicability across diverse software systems and repair scenarios. Methodologically, these approaches can be classified into heuristic, rule-based, or search-based, and static or dynamic analysis-based techniques. However, these methods often struggle with generalization (they rely on project-specific data), scalability, and broad applicability. They are typically tailored to specific programming languages or focus narrowly on particular types of test repairs, limiting their application. For instance, \textit{ReAssert}~\cite{daniel2009reassert} focuses solely on repairing test oracles (assertions). While static or dynamic analysis-based techniques, such as symbolic execution in \textit{TRIP}~\cite{li2019intent}, involve a more generic approach with a broader scope, these methods are often challenging to implement for different programming languages and are expensive to scale for large codebases. From an evaluation standpoint, existing studies rely on benchmarks that are limited in size and diversity, raising concerns about the validity of their evaluations. For example, \textit{TRIP}~\cite{li2019intent} is evaluated on only 91 broken tests across four projects. Further, most of these studies are not fully reproducible because their source code or benchmarks are not publicly accessible.}

A recent adaptation of language models for source code shows satisfactory results for different software engineering tasks such as code generation, code completion, and program repair~\cite{hou2023llm4se}. Motivated by these studies, we propose \approachName{} (\approachNameAbbr{})~\cite{targetGitHub}, an approach that leverages language models to automatically repair broken test cases. Our solution is based on the insight that certain modifications in the source code of a SUT warrant corresponding updates in the associated test case source code. Thus, we formulate the test repair problem as a language translation task via a two-step approach. The first step focuses on identifying and collecting essential data that characterize the test breakage. The second step utilizes the gathered information to shape inputs and outputs for fine-tuning a given language model for the test repair task.

We conduct an extensive experimental analysis and assess \approachName{} by creating \benchmarkName{}~\cite{tarbenchFigshare}, a benchmark comprising 45,373 broken test repairs across 59 distinct projects, making it by far the most comprehensive benchmark to date in this application context. We demonstrate that fine-tuning a language model using \approachName{} for repairing broken tests achieves an exact match accuracy (EM) of 66.1\% and a plausible repair accuracy (PR) of 80\%. The EM denotes repairs perfectly matching the ground truth, while the PR denotes repairs that execute successfully. \approachName{} significantly outperforms the baseline, which is the best language model when it is not fine-tuned with crucial context information for the test repair task, by a substantial margin of 37.4 EM percentage points. To maximize the practicality of \approachName{}, we introduce and evaluate a model to predict the reliability of repairs generated by \approachName{}, helping developers to decide whether to trust them or not. Furthermore, we investigate the generalizability of \approachName{} by showing that a model fine-tuned on specific projects can be effectively applied to other software projects with an acceptable margin of EM loss. This is in contrast to existing work on broken test repair that necessitates project-based analysis.

Overall, the main contributions of this work are as follows.
\begin{itemize}
    \item \textbf{\approachName{}: An automatic broken test case repair technique using language models.} 
    \approachName{}'s fine-tuned language model achieves an EM and PR of 66.1\% and 80\%, respectively, a significant achievement given the state of the art. This outperforms the application of existing language models without leveraging test repair context data, which only yield an EM of 28.7\%. \TR{R3.1}{Moreover, \approachName{} does not have the methodological limitations of existing studies discussed above, as it is not restricted to specific programming languages, and its support for repair types is not limited.}
    
    \item \textbf{\benchmarkName{}: A large and comprehensive benchmark of broken tests and their repair.} 
    \TR{R3.1}{As previously discussed, existing benchmarks in the literature are limited in both size and diversity, making them insufficient for effectively training language models. To overcome this limitation, we developed a tool capable of extracting broken tests and their corresponding repairs from open-source projects.} Using this tool, we constructed \benchmarkName{}, a comprehensive and high-quality benchmark with including 45,373 instances, which serves as the foundation of our study. We believe that \benchmarkName{} will be valuable to researchers in this field, and we have made both the tool and dataset publicly available for future use\footnote{\approachName{}: \underline{\url{https://github.com/Ahmadreza-SY/TaRGet}}}\textsuperscript{,}\footnote{\benchmarkName{}: \underline{\url{https://doi.org/10.6084/m9.figshare.25008893}}}.

    \item \TR{R3.16}{\textbf{Empirical evaluation of \approachName{}.}}
    The study addresses three research questions that concern evaluating \approachName{} using different configurations, analyzing instances where \approachName{} fail to generate a correct repair, investigating ways to maximize the practical usage of \approachName{} by predicting whether repairs can be trusted, and assessing the generalizability of \approachName{} and the amount of training data needed to fine-tune it for test repair.
\end{itemize}

The rest of this paper is organized as follows. Section~\ref{sec:background} introduces the foundational concepts which our work relies on. Following that, Section~\ref{sec:approach} describes our approach in detail. In Section~\ref{sec:study}, we introduce our benchmark and present the results of our experimental study, addressing three key research questions. Section~\ref{sec:related-work} examines and compare existing studies relevant to our work. Lastly, Section~\ref{sec:conclusion} summarizes our findings and outlines potential paths for future enhancements.

\section{Background}
\label{sec:background}
We address the problem of repairing broken test cases by utilizing language models. This is motivated by the idea that automated program repair, which is similar to automated test repair, can be approached by translating faulty code snippets into repaired code snippets~\cite{zhang2023survey}. The rest of this section will elaborate on this idea and act as a primer on the concepts of language models, code language models, pre-training, fine-tuning, and other concepts upon which our work relies.

\subsection{Language Models}
Language models aim to perform language-based tasks by modeling the likelihood of token sequences in order to predict an output, which is often a generated sequence of tokens~\cite{zhao2023survey}. Code Language Models (CLMs) are a specialized category of language models designed for code-related tasks, forming the focus of our study. CLMs  are often trained on multiple programming languages alongside natural language texts \cite{ahmad2021unified}, \cite{wang2023codet5p}, \cite{nijkamp2022codegen}. Consequently, they are well suited for code-related tasks, such as code summarization, code generation, program repair, and code translation.

\subsubsection{Language Model Architecture}
The architecture of language models are largely inspired by the pairing of a self-attention layer with a feed forward network, originally proposed by Vaswani et al.~\cite{vaswani2017attention}. These self-attention layer and feed forward network pairs are then used to construct encoders and decoders, which can be combined to construct language models. There are two variations of language model architectures that are relevant to this study: encoder-decoder models, and decoder-only models. Encoder-only models also exist but they are not utilized in our work, since they require an additional decoder to perform generative tasks \cite{zhang2023survey}, and are thus not described here.  

Encoder-decoder models are models which use both an encoder and a decoder. The sequence-to-sequence Transformer \cite{vaswani2017attention} is an example of such a model, and is the basis on which many modern language models were built. In an encoder-decoder model, the encoder takes the input token sequence, and converts it into a hidden, intermediary representation, which is then used by the decoder to generate the output token sequence \cite{hou2023llm4se}, \cite{zhang2023survey}.

The encoder generally consists of several layers, each of which has two sub-layers: a multi-head self-attention mechanism, and a fully connected feed-forward network. The decoder is also constructed of several layers, each of which has three sub-layers: a masked multi-head attention mechanism, a multi-head self-attention mechanism, and a fully connected feed-forward network \cite{vaswani2017attention}. The multi-head self-attention mechanism performs linear transformations on the inputs, resulting in several projections of the the input, which are recombined to give the final output \cite{vaswani2017attention}. The decoder generates the final output by building a sequence of tokens one token at a time in an auto-regressive process, selecting the next output token based on the model input and the current, incomplete output sequence. An example of an encoder-decoder model that we use in this study is CodeT5+~\cite{wang2023codet5p}.

Decoder-only models are models which only have a decoder with no encoder, as the name suggests. These models start from an initial state, and gradually build an output sequence of token, attending to the previously generated output tokens \cite{hou2023llm4se}, \cite{zhang2023multilingual}. The decoders in this type of model work similarly to those used in the encoder-decoder architecture. They rely on an ability to understand the target language and its nuances \cite{hou2023llm4se}. An example of a decoder-only model that we use in this study is CodeGen~\cite{nijkamp2022codegen}.

\subsection{Pre-Training, Fine-Tuning, and Inference of Code Language Models}
It is common for language models to be pre-trained on large corpora of general-purpose data for a variety of tasks, resulting in a general model that can then be adapted to specific down-stream tasks \cite{zhou2023comprehensive}. For language models, there are five general types of pre-training tasks: mask language modeling, denoising autoencoder, replaced token detection, next sentence prediction, and sentence order prediction \cite{zhou2023comprehensive}. These pre-trained models are then adapted to specific tasks through the process of fine-tuning, during which the model is trained on additional data which is focused on the tasks of interest. 

Fine-tuning a model is simply performing further training on a pre-trained model so that it learns to perform specific tasks. Fine-tuning a model this way has benefits, as the model retains the vocabulary and any insights it learned during pre-training, and thus can be trained on a more limited dataset \cite{zhang2023survey}. Fine-tuning is commonly performed with the goal of minimizing the model's cross-entropy loss on a specific task \cite{ye2022neural}. This is done by training the decoder on iterations of the output sequence. The decoder considers the input representation and the previously generated output sequence to determine the next token in the output sequence. The loss is determined based on what token the decoder determines is next in the sequence, and is minimal when the decoder selects the correct token.

As a practical example of pre-trained models being fine-tuned, consider the PLBART CLM, which was pre-trained on a corpus of data to perform the task of denoising both natural language and programming languages (Java and Python in particular) \cite{ahmad2021unified}. This pre-trained model was then fine-tuned several different times to perform several tasks, including code generation, code summarization, and code translation, amongst other tasks \cite{ahmad2021unified}.

At the inference stage, given that the model output can consist of any sequence of tokens from the target language, the output space is quite large. To address this, generative models use different techniques to strategically select the output. In line with previous program repair studies~\cite{zhang2023survey}, we adopt the beam search ranking strategy to prioritize the model output. The beam search algorithm operates by maintaining a set of $k$ sequences, where $k$ refers to the beam size. At each generation step, the decoder predicts the probability distribution of tokens in the vocabulary based on the previous tokens for each of the $k$ sequences. Next, the beam search selects the top-$k$ tokens with the highest scores. The score for a token is calculated by multiplying its probability with the probabilities of the previous tokens in the sequence. This search process continues until either the end-of-sequence (EOS) token is encountered or a maximum output sequence length is reached.

\section{Approach: \approachName{}}
\label{sec:approach}

\TR{R1.2 R3.16}{Let $V_i$ and $V_{i+1}$ denote two consecutive versions of a SUT. Let $T_i$ denote a test case that is broken due to the code changes introduced in $V_{i+1}$ relative to $V_i$, and let $L$ be the set of breakage lines, which are the specific lines in $T_i$ that require repair. We define our approach, \approachName{}, as follows.
\[\text{\approachName{}}(T_i, L, V_i, V_{i+1}) \rightarrow T_{i+1}\]
where $T_{i+1}$ is the repaired version of $T_i$.}
\TR{R1.1}{Note that we distinguish between \textit{error lines} and \textit{breakage lines}. Error lines are the lines in $T_i$ where compile or runtime errors occur, whereas breakage lines refer to the lines containing the root cause of the test case breakage and where repairs should be made. These two sets of lines may not necessarily overlap.}

\begin{figure}[tbp]
    \centering
    \includegraphics[width=0.8\columnwidth]{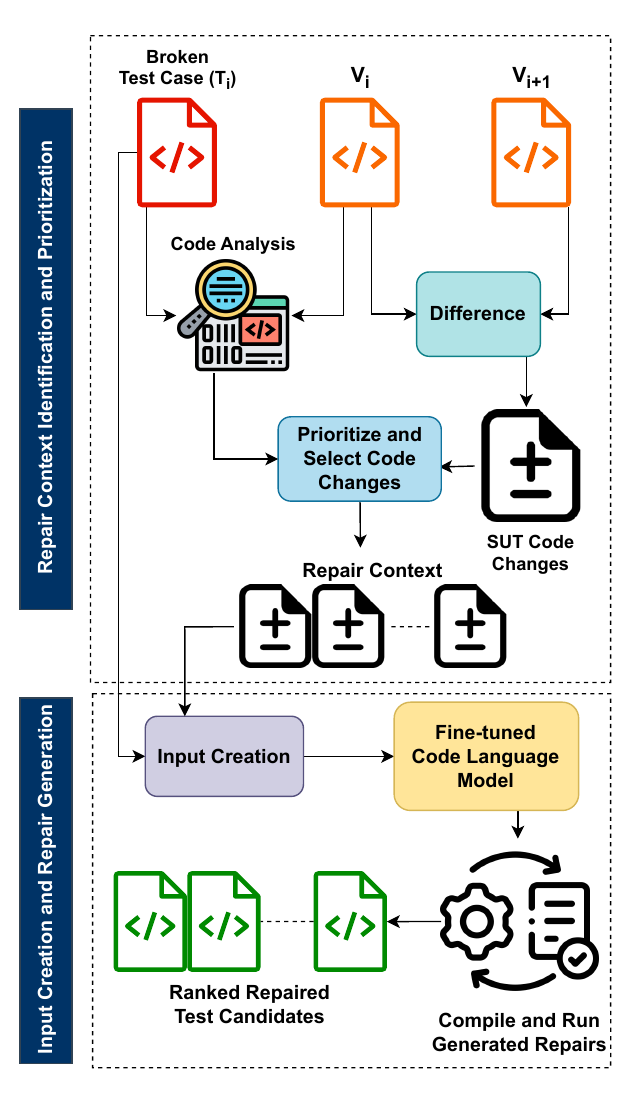}
    \caption{An overview of \approachName{} given a broken test case and two versions ($V_i$ and $V_{i+1}$) of the system under test (SUT).}
    \label{fig:approach}
\end{figure}

Figure \ref{fig:approach} presents an overview of \approachName{}, encompassing two high-level steps: (1) repair context identification and prioritization, and (2) application of CLMs to repair tests. 
The first step focuses on identifying and gathering essential data that characterize the breakage of tests. The second step uses the collected information from the first step to shape the inputs and outputs for utilizing a given CLM (fine-tuning and inference), specifically for test repair. It also addresses the generation of test repairs. In this section, we discuss the details of each step.

Figure~\ref{fig:repair-example} illustrates an example of test repair for the \textit{testDeposit} test case, testing the \textit{deposit} method within the \textit{BankAccount} class. Figures~\ref{fig:re-sut} and \ref{fig:re-test} display the original source code for \textit{deposit} and \textit{testDeposit}, respectively. Also, Figure~\ref{fig:re-changes} contains modifications to the \textit{BankAccount} class, introducing currency support. These changes include adding a new field---\textit{currency}---to the class, and updating the account balance based on the exchange rate during each deposit. These changes caused  \textit{testDeposit} to fail at lines 3 and 4 in Figure~\ref{fig:re-test}, indicating the breakage location. Consequently, as shown in Figure~\ref{fig:re-repair}, \textit{testDeposit} is repaired by including the currency value in both the \textit{BankAccount} constructor and the \textit{deposit} method. This example serves as a running example of test repair throughout the paper, and we will illustrate \approachName{} using this example whenever possible.

\begin{figure*}[tbp]
    \centering
    \begin{subfigure}{0.49\linewidth}
        \centering
        \begin{tcolorbox}[codebox]
        \begin{lstlisting}[language=Java,gobble=10,numbers=left]
            public class BankAccount {
            <@\quad@>private int balance = 0;
            <@\quad@>public int getBalance() { return balance; }
            <@\quad@>public int deposit(int amount) {
            <@\quad\quad@>balance += amount;
            <@\quad@>}             
            }
        \end{lstlisting}
        \end{tcolorbox}
        \subcaption{System Under Test ($V_i$)}
        \label{fig:re-sut}
    \end{subfigure}
    \begin{subfigure}{0.49\linewidth}
        \centering
        \begin{tcolorbox}[codebox]
        \begin{lstlisting}[language=Java,gobble=10,numbers=left]
            @Test
            void testDeposit() {
            <@\quad@>Bank account = new BankAccount();
            <@\quad@>account.deposit(500);
            <@\quad@>assertEquals(500, account.getBalance());
            }
        \end{lstlisting}
        \end{tcolorbox}
        \subcaption{Test Case ($T_i$)}
        \label{fig:re-test}
    \end{subfigure}

    \vspace{6pt}

    \begin{subfigure}{0.49\linewidth}
        \centering
        \begin{tcolorbox}[codebox]
        \begin{lstlisting}[language=Java,gobble=10,numbers=left,basicstyle=\tiny\ttfamily,numberstyle=\tiny,style=codeHighlighting]
            public class BankAccount {
            <@\quad@>private int balance = 0;
            <@\quad@>@@+ private final String currency;@@
            <@\quad@>@@+ public BankAccount(String currency) { this.currency = currency; }@@
            <@\quad@>public int getBalance() { return balance; }
            <@\quad@>&- public int deposit(int amount) {&
            <@\quad\quad@>&- balance += amount;&
            <@\quad@>`+ public int deposit(int amount`@@, String depositCurrency @@`) {`
            <@\quad\quad@>`+ balance += amount `@@* Exchange.getRate(depositCurrency, currency)@@`;`
            <@\quad@>}        
            }
        \end{lstlisting}
        \end{tcolorbox}
        \subcaption{Code changes resulting in $V_{i+1}$}
        \label{fig:re-changes}
    \end{subfigure}
    \begin{subfigure}{0.49\linewidth}
        \centering
        \begin{tcolorbox}[codebox]
        \begin{lstlisting}[language=Java,gobble=10,numbers=left,style=codeHighlighting]
            @Test
            void testDeposit() {
            <@\quad@>&- Bank account = new BankAccount();&
            <@\quad@>&- account.deposit(500);&
            <@\quad@>`+ Bank account = new BankAccount(`@@"USD"@@`);`
            <@\quad@>`+ account.deposit(500`@@, "USD"@@`);`
            <@\quad@>assertEquals(500, account.getBalance());
            }
        \end{lstlisting}
        \end{tcolorbox}
        \subcaption{Test Case Repair resulting in $T_{i+1}$}
        \label{fig:re-repair}
    \end{subfigure}
    \caption{Test case repair example}
    \label{fig:repair-example}
\end{figure*}

\TRR{C1.1}{Admittedly, an ideal automated test repair solution should address both breakage localization and repair, which are both challenging tasks. As shown in Section~\ref{breakagelines}, breakage localization, an active area of research often termed "fault localization", is not straightforward. \approachName{} contributes to a comprehensive automated solution, by effectively addressing the automated repair problem.}

\TRRC{Even when knowing the breakage lines, determining the appropriate repair is often complex. In the example shown in Figure~\ref{fig:compile_example}, although the breakage line (line 9) is given, developers face multiple repair options, necessitating careful consideration. Furthermore, our analysis of \benchmarkName{} in Section~\ref{rq2.1} highlights cases with a large search space for potential repairs. Finally, \approachName{}, regardless of repair complexity, provides automation which reduces the time required to interpret breakage lines, test code, and SUT code, alleviating the repetitive and tedious nature of the task.}

\subsection{Repair Context Identification and Prioritization} \label{repair-context}
Merely examining the test case code and breakage location without considering the changes in the SUT is insufficient to repair a broken test. It is essential to examine all changes made to the SUT, and detect the ones that led to the test case breakage. We refer to these changes in the SUT source code as the \textit{repair context}. Adding additional context is also a common practice when language models are applied to automated program repair, where the context will consist of code related to the faulty system code~\cite{zhang2023survey}.

While we employ a comprehensive approach to gather all relevant information for the repair context, it is important to note that current CLMs come with limitations regarding input size, also known as \textit{context window}. This necessitates a thoughtful selection of the repair context for input. Therefore, we introduce prioritization techniques to ensure that CLMs receive the most crucial information. 

\TR{R1.5}{While test repair context identification and fault localization share similarities, a critical distinction lies in the latter's reliance on oracles; fault localization depends on test suites to identify faults, whereas our context lacks a reliable oracle. Moreover, fault localization aims to pinpoint faulty statements, while our approach focuses on identifying correct code changes that cause obsolete test cases to fail. Given these differences, we opted for simple and widely validated heuristics based on static code analysis and text similarity to identify repair contexts, which we describe in this section.}

\TR{R1.2}{We recognize that SUT code changes alone may sometimes be insufficient to repair a test case. However, since SUT changes are the root cause of the test case breakage, they inherently contain useful information and are relevant. The extent to which additional repair context beyond SUT changes is required remains uncertain. Furthermore, the unchanged project-level context is significantly larger than the changed context, making the identification and selection of the most relevant context a challenging task. While this issue lies beyond the scope of our study, it represents a promising avenue for future research.}

In the following sections, we shed light on the details of \approachName{} for repair context identification and prioritization.

\subsubsection{Repair Context Identification}
\label{sec:repair-context}
First, we automatically identify the information linked with the repair context of a broken test case by following the steps below,  assuming (1) a broken test case $T_i$ is caused by changes made to the SUT from version $V_i$ to $V_{i+1}$, and (2) a \textit{hunk} defines a specific section within a source code file, consisting of one or more adjacent lines that have been deleted, added, or changed.

\textbf{Identifying Method- and Class-Level Hunks.}
\TR{R1.2 R3.3}{
We identify method-level hunks, denoted as $M$, by comparing the code of methods within the same class between $V_i$ and $V_{i+1}$. Changes that occur within the class but outside of methods are identified as class-level hunks, denoted as $C$. When comparing source files, even if a change spans multiple methods, we treat it as separate hunks, each corresponding to a single method. According to this definition, class-level hunks do not intersect with method-level hunks ($M \cap C = \emptyset$). We identify methods using their fully qualified names, which include the package name, class name, method name, and argument types. If a method's name changes, we match it with the most similar fully qualified name within the same class, using the \textit{Jaro–Winkler} string distance metric~\cite{cohen2003jarowinkler} to measure similarity. This measure is effective for matching strings with minor variations, and it emphasizes similarities at the beginning of the string, making it ideal for minor method name changes or method signatures with argument changes at the end of the string.}

\TRC{In the context of the running example (Figure~\ref{fig:re-changes}), we have identified two hunks within the \textit{BankAccount} class: (1) $h_1$, which includes two added lines (lines 3 and 4), and (2) $h_2$, which consists of two deleted lines (lines 6 and 7) and two added lines (lines 8 and 9). The method-level hunk set of the running example is $M = \{h_2\}$, and the class-level hunk set is $C = \{h_1\}$.}

\textbf{Generating Method and Class Level Call Graphs.} 
We use the Spoon library~\cite{spoon} to conduct static code analysis on $V_i$, generating two call graphs, namely method-level ($G_m$) and class-level ($G_c$) call graphs, for $T_i$. $G_m$ encompasses all SUT methods or constructors that are directly or indirectly invoked by $T_i$. In this graph, $T_i$ acts as the root, and each method invocation forms an edge that connects the calling method to the called method. In the context of the running example, the root node of $G_m$ is \textit{testDeposit()} method. This root node directly connects to three nodes: \textit{BankAccount.deposit(int)}, \textit{BankAccount()}, and \textit{BankAccount.getBalance()}.

Similarly, $G_c$ encompasses all classes that are directly or indirectly used by $T_i$. Similar to $G_m$, $T_i$ is the root of $G_c$. However, in $G_c$, non-root nodes represent classes, and each usage of a class (i.e., calling a method or a constructor of the class) is considered as an edge connecting the using node to the used class. The $G_c$ of the running example only has the test as the root node connected to the \textit{BankAccount} class as the non-root node.

\textbf{Creating the Repair Context Sets.}
\TR{R1.2 R3.3}{
We create three sets of hunks to serve as the repair context for our approach: $R$, $R_m$, and $R_c$. 
$R$ includes all changes in the SUT, including both method-level and class-level hunks, and is defined as:
\[R = M \cup C\]}
\TRC{$R_m \subset R$ is a set of method-level hunks, where each hunk corresponds to a method that is directly or indirectly called by $T_i$, as indicated by $G_m$. It is defined as:
\[R_m = \{h \in M\,|\,h \text{ is covered by } T_i \text{ in } G_m\}\]}
\TRC{$R_c \subset R$ is a set of both method-level and class-level hunks that are within classes covered by $T_i$, as indicated by $G_c$. However, $R_c$ specifically excludes any hunks already in $R_m$. It is defined as:
\[R_c = \{h \in (M \cup C)\,|\, h \notin R_m \wedge h \text{ is covered by } T_i \text{ in } G_c\}\]}

In the context of the running example (Figure~\ref{fig:re-changes}), and based on the two hunks identified earlier, $R = \{h_1, h_2\}$. Since the \textit{BankAccount.deposit(int)} method is included in the $G_m$ of \textit{testDeposit()}, $R_m = \{h_2\}$, as $h_2$ corresponds to the \textit{deposit} method. Furthermore, since the \textit{BankAccount} class is part of the $G_c$ of \textit{testDeposit()}, $R_c = \{h_1\}$, as $h_1$ is within that class and is not part of $R_m$.

\TRR{C3.1}{We utilize the repair context sets ($R$, $R_m$, and $R_c$) and the call graphs ($G_m$ and $G_c$) to prioritize and select the hunks most relevant to the repair task. These elements enable us to identify the structural code relationships between the test cases and the SUT. Further details on the hunk prioritization process are provided in the following section.}

\subsubsection{Repair Context Prioritization}
\label{sec:hunkpriority}
As discussed, CLMs have limitations on their input size, which entail the careful selection of the repair context ($R$) as input. In certain scenarios, the token size of $R$ may exceed the input size limit. Also, CLMs may produce different outputs based on the order of the repair context provided in the input. Since exploring all possible orderings is computationally infeasible, we prioritize placing the information that we consider most important at the beginning of the input sequence. Therefore, we employ a prioritization strategy for the hunks in $R$ to select the top $k$ most relevant hunks ($1 \leq k < |R|$) and create $R^\prime$. We define the hunk prioritization criteria in the following.

\textbf{Call Graph Depth.}
To determine the call graph depth ($d$) for a given hunk \TR{R1.2}{$h \in (R_m \cup R_c)$}, we examine its associated node $n$ within the call graph $G$ ($G_m$ or $G_c$) related to the broken test case $T_i$. The value of $d$ is calculated by counting the number of edges between the root node of $G$ and $n$. When $d$ equals 1, it shows that the code within $h$ is directly invoked by $T_i$ and any modifications in $h$ are highly relevant to $T_i$. \TR{R3.4}{Thus, lower call graph depth values indicate a closer relationship and are given higher priority, with $1$ being the best value (direct relationship) for $d$. Higher values refer to indirect relationships, making them less relevant.} For the running example, both $h_1$ and $h_2$ have a $d$ value of 1. This corresponds to the \textit{deposit} method and the \textit{BankAccount} class, both directly accessed from the \textit{testDeposit()} test case.

\textbf{Context Type.}
\TR{R1.2}{The context type ($CT$) for $h$ determines whether $h \in R_m$ or $h \in R_c$. If $h$ belongs to $R_m$, $CT$ represents a \textit{method}; if $h$ belongs to $R_c$, $CT$ represents a \textit{class}.} When selecting hunks based on context types for effective test case repair, hunks with the \textit{method} context type offer greater potential value compared to those with the \textit{class} context type. This is because method-level relations are more fine-grained, reducing the selection of hunks that are irrelevant to the test code. Therefore, we assign a higher priority to hunks with a \textit{method} context type. In the running example, $h_2$ has a method context type, while $h_1$ has a class context type.

\textbf{TF-IDF Similarity.}
In addition to the structural associations observed between the test case and the SUT, there are textual similarities among variables, methods, classes, and literal values that are not captured by call graphs. Thus, these similarities can  play a crucial role in identifying the most relevant repair context. We measure text similarity by comparing the changed lines of the hunks in the SUT against (1) $L$, the breakage lines of $T_i$ (\textit{Breakage TF-IDF Similarity}), or (2) the complete code of $T_i$ (\textit{Test TF-IDF Similarity}), using the TF-IDF (term frequency-inverse document frequency) algorithm \cite{tfidf}. First, we use the tokenizer of the code language model to convert the code to tokenized vectors. This tokenization process produces $\left|R\right|$ documents, each corresponding to a hunk $h \in R$, alongside an additional document for the selected lines of $T_i$ (whether breakage lines or all lines). Then, for each document, TF-IDF creates a normalized vector of vocabulary size (the count of unique tokens) by evaluating the significance of the document's individual tokens. This evaluation is based on token frequency across all documents and its occurrence within individual documents. We compute the cosine similarity between the TF-IDF vector of each $h \in R$ and the selected lines of $T_i$, prioritizing the hunks based on the highest similarity to the lowest.

\subsection{Input Creation and Repair Generation}

\begin{table*}[tbp]
\centering
\resizebox{\linewidth}{!}{%
\begin{tabular}{llllll}
\toprule
\multicolumn{2}{c}{Input-output Format ($IO$)}      & Repair Context & Hunk Prioritization                                          & Hunk Represenation & Output        \\
\midrule
$IO_1$ & Base                                & $R_m \cup R_c$ & CGD, CT, BrkSim      & Line-level         & Code Sequence \\
$IO_2$ & Text-based Similarity Hunk Ordering & $R$            & BrkSim, Rep, TSim    & Line-level         & Code Sequence \\
$IO_3$ & Word-level Hunk Representation      & $R$            & BrkSim, Rep, TSim    & Word-level         & Code Sequence \\
$IO_4$ & Edit Sequence Output                & $R$            & BrkSim, Rep, TSim    & Word-level         & Edit Sequence \\
\bottomrule
\end{tabular}%
}
\caption{Overview of input and output formats ($IO$s) in test repair generation models. Call Graph Depth (CGD), Context Type (CT), Breakage TF-IDF Similarity (BrkSim), Repetition (Rep), and Test TF-IDF Similarity (TSim) are the characteristics used for hunk prioritization. For detailed definitions of these characteristics and others, please refer to Section~\ref{sec:approach}.}
\label{tab:io}
\end{table*}

At this stage, we have gathered all the necessary information required to create the input and the output to fine-tune a CLM and generate repairs for broken tests. There is no single way for shaping the input and output format ($IO$) for CLMs and the selection of a suitable $IO$ for CLMs necessitates experimental analysis. Thus, in this work, we explore four different approaches to shape the $IO$ of test case repairs. An overview of the characteristics of the $IO$s is shown in Table~\ref{tab:io}. In the following, we first provide details for the $IO$s that we explored, explaining the input's repair context, the hunk prioritization and representation, and the output. Finally, we explain the generation process of test repairs.

\subsubsection{The Input}
\label{input}
For $IO_1$ (\textit{Base}), we select the repair context from $R_m \cup R_c$ and adopt three criteria sorted by their respective priorities: \textit{Call Graph Depth}, \textit{Context Type}, and \textit{Breakage TF-IDF Similarity}. As discussed, call graph depth and context type specifically capture the structural code relations between the test code and the repair context, while TF-IDF similarity captures the text-based code relations. We integrated the three criteria to differentiate between hunks in cases where two hunks share the same value for one criterion. Specifically, when the call graph depth and context type of two hunks are equal, their prioritization is determined by their similarity criterion. 
We used this order under the assumption that structural code relations are consistently accessible, given the inherent nature of programming languages. However, text-based code relations depend on developers' coding styles and the use of meaningful names across various sections of the code. Consequently, structural relations generally serve as a more dependable indicator of associations between hunks and test code compared to text-based relations.

For $IO_2$ (\textit{Text-based Similarity Hunk Ordering}), $IO_3$ (\textit{Word-level Hunk Representation}), and $IO_4$ (\textit{Edit Sequence Output}), we adopt three different criteria sorted by their respective priorities: \textit{Breakage TF-IDF Similarity}, \textit{Repetition}, and \textit{Test TF-IDF Similarity}. This prioritization emphasizes text-based similarity between the test code and the repair context. The repetition criterion measures how frequently a hunk is repeated within the repair context. When there are multiple occurrences of the same change across the source code, it is even more likely to be valuable for repairing the broken test case. 

The repair context selection for $IO_2$, $IO_3$, and $IO_4$ is not limited to $R_m$ or $R_c$; instead, we select the hunks from $R$. We do this to mitigate potential imprecision in using call graphs generated through static analysis. Unlike code execution and dynamic analysis, static analysis can introduce imprecision in type or method inference as well as in pointer analysis. Additionally, we aim to investigate scenarios where SUT changes beyond the scope of call graphs are required for repairing the test.
\TR{R1.6}{Including all hunks may introduce irrelevant repair context. We mitigate this issue by prioritizing hunks based on their similarity to the test code. Additionally, removing crucial context can be more detrimental than including irrelevant context. Therefore, we experimented with a more inclusive repair context selection that involves all hunks.}

For all $IO$s, the input is created per each test case $T_i$ as a sequence consisting of two parts: the test context ($TC$) and the repair context ($RC$). Each part is prefixed with a special token. The input sequence $I$ is thus formatted as follows:
\begin{flalign*}
I =& 
\ \specialToken{testContext}\ TC
\ \specialToken{repairContext}\ RC
\end{flalign*}
The special tokens \specialToken{testContext} and \specialToken{repairContext} mark the start of their respective parts, indicating their position within the input. The $TC$ consists of the complete source code of the broken test method, with the breakage lines enclosed by the two special tokens \specialToken{breakage} and \specialToken{/breakage}. Meanwhile, the $RC$ consists of code hunks from $R^\prime$ (defined in Section~\ref{sec:hunkpriority}), arranged in order of priority as discussed above. In Figure~\ref{fig:io-input}, the lines 1-9 show the $TC$ input format, and lines 10-17 show the $RC$ for our running example.

\begin{figure}[tbp]
    \centering    
    \begin{subfigure}{0.99\linewidth}
        \begin{tcolorbox}[codebox]
        \begin{lstlisting}[language=Java,gobble=10,numbers=left]
            <@\textcolor{javagreen}{[<TESTCONTEXT>]}@>
            @Test
            public void test() {
            <@\quad@><@\textcolor{javagreen}{[<BREAKAGE>]}@>
            <@\quad@>BankAccount account = new BankAccount();
            <@\quad@>account.deposit(500);
            <@\quad@><@\textcolor{javagreen}{[</BREAKAGE>]}@>
            <@\quad@>assertEquals(500, account.getBalance());
            }
            <@\textcolor{javagreen}{[<REPAIRCONTEXT>]}@> 
            <@\textcolor{javagreen}{[<HUNK>]}@> 
            public int deposit(int amount <@\textcolor{javagreen}{[<ADD>]}@> , String depositCurrency <@\textcolor{javagreen}{[</ADD>]}@>) {
            <@\quad@>balance += amount <@\textcolor{javagreen}{[<ADD>]}@> * Exchange.getRate(depositCurrency, currency) <@\textcolor{javagreen}{[</ADD>]}@> ;
            } <@\textcolor{javagreen}{[</HUNK>]}@>
            <@\textcolor{javagreen}{[<HUNK>]}@> <@\textcolor{javagreen}{[<ADD>]}@> 
            private final String currency;
            public BankAccount(String currency) { this.currency = currency; } <@\textcolor{javagreen}{[</ADD>]}@> <@\textcolor{javagreen}{[</HUNK>]}@>
        \end{lstlisting}
        \end{tcolorbox}
        \subcaption{Input Format}
        \label{fig:io-input}
    \end{subfigure}

    \vspace{6pt}
    
    \begin{subfigure}{0.99\linewidth}
        \begin{tcolorbox}[codebox]
        \begin{lstlisting}[gobble=12,numbers=none,xleftmargin=0pt]
            BankAccount account = new BankAccount("USD");
            account.deposit(500, "USD");
        \end{lstlisting}
        \end{tcolorbox}
        \subcaption{Output Format}
        \label{fig:io-output}
    \end{subfigure}
    \caption{Example input format with word-level hunk representation and the expected code sequence output.}
    \label{fig:input-output}
\end{figure}

The $TC$ part is consistent across all $IO$s, while the hunk representation in the $RC$ varies. For $IO_1$ and $IO_2$, we use a line-level hunk representation, whereas for $IO_3$ and $IO_4$, we use a word-level representation. In both line-level and word-level cases, the hunk is enclosed with the special tokens \specialToken{hunk} and \specialToken{/hunk}.

For line-level representation, the sequence starts with \specialToken{del}, listing all deleted lines of the hunk, followed by \specialToken{add}, listing all added lines. If a hunk solely comprises deleted or added lines, only the existing set, along with its corresponding special token, is included. Figure~\ref{fig:line-level-hunk} illustrates the line-level representation of $h_2$.

\begin{figure}[tbp]
    \centering
    \begin{subfigure}{\linewidth}
        \begin{tcolorbox}[codebox]
        \begin{lstlisting}[language=Java,gobble=12,numbers=none,xleftmargin=0pt]
            <@\textcolor{javagreen}{[<HUNK>]}@> <@\textcolor{javagreen}{[<DEL>]}@> 
            public int deposit(int amount) { 
            <@\quad@>balance += amount;
            <@\textcolor{javagreen}{[<ADD>]}@> 
            public int deposit(int amount, String depositCurrency) {
            <@\quad@> balance += amount * Exchange.getRate(depositCurrency, currency); <@\textcolor{javagreen}{[</HUNK>]}@>
        \end{lstlisting}
        \end{tcolorbox}
    \end{subfigure}
    \caption{Example of the line-level representation of a code hunk ($h_2$ in our running example) in the input.}
    \label{fig:line-level-hunk}
\end{figure}

In the line-level representation, deleted and added lines often share many tokens, causing redundancy in the input and requiring the CLM to discern changed tokens. In contrast, the word-level representation combines deleted and added lines, enclosing only the changed words (tokens) with special tokens. This representation uses special tokens to enclose deleted words within \specialToken{del} and \specialToken{/del} and added words within \specialToken{add} and \specialToken{/add} for each group of changed words. Lines 11-14 and lines 15-17 in Figure~\ref{fig:io-input} show the word-level representations of hunks $h_2$ and $h_1$, respectively.

\subsubsection{The Output}\label{output}
For $IO_1$, $IO_2$, and $IO_3$, the code sequence of the repaired lines, derived from applying changes to the breakage lines, constitute the expected output of the CLM. Figure~\ref{fig:io-output} shows the code sequence output of our running example. Figure~\ref{fig:io-output} is a code transformation of lines 5 and 6 in Figure~\ref{fig:io-input}.

For $IO_4$, we use the edit sequence encoding as the output. This is a series of replacements that, when applied to the breakage lines of code, results in the repaired lines, and is derived from the unambiguous edit sequence formulation proposed by Zhang et al.~\cite{zhang2023multilingual}. They found that modeling code changes as edit sequences significantly outperforms state-of-the-art approaches for the multilingual code co-evolution task. Given the similarity between the task they addressed and our test case repair task in this study, we decided to experiment with edit sequences. \TR{R3.7}{This output format includes solely the edit sequence, which is then applied to the test breakage lines to create the repair candidate.} The key component of this formulation is selecting a sequence of tokens to be replaced from the breakage lines, such that the selected sequence is unique within the breakage lines. This ensures that each replacement can be performed without needing additional location information. \TR{R3.7}{An edit sequence is assumed invalid and discarded if the sequence of tokens to be replaced is not uniquely identifiable, as this introduces ambiguity.} Each replacement $E$ in an edit sequence with the following format:
\begin{flalign*}
E =& 
\ \specialTokenL{replaceOld}\ s
\ \specialTokenL{replaceNew}\ n
\ \specialTokenL{replaceEnd}\
\end{flalign*}

The special tokens \specialTokenL{replaceOld}, \specialTokenL{replaceNew}, and \specialTokenL{replaceEnd} define the replacement, indicating its components. Since an edit sequence can consist of multiple replacements, \specialTokenL{replaceEnd} is used to indicate that the preceding replacement is completely defined. Component $s$ indicates the target token sequence to be replaced, which must be uniquely identifiable within the breakage lines of code. Component $n$ indicates the token sequence that $s$ is to be replaced with. A simple example of an edit sequence encoding can be seen in Figure~\ref{fig:edit-seq-base}.

\begin{figure}[tbp]
    \centering
    \begin{subfigure}{\linewidth}
        \centering
        \small Ground Truth Test Repair
        \begin{tcolorbox}[codebox]
        \begin{lstlisting}[language=Java,gobble=10,numbers=left,style=codeHighlighting]
            &- account.deposit (&^amount^&, "EUR" ) ;&
            `+ account.deposit (`@@new Money ( amount )@@`, "EUR" ) ;`
        \end{lstlisting}
        \end{tcolorbox}
        \small Edit Sequence Output
        \begin{tcolorbox}[codebox]
        \begin{lstlisting}[gobble=12]
            <@\textcolor{javagreen}{[<replaceOld>]}@> amount 
            <@\textcolor{javagreen}{[<replaceNew>]}@> new Money ( amount ) 
            <@\textcolor{javagreen}{[<replaceEnd>]}@>
        \end{lstlisting}
        \end{tcolorbox}
        \caption{A simple edit sequence where the changed tokens are uniquely identifiable.}
        \label{fig:edit-seq-base}
    \end{subfigure}
    \par\bigskip
    \begin{subfigure}{\linewidth}
        \centering
        \small Ground Truth Test Repair
        \begin{tcolorbox}[codebox]
        \begin{lstlisting}[language=Java,gobble=10,numbers=left,style=codeHighlighting]
            &- &^BankAccount^& account = new &^BankAccount^&( ) ;&
            `+ `@@ChequingAccount@@` account = new `@@ChequingAccount@@`( ) ;`
            `+ `@@account.setCurrency ( "USD" ) ;@@
        \end{lstlisting}
        \end{tcolorbox}
        \small Edit Sequence Output
        \begin{tcolorbox}[codebox]
        \begin{lstlisting}[gobble=10]
            <@\textcolor{javagreen}{[<replaceOldKeepAfter>]}@> BankAccount account 
            <@\textcolor{javagreen}{[<replaceNewKeepAfter>]}@> ChequingAccount account
            <@\textcolor{javagreen}{[<replaceEnd>]}@>
            <@\textcolor{javagreen}{[<replaceOldKeepBefore>]}@> new BankAccount 
            <@\textcolor{javagreen}{[<replaceNewKeepBefore>]}@> new ChequingAccount 
            <@\textcolor{javagreen}{[<replaceEnd>]}@>
            <@\textcolor{javagreen}{[<replaceOldKeepBefore>]}@> ; 
            <@\textcolor{javagreen}{[<replaceNewKeepBefore>]}@> ;
            account.setCurrency ( "USD" ) ;
            <@\textcolor{javagreen}{[<replaceEnd>]}@>
        \end{lstlisting}
        \end{tcolorbox}
        \caption{\TRC{A complex edit sequence requiring extra tokens for unique identification, also showing the handling of new lines.}}
        \label{fig:edit-seq-keep-before}
    \end{subfigure}
    \caption{Edit sequence output encoding examples}
    \label{fig:edit-seq-example}
\end{figure}

On top of \specialTokenL{replaceOld} and \specialTokenL{replaceNew}, we used additional pairs of special tokens to give hints about the replacement when the sequence of changed tokens is not uniquely identifiable within the breakage lines of code. To find a unique $s$ within the breakage lines, additional tokens may need to be prepended and/or appended to the changed code. In these cases, $s$ and $n$ will have a shared set of tokens that remain the same throughout the edit sequence, and the special replace tokens indicate the relative position of the shared tokens. An example of this scenario can be seen in Figure~\ref{fig:edit-seq-keep-before}. 

\TR{R1.3 R3.7}{The example displays a code change where the \textit{account} variable's type is changed to \textit{ChequingAccount}. This requires updating the \textit{BankAccount} text in two places, necessitating a clear definition of where each replacement should occur. As each replacement must identify a unique sequence of tokens, both instances of replacing the \textit{BankAccount} token require additional tokens to ensure uniqueness. Since a replacement specifying \textit{\specialTokenL{replaceOld} BankAccount \specialTokenL{replaceNew} ChequingAccount \specialTokenL{replaceEnd}} would be ambiguous (as it could be applied to multiple locations and it is unclear which locations it should be applied to), it is considered to be invalid. Instead, the two instances of \textit{BankAccount} are each specified with a unique replacement. In the first instance, the \textit{account} token follows \textit{BankAccount} to form a uniquely identifiable sequence, using the special tokens \specialTokenL{replaceOldKeepAfter} and \specialTokenL{replaceNewKeepAfter} to indicate that a sub-sequence of the end part remains unchanged. This results in a replacement of  \textit{\specialTokenL{replaceOldKeepAfter} BankAccount account \specialTokenL{replaceNewKeepAfter} ChequingAccount account \specialTokenL{replaceEnd}}, which is a valid replacement since it can only be applied to one location in the test. Similarly, the second replacement of the \textit{BankAccount} token includes the \textit{new} token before \textit{BankAccount} to create a distinct sequence, using the special tokens \specialTokenL{replaceOldKeepBefore} and \specialTokenL{replaceNewKeepBefore} to signify that a sub-sequence of the beginning part remains unchanged. These special tokens also guide the insertion of new statements. Since new statements do not change any old tokens, the edit sequence replaces the end of the preceding statement (the semicolon) with the semicolon followed by the new line.}

\TRC{It should be noted that while a valid edit sequence can be constructed using only \specialTokenL{replaceOld} and \specialTokenL{replaceNew}, the other special tokens (such as \specialTokenL{replaceOldKeepBefore}) are used to delineate between location indicators and actual repairs. These special tokens are used with the goal of helping the model focus on the tokens which comprise the actual repair, rather than the tokens which are used to indicate the repair's location. As an example, \textit{\specialTokenL{replaceOldKeepAfter} BankAccount account \specialTokenL{replaceNewKeepAfter} ChequingAccount account \specialTokenL{replaceEnd}} is used over \textit{\specialTokenL{replaceOld} BankAccount account \specialTokenL{replaceNew} ChequingAccount account \specialTokenL{replaceEnd}} as the former specifies that the \textit{account} token is not significant to the repair and is only a location marker, whereas the latter can be interpreted as indicating that \textit{account} is an important part of the repair.}\vspace{1cm}

\subsubsection{\TRC{Input-Output Format Selection}}

\begin{table}[tbp]
    \resizebox{\linewidth}{!}{%
    \begin{tabular}{llll}
        \toprule
        \multicolumn{2}{c}{Parameter} & ID & Values\\
        \midrule
        \multirow{2}{*}{\textit{RC}} & \multirow{2}{*}{Repair Context} & \textit{$rc_1$} & Covered hunks ($R_m \cup R_c$)\\
        \cmidrule{3-4}
         & & \textit{$rc_2$} & All hunks ($R$)\\
        \midrule
        \multirow{2}{*}{\textit{HP}} & \multirow{2}{*}{Hunk Prioritization} & \textit{$hp_1$} & Sorts by CGD, CT, BrkSim\\
        \cmidrule{3-4}
         & & \textit{$hp_2$} & Sorts by BrkSim, Rep, TSim\\
        \midrule
        \multirow{2}{*}{\textit{HR}} & \multirow{2}{*}{Hunk Representation} & \textit{$hr_1$} & Line-level\\
        \cmidrule{3-4}
         & & \textit{$hr_2$} & Word-level\\
        \midrule
        \multirow{2}{*}{\textit{O}} & \multirow{2}{*}{Output} & \textit{$o_1$} & Code sequence\\
        \cmidrule{3-4}
         & & \textit{$o_2$} & Edit sequence\\
        \bottomrule
    \end{tabular}
    }
    \centering
    \caption{\TRC{The parameters used to define $IO$s in Table~\ref{tab:io}. An $IO$ is characterized by one value for each parameter.}}
    \label{table:inputParamOptions}
\end{table}

\TR{R1.4 R3.5}{In Sections~\ref{input} and \ref{output}, we described the four $IO$s selected for our experiments. As outlined in Table~\ref{tab:io}, each $IO$ is defined by four parameters, with each parameter having two possible values, as shown in Table~\ref{table:inputParamOptions}. In this section, we explain (1) the rationale behind selecting these four parameters and assigning two values to each, and (2) why, out of the 16 possible combinations, we chose these four $IO$s. For clarity, we refer to the values for each parameter using the IDs specified in Table~\ref{table:inputParamOptions}. For instance, using covered hunks as the repair context is denoted as $rc_1$. For example, $IO_1$ is defined by $(rc_1, hp_1, hr_1, o_1)$.
}

\TRC{Fine-tuning CLMs on 36k training instances is resource-intensive, particularly when it involves multiple iterations for different parameter value combinations. This makes it necessary to pre-select parameter value combinations that are likely to be effective. The performance of language models during fine-tuning depends significantly on the input context, the expected output, and the representation of this information. We carefully considered these aspects and identified four key parameters: the provided context ($rc$), the order of the context ($hp$), the representation of the context ($hr$), and the expected output ($o$). The rationale for the selection of each parameter is detailed in Sections \ref{input} and \ref{output}. Also, formatting inputs or outputs based on the selected parameter values is computationally efficient, since it is relying on static analysis and straightforward text processing, while leading to diverse formats.}

\TRC{Not all 16 parameter combinations are feasible. $(rc_2, hp_1)$ cannot be used together because some of the hunks in $R$, derived from using $rc_2$, are not covered by the test's call graph, making them impossible to prioritize with $hp_1$. This constraint excludes four combinations.
While using $(rc_1, hp_2)$ is technically feasible, its effectiveness is limited. $hp_2$ prioritizes hunks based on text similarity, which is designed to explore and prioritize all hunks effectively. However, using it to prioritize only the hunks covered by the test method's call graph contradicts its primary purpose, leading us to exclude combinations involving these two values. As a result, eight combinations remained to investigate.}

\TRC{Of these remaining eight, to reduce the enormous cost of our experiments, four were eliminated based on preliminary experimental results, which were done on a smaller dataset. For instance, $IO_2 (rc_2, hp_2, hr_1, o_1)$ showed improved performance over $IO_1 (rc_1, hp_1, hr_1, o_1)$, leading to the exclusion of the remaining formats that utilized $(rc_1, hp_1)$. Similarly, $IO_4 (rc_2, hp_2, hr_2, o_2)$ showed decreased performance compared to $IO_3 (rc_2, hp_2, hr_2, o_1)$, leading us to discard other formats utilizing $o_2$. Consequently, our experiments ended up covering four parameter value combinations.}

\subsubsection{Repair Generation}
We utilize CLMs with both encoder-decoder and decoder-only transformer architectures \cite{vaswani2017attention} that are pre-trained on code (e.g., CodeT5+~\cite{wang2023codet5p}). In both architectures, the decoder generates output tokens in an auto-regressive manner. To teach test repair patterns and repair generation to the model, we fine-tune CLMs using a test repair training set. We formulate the fine-tuning task as a neural machine translation task where the broken test case along with the required context are translated into the repaired test case. During fine-tuning, the model is presented with the inputs (broken test code and repair context) and their corresponding outputs (repaired test code). The model learns to minimize the difference between its predicted repairs and the ground truth repairs in the training data. Subsequently, we employ the fine-tuned model to generate repairs for broken test cases. Language models have a predefined vocabulary comprising various tokens. Therefore, the input source code goes through tokenization before being provided as input to the model, and the resulting outputs are also in token format.

As described in Section~\ref{sec:background}, we use the beam search ranking strategy to generate candidate test case repairs. Finally, we replace the breakage lines of the original test code with the candidate repair patches. We compile and run the repaired test cases against the new version of the SUT ($V_{i+1}$). By following this process, we identify the plausible test repairs, i.e., candidate test repairs that compile and pass. These plausible test repairs are then recommended as the outcome of \approachName{}.

\section{Study Design and Results}
\label{sec:study}
In this section, we present the experimental study to evaluate our automatic test case repair technique leveraging language models (\approachName{}) and answer the following research questions.

\textbf{RQ1.1} How do different combinations of pre-trained code language models (CLMs) and input formats perform in repairing test cases using \approachName{}?

\TR{R3.8}{As outlined in Section~\ref{sec:approach}, our methodology employs multiple input-output formats ($IOs$) to adapt CLMs for test case repair. RQ1.1 seeks to evaluate the effectiveness of different $IOs$ across three selected CLMs, with the goal of identifying the conditions under which our approach yields the best results. This evaluation is critical to understanding the adaptability and effectiveness of our approach in various contexts. Furthermore, the best configuration identified in RQ1.1 is subsequently applied in the remaining RQs to investigate and discuss other aspects of the approach.}

\TR{R1.12}{\textbf{RQ1.2} How does the best configuration of \approachName{} from RQ1.1 perform against baselines?}

\TR{R3.8}{RQ1.2 investigates whether our proposed approach, in its best configuration, offers any advantages over one relevant existing study and two simple baselines. The goal of this comparison is to thoroughly evaluate the effectiveness and practicality of our approach in relation to baseline methods, complementing its assessment in RQ1.1.}
    
\textbf{RQ2.1} How effective is \approachName{} at repairing test cases based on the test repair characteristics found in our benchmark (\benchmarkName{})?

\TR{R3.8}{Test breakages can arise from a variety of causes. RQ2.1 seeks to examine whether \approachName{}'s effectiveness varies depending on the specific type of test breakage being repaired. Understanding this is crucial for the practical application of our approach, as it allows us to pinpoint the scenarios where it performs optimally and highlights challenging areas to direct future work in addressing diverse test failure scenarios.}

\textbf{RQ2.2} Can we accurately predict, at the time of repair, the situations in which \approachName{} cannot effectively repair a test in practice?

\TR{R3.8}{Although a perfect solution for test repair may be achievable in the future, we have not currently achieved it yet. Consequently, being able to predict when \approachName{} will deliver highly accurate repairs and when it might not is essential for its effective use. This predictive capability would help practitioners avoid unnecessary effort by suggesting them to avoid low-quality repairs. RQ2.2 seeks to explore whether, for a given broken test case, we can predict the model's ability to generate correct repairs without executing the model itself.}
    
\textbf{RQ3.1} What is the impact of the amount of data available for fine-tuning on test repair performance?

\TR{R3.8}{CLMs require extensive datasets for effective pre-training. Although fine-tuning these models requires significantly less data, the quality and relevance of the data become critical, making the process of collecting a suitable fine-tuning dataset both resource-intensive and costly. Understanding the impact of the size of the fine-tuning dataset on model accuracy is important for performing a cost-benefit analysis. RQ3.1 specifically investigates how varying the size of the fine-tuning dataset impacts model performance.}

\textbf{RQ3.2} How does the performance of the fine-tuned model generalize to unseen projects?

\TR{R3.8}{RQ3.2 explores whether \approachName{} can effectively generalize to repair tests from projects whose data was not utilized during fine-tuning. Addressing this question will show whether \approachName{} can be applied to new projects, thereby reducing or potentially eliminating the need to gather project-specific historical data for fine-tuning.}

In the following, we describe the evaluation metrics we employ and introduce \benchmarkName{}. We then elaborate on the experimental process, present the results, and finally discuss their practical implications.

\subsection{Evaluation Metrics}
Based on existing studies on code generation tasks~\cite{zeng2022codelmstudy,lu2021codexglue,wang2021codet5}, we adopt four evaluation metrics to assess the performance of test case repair generation. Specifically, we rely on: exact match accuracy (EM), BLEU~\cite{papineni2002bleu}, CodeBLEU~\cite{ren2020codebleu}, and plausible repair accuracy (PR). These metrics evaluate the quality of generated repair candidates by comparing them with the ground truth and assessing their execution results. In the context of test case repair, we define each metric based on the assumption that our evaluation set ($E$) contains the broken test cases, for each $t \in E$, the ground truth repair equals to $gt(t)$, and for each $t \in E$, our model generates $k$ repair candidates, i.e., $C_t = \{c_1, c_2, \ldots, c_k\}$.

\textbf{Exact Match Accuracy (EM).}
The EM metric~\cite{hou2023llm4se}, also known as perfect accuracy, refers to the percentage of test cases that include at least one repair candidate with source code exactly matching the ground truth. The EM metric is defined as follows:
\begin{equation*}
\text{EM} = \frac{1}{|E|} \sum_{t \in E} \begin{cases} 1, & \text{if } gt(t) \in C_t \\ 0, & \text{otherwise} \end{cases}
\end{equation*}

While the EM metric is very strict, it fails to consider repair candidates that have similar semantics but differ in syntax. Therefore, we also adopted other metrics to provide a comprehensive evaluation.

\textbf{BLEU and CodeBLEU.}
The BLEU metric~\cite{papineni2002bleu} was originally designed for evaluating machine translation tasks. It measures the extent of overlapping n-grams between a candidate translation and a set of reference translations. However, BLEU falls short in evaluating code-related tasks as it fails to account for the tree structure and logic inherent in code. To address this limitation, Ren et al. introduced CodeBLEU~\cite{ren2020codebleu}, which extends BLEU by incorporating code-specific information into its measurements.
Both BLEU and CodeBLEU \TR{R3.16}{are} computed based on a single repair candidate per test case. Thus, for each $t \in E$, we choose the optimal repair candidate from the available $k$ candidates. The best repair candidate is determined either by an exact match with the ground truth or, if no exact match exists, by selecting the candidate with the highest score in the beam search ranking. It is worth noting that we use the BLEU-4 variation, which is widely adopted and is the geometric mean of all n-gram precision values up until 4-gram~\cite{papineni2002bleu}. Throughout the remainder of this paper, any references to BLEU will be referring to this variant.

\textbf{Plausible Repair Accuracy (PR).}
\TR{R1.15}{The PR metric~\cite{zhang2023survey} evaluates the effectiveness of test case repairs by evaluating the percentage of test cases that contain at least one repair candidate that compiles successfully and passes on the updated SUT version.  In other words, a repair is considered successful if it avoids both compile-time and runtime errors, and the test execution logs confirm a pass. This metric offers a practical assessment of the syntactic and, to some extent, semantic correctness of the repair by validating it within the real execution environment.} The PR metric is defined as follows:
\begin{equation*}
\text{PR} = \frac{1}{|E|} \sum_{t \in E} \begin{cases} 1, & \exists c \in C_t : c \text{ compiles and passes} \\ 0, & \text{otherwise} \end{cases}
\end{equation*}

\subsection{Benchmark: \benchmarkName{}}
\label{benchmark}
Current research on test repair lacks a public, sufficiently large and diverse benchmark for evaluating and comparing their methods. In short, existing test repair benchmarks suffer from their small size, limited scope regarding repair categories, and their lack of reproducibility and reusability. We discuss these limitations in great detail in Section~\ref{sec:related-work}. To enable high-quality research on test repair, we have undertaken the major initiative to construct a large, high-quality test repair benchmark, which we name \benchmarkName{}~\cite{tarbenchFigshare}.

\begin{figure*}[tb]
    \centering
    \includegraphics[width=0.95\textwidth]{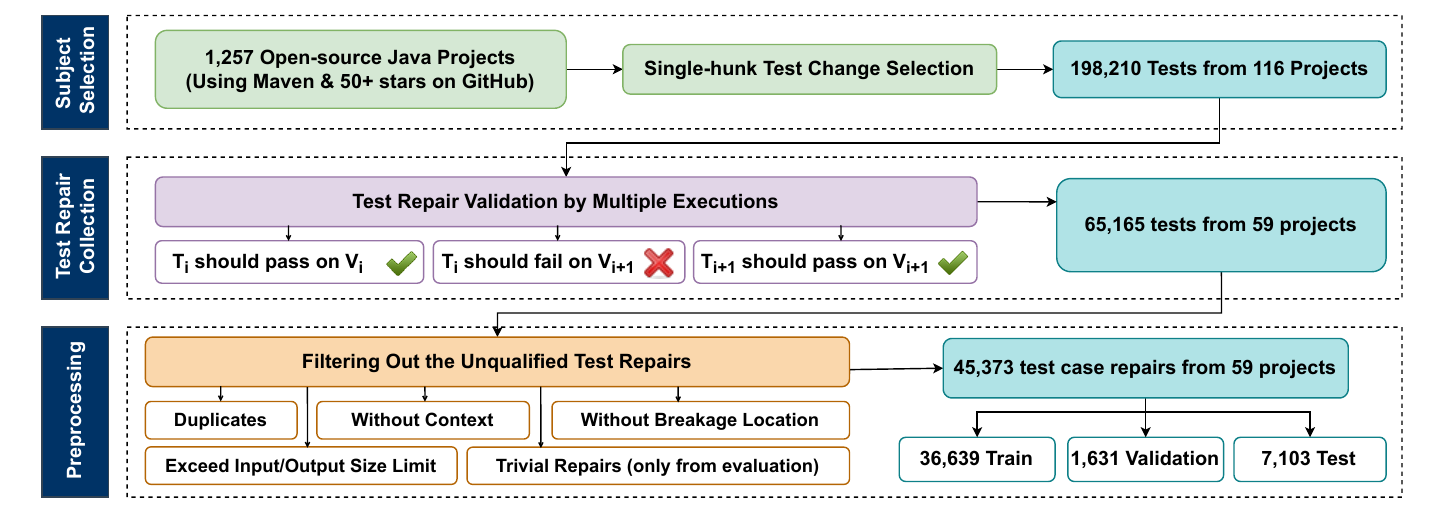}
    \caption{Overview of the creation steps of our test case repair benchmark (\benchmarkName{}).}
    \label{fig:repair-collection}
\end{figure*}

To create the dataset used in this study, we followed a multi-step process, which involved the selection of subjects, test repair collection, data preprocessing, and data splitting. An overview of the dataset creation steps is shown in Figure~\ref{fig:repair-collection}. We provide the details for each step in this section.

\subsubsection{Subject Selection}
\label{benchmark_subjects}
To initiate the dataset creation process for this study, we began by following these steps to select the study's subjects.

\textbf{Selecting Projects.}
We began by using the Java projects from CodeSearchNet~\cite{husain2020codesearchnet}, which provided us with 4,662 Java projects. CodeSearchNet's projects, while used for code retrieval tasks, were selected from a comprehensive search of open-source projects to meet quality criteria, thus forming a good initial set for our purpose.

Next, we fetched the projects from GitHub and selected the ones using Maven with a minimum of 50 stars. GitHub hosts a vast number of repositories, and many of them may not be actively maintained or may not have a clear purpose. Therefore, filtering by stars helps \TR{R3.16}{to} reduce this kind of noise. This step led to a selection of 1,257 projects.

\textbf{Selecting Test Cases.}  
\TR{R1.7}{We analyzed each commit, $C$, to identify code changes within unit test methods across all projects. Recall from Section~\ref{sec:approach} that we defined  $V_i$ and $V_{i+1}$ to represent two consecutive versions of the SUT, where $T_i$ is a broken test case in $V_i$, and $T_{i+1}$ is its repaired version in $V_{i+1}$. Assuming $V_{i+1}$ corresponds to the code at commit $C$ and $V_i$ to $C$'s parent, we examined the code changes in $C$. We then selected all available $(T_i, T_{i+1})$ pairs in $C$ as potential test case repairs, and automatically matched $T_i$ with $T_{i+1}$ using their fully qualified names, as outlined in Section~\ref{sec:repair-context}.}
We selected test method changes with only one hunk (single-hunk method changes), where the code change is confined to a single chunk of code. This step led to the selection of 116 projects with 198,210 test method changes.

\TR{R1.2}{
Our motivation for selecting single-hunk test method changes (potential test repairs) is to facilitate the learning process of CLMs. Single-hunk changes represent isolated modifications that can be easily associated with specific intentions, providing clarity and focus for learning. This approach is also a common practice in most studies~\cite{huang2023aprsurvey, ye2022neural, zhu2021syntax, li2020dlfix, chen2019sequencer}, especially in the program repair domain.
Note that single-hunk test repairs are not necessarily simple. For example, a multi-hunk test repair might involve a simple renaming of a method repeated multiple times across the test method. In contrast, a single-hunk test repair might involve adding a zero at the end of a literal integer, which requires a deep understanding of the SUT. Therefore, the number of hunks in a test repair does not necessarily correlate with its complexity.}

\TRC{While our focus is on single-hunk test repairs, our approach can be adapted to handle multiple hunks by either iteratively addressing each one or repairing them in one step by marking multiple breakage locations in the input. However, extending support to multi-hunk repairs requires additional effort to modify our single-hunk technique and gather relevant multi-hunk data. We believe that expanding to multi-hunk repairs is a valuable direction for future research.}

\TR{R3.12}{Note that when collecting test repairs from projects, we focus on the default branch rather than examining every branch. We begin with the most recent commit on the default branch and work backwards through its commit history, inspecting all reachable commits. This approach is based on the observation that most projects maintain a primary branch-often named \textit{``main''} or \textit{``master''}—into which most changes are merged after successful code reviews and CI/CD checks. Thus, changes in the default branch have typically undergone thorough review processes and automated testing, ensuring their reliability.}

\subsubsection{Test Case Repair Collection}
\label{sec:testRepairCol}
Not all test method changes are valid test repairs since they can be test refactorings or unsuccessful repair attempts. Following the selection of projects, we proceeded to analyze the commits associated with the test method changes to identify valid test repairs. To determine whether a test method change is a test repair and to ensure the quality of our data, we conducted three test executions
\footnote{To execute the test cases, we use Maven version 3.6.3 and Java Development Kit (JDK) versions 1.8.0\_192, 11.0.16\_8, or 17.0.2, depending on the compiler version specified in the project's pom.xml file.}.

\TR{R1.8}{To determine the result of each test execution, we analyzed the Maven logs and categorized executions as either valid or invalid. Valid executions are those where the test passes or fails due to issues within the test code itself. Invalid executions result from factors unrelated to the test code. We identified valid executions based on the three Maven log patterns outlined in Table~\ref{tab:valScenarios}, which cover successful executions and failures due to compilation or runtime errors within the test code. Any execution log that does not match these patterns is considered an invalid execution and is discarded. Examples of such invalid scenarios include dependency resolution failures and compilation or runtime errors external to the test code.}
\begin{table}[tbp]
\centering
\resizebox{0.95\linewidth}{!}{
\begin{tabular}{l}
\toprule
\textbf{Valid Text Execution Scenarios} \\
\midrule
\textbf{1. Successful test case execution} \\
\small \textbf{Pattern:} \\
\footnotesize \texttt{"Tests run: 1, Failures: 0, Errors: 0, Skipped: 0"} \\
\small \textbf{Example:} \\
\footnotesize \texttt{[INFO] Running TestClass.testMethod} \\
\footnotesize \texttt{[INFO] Tests run: 1, Failures: 0, Errors: 0, Skipped: 0} \\
\footnotesize \texttt{[INFO] BUILD SUCCESS} \\
\midrule
\textbf{2. Compilation error within the test case} \\
\small \textbf{Pattern:} \\
\footnotesize \texttt{"COMPILATION ERROR"} \textbf{or} \texttt{"Compilation failure"} \\
\footnotesize \textbf{And} \texttt{"[ERROR] /.+/<test\_rel\_path>:[(\textbackslash d+),\textbackslash d+].*"} \\
\small \textbf{Example:} \\
\footnotesize \texttt{[ERROR] COMPILATION ERROR :} \\
\footnotesize \texttt{[ERROR] /path/to/TestClass.java:[25,13] cannot find symbol} \\
\midrule
\textbf{3. Runtime failure within the test case} \\
\small \textbf{Pattern:} \\
\footnotesize \texttt{"[ERROR] <test\_class>.<test\_method>:(\textbackslash d+)"} \textbf{or} \\
\footnotesize \texttt{"at .+<test\_class>.<test\_method>(<test\_class>.java:(\textbackslash d+))"} \\
\small \textbf{Example:} \\
\footnotesize \texttt{[ERROR] TestClass.testMethod:15 NullPointerException} \\
\footnotesize \texttt{[ERROR] at TestClass.testMethod(TestClass.java:15)} \\
\bottomrule
\end{tabular}
}
\caption{\TRC{Valid test execution scenarios identified by Maven log patterns, with examples showing successful execution, compilation errors, and runtime failures in test case code.}}
\label{tab:valScenarios}
\end{table}

We conducted the following three types of test executions to identify test repairs among the test method changes.

\textbf{The Test Case Initially Passes.} We executed $T_i$ and verified its successful execution against $V_i$. In the event of $T_i$ failing, we excluded the test, because if $T_i$ is already failing before any change to the SUT, even if the commit repairs the test, we are unable to extract the necessary repair context for our approach to function effectively. This step resulted in the exclusion of 91,059 tests.

\TR{R3.10}{We conducted a detailed analysis of the test execution results during this initial phase. The failing executions were categorized into four groups: (1) compile or runtime errors from the SUT, accounting for 23\% of cases; (2) compile or runtime errors from the test cases, representing 16\%; (3) dependency errors related to libraries or plugins, despite using Maven and the correct Java version, which made up 33\%; and (4) unknown failures, where the cause could not be automatically identified from the execution logs, contributing 28\%. Notably, only 16\% of the failures were directly linked to issues with the test cases. The majority were caused by environment-related problems or faults in the SUT, potentially due to ongoing development.}
    
\textbf{SUT Changes Break the Original Test Case.} We subsequently executed $T_i$ against $V_{i+1}$ and assessed whether it failed specifically due to problems in the test code. In cases where the test did not fail, \TR{R1.8}{or its execution was found to be invalid as previously described,} we excluded the test. This decision was based on our aim to identify valid broken test cases. This step resulted in the exclusion of 33,977 tests.
    
\textbf{The Test Case Change Repairs the Broken Version.} We executed $T_{i+1}$ against $V_{i+1}$ and verified whether $T_{i+1}$ passed, ensuring that the commit repaired the test. In cases where $T_{i+1}$ did not pass, indicating that the changes did not repair the test, we excluded the test. This step resulted in the exclusion of 1,592 tests.

After applying the above-mentioned analysis based on the test method changes, we ended up with 71,582 test case repairs across 112 projects. \TR{R1.9 R3.11}{We found that, in certain projects, significant test repair activity was undetected by our approach. This determination was made through a two-step process:}

\TRC{\textbf{Test Case Execution Analysis.} We examined the test cases to identify those that did not meet the first two conditions of the test executions ($T_i$ passing on $V_i$ and $T_i$ failing on $V_{i+1}$). When a substantial percentage of potential test repairs (test method changes) in a project did not meet these conditions, it suggested limitations such as the need for a specialized testing environment or unresolved dependencies by Maven. Projects where over 50\% of test method changes did not meet these conditions were skipped. This criterion led to the exclusion of 55 projects.}

\TRC{\textbf{Test Repair and Commit Count.} From the 55 skipped projects, we reviewed the number of test repairs and commits. We selected two projects for inclusion in our dataset due to their significantly higher numbers of valid test repairs and commits, averaging 2,500 and 450, respectively, compared to averages of 120 and 25 in the other 53 skipped projects.}

\TRC{To summarize, this process ensured that only projects with detectable and significant test repair activities were included in our dataset. Due to potential instability in testing or the need for special testing environments, projects with lower detectable test activities are expected to lead to many experimental problems. Indeed, using these projects would contribute a limited number of test cases while requiring the execution of test cases in a specialized environment that we may not have access to. In this step, 53 projects, which had a total of 6,417 test repairs, were skipped, while 65,165 test repairs across 59 projects were retained.}

To complete our benchmark, we also needed to collect the changes in the SUT for creating the repair context. Therefore, for each test case repair commit, we collected all the changes in the SUT source files, i.e., files that do not include ``\textit{src/test}'' in their path and do not include the JUnit test annotation (\textit{@Test}). In collecting SUT changes, we considered code changes within the existing source files of the SUT. We excluded changes that involved the addition or removal of an entire file. The rationale behind this exclusion is that complete file additions or removals typically result in lengthy code sequences, with most of the code in these files being irrelevant for repairing test cases. Enriching the repair context with relevant code from complete file additions or removals is a task that can be explored in future work.

\subsubsection{Preprocessing and Splitting}
\label{benchmark_preprocessing}
In the next step, we conducted preprocessing to eliminate redundant repairs, large tests which do not fit the model's input, and repairs for which we could not obtain complete information, as detailed below. The following preprocessing steps were applied to the dataset derived from the previous section, which included 65,165 test repairs across 59 projects.

\textbf{Duplicate Test Repairs.} Identical test repairs may exist in different commits or branches. \TR{R3.12}{Since we analyze every commit as detailed in Section~\ref{benchmark_subjects}, we may encounter duplicate test repairs, as a test repair may appear both in a merge commit and in the original commit from its source branch.} This step resulted in exclusion of 1,039 tests.

\textbf{Empty Repair Context.}epair context for a test repair when the Spoon library, a static code analysis tool~\cite{spoon} we utilized, failed to process the project's code. We excluded such cases as our approach relies on having a repair context to provide to our model. This step resulted in the exclusion of 3,748.

\textbf{Absence of Test Breakage Location.} In accordance with our problem definition outlined in Section~\ref{sec:approach}, we assume that the test breakage location is included in the input. During our data analysis, we found that, in certain cases, the test breakage location could not be identified. Specifically, in 1,248 tests, the test repair consisted solely of adding new lines. In other words, the repair introduced new lines of code to address the problem without changing any of the existing lines from the test that was initially broken. As a result, the breakage location remains undetermined in these cases. Furthermore, we identified 19 test cases in which the breakage location could not be determined using the Spoon library and, in 7 other test cases, the broken lines were actually comments. As a result of these findings, we excluded a total of 1,274 tests in this preprocessing step.

\textbf{Maximum Input and Output Length.} We set the maximum input length to 512 tokens and the maximum output length to 256 tokens. While the language models we use can handle larger inputs and outputs, we were constrained by available computing resources (Section~\ref{sec:implementation}). Also, we used the 512 limit to be consistent across models as the maximum token limit varies across models. We excluded test cases that exceeded the limit, considering both the test code and at least one code change hunk in the SUT. It is worth noting that with the expected advancements in language models and hardware, this practical limitation and constraint is going to become less significant in the future. This step resulted in excluding 5,383 test cases.

\subsubsection{\TRC{Quality Checking}}
\label{sec:tarbench_quality}
\TR{R1.10}{In addition to analyzing test executions (as detailed in Section~\ref{sec:testRepairCol}), to significantly enhance data quality by only including passing test cases or those that fail due to problems in the test code, we implemented two additional criteria to further improve data quality and remove low quality repairs and noise.}

\textbf{Test Refactoring and Relocation.} We excluded cases where the repair only involved the relocation or refactoring of test code, with no changes to the SUT. Examples include renaming test utility methods or classes and upgrading testing library versions, such as JUnit. These cases occur when all changes in the commit are confined to test source code, specifically files containing ``\textit{src/test}'' in their path. \TRR{C1.2}{A manual review of randomly-selected cases confirmed that such test repairs were consistently unrelated to the SUT. Consequently, they were classified as refactoring activities and treated as noise in our dataset.} This step resulted in the exclusion of a total of 6,302 tests.

\TRR{C1.2}{While our heuristic may overlook some valid test repairs---such as cases where production code changes occur in one commit and corresponding test repairs happen in subsequent commits---it achieves a balance between minimizing noise and maximizing repair instance count. Given the relatively small proportion of test-change-only instances (6k out of 65k) and the complexity associated with more nuanced heuristics, this approach offers a practical trade-off for this study.}

\TRC{\textbf{Trivial Test Repairs.} Trivial repairs refer to repairs that can be accomplished with development tools, such as code editors or IDEs. We define trivial repairs as repairs that only include the renaming of class or method names. We designed a simple technique to automatically detect trivial repairs. First, we detect renamed classes and methods in the SUT using RefactoringMiner~\cite{refactoringMiner}. Second, we identify the classes and methods used in the broken test case using Spoon~\cite{spoon}. Finally, if there is at least one common class or method name between the first and second steps, we identify the test repair as trivial.}

\TRC{Trivial repairs fail to capture the challenges of the test case repair problem. Thus, we removed them from our evaluation set to obtain more realistic results based on substantial and complex test case repairs and prevent the risk of artificially inflating our performance metrics due to trivial repairs.}

\TR{R1.13}{Though excluded for evaluation purposes, during our preliminary experiments, we observed that including trivial repairs in the training set enhanced our model's performance. We provide additional details to support the effectiveness of keeping trivial repairs in the training set in the results section. Trivial repairs, although simple, provide valuable information about basic patterns and structures in the test and production code that the model needs to learn. By including these repairs, the model gains a foundational understanding of syntax and semantics, which can enhance its capability to handle more complex repairs.}

\TRC{We identified 9,242 trivial test repairs in our dataset, removed 2,046 of them from the evaluation set, and kept 7,196 in the training set. We will provide further details on the training and evaluation sets in the following section.}

\subsubsection{Data Splitting}
Following the above steps, we divided the data into the three splits: the training set (80\%), the validation set (5\%), and the test set (15\%). To ensure a fair and realistic evaluation, it is crucial that the more recent test case repairs are not exposed to the model during training. This is important because newer code may contain the ground truth of older test repairs. Thus, for each project, we kept older commits in the training set and included newer commits in both the validation and test sets. We use the validation set to determine when to stop training, preventing the model from overfitting.

Following the completion of all steps, \benchmarkName{} includes 45,373 test repairs across 59 distinct projects, making it notably larger and more diverse than the largest existing test repair dataset in the literature, which only consists of 235 test repairs across 14 projects (see Section~\ref{sec:related-work}). In this dataset, we have \textbf{36,639} training instances, \textbf{1,631} validation instances, and \textbf{7,103} test instances.

\subsubsection{\TRC{Analysis of Breakage and Error Lines}}
\label{breakagelines}
\TR{R1.1}{While tools such as IDEs and CI systems can capture error locations in test code, identifying the actual breakage lines that lead to those errors is not always straightforward. We analyzed the test repairs in our benchmark to clarify the distinction between breakage lines and error lines, which are defined in Section~\ref{sec:approach}. These two sets of lines do not always overlap, as repairs may occur in locations different from where the errors originated. The results of this analysis are presented in Table~\ref{table:breakageExceptionOverlap}. The dataset consists of 82.1\% compilation failures and 17.9\% runtime failures, reflecting two primary types of test failures.}

\begin{table}[tbp]
    \resizebox{\linewidth}{!}{%
    \begin{tabular}{lccc}
        \toprule
        Case & Compile (\%) & Runtime (\%) & Total (\%) \\
        \midrule
        No Intersection & 37.0 (13786) & 60.3 (4899) & 41.2 (18685)  \\
        Exact Match & 55.3 (20608) & 26.0 (2110) & 50.1 (22718)  \\
        Some Intersection & 7.7 (2851) & 13.7 (1119) & 8.7 (3970)  \\
        \bottomrule
    \end{tabular}
    }
    \caption{\TRC{Breakdown of the intersections between the breakage lines and error lines. The values in brackets indicate the count of instances for each cell.}}
    \label{table:breakageExceptionOverlap}
\end{table}

\TRC{Regarding the relationship between the error lines and the breakage lines, in 50.1\% of the failures, the breakage lines exactly match the error lines. In the remaining cases, however, they do not match perfectly: 41.2\% of the failures show no intersection between breakage and error lines, while 8.7\% have a partial intersection without a complete match. This indicates that in nearly half of the dataset, identifying breakage lines is not a straightforward task, specifically for IDEs or developers, as error and breakage lines do not fully match. Furthermore, we found that the breakage lines and error lines match exactly in 26\% of the runtime error instances and 55.3\% of the compilation error instances. This demonstrates that identifying breakage lines associated with runtime errors is more challenging than with compilation errors, though it is not always straightforward for compilation errors either.}

\TRC{The analysis highlights that identifying breakage lines is inherently complex and cannot be solely determined by error lines. This suggests that test repair tools must employ more sophisticated techniques, beyond relying on error line information alone, to accurately pinpoint breakage locations. This is particularly crucial for addressing runtime failures, where the challenge is more noticeable.}

\subsection{Test Repair Performance of CLMs and Input Formats (RQ1.1)}
\label{rq1}
\begin{table*}[tbp]
\centering
\resizebox{\textwidth}{!}{%
\begin{tabular}{llrrrrrrrrrrrr}
\toprule
 \multicolumn{2}{c}{\multirow{2}{*}{Input-output Format ($IO$)}} & \multicolumn{3}{c}{EM} & \multicolumn{3}{c}{PR} & \multicolumn{3}{c}{CodeBLEU} & \multicolumn{3}{c}{BLEU} \\
\cmidrule(lr){3-5} \cmidrule(lr){6-8} \cmidrule(lr){9-11} \cmidrule(lr){12-14}
 & & CT5+ & PLB & CG & CT5+ & PLB & CG & CT5+ & PLB & CG & CT5+ & PLB & CG \\
\midrule
$IO_1$ & Base & 61.0 & 54.0 & 51.8 & 77.2 & 78.0 & 73.8 & 76.7 & 76.4 & 72.4 & 77.7 & 77.8 & 73.1 \\
$IO_2$ & Text-based Similarity Hunk Ordering & \textbf{66.1} & 57.9 & 57.7 & 80.0 & 79.2 & 77.0 & \textbf{79.3} & 77.9 & 74.7 & \textbf{80.0} & 79.2 & 75.7 \\
$IO_3$ & Word Level Hunk Representation & 63.2 & 59.1 & 58.1 & \textbf{80.2} & 79.2 & 77.0 & 77.2 & 78.4 & 75.6 & 78.3 & 79.6 & 76.4 \\
$IO_4$ & Edit Sequence Output & 52.9 & 47.6 & 45.7 & 66.4 & 64.4 & 56.9 & 66.8 & 61.2 & 56.9 & 67.5 & 61.8 & 57.0 \\
\bottomrule
\end{tabular}
}
\caption{Comparison of test case repair performance among the CodeT5+ 770M (CT5+), PLBART Base 140M (PLB), and CodeGen Multi 350M (CG) code language models, and across four different input-output formats.}
\label{tab:rq1_performance}
\end{table*}

In recent years, CLMs (language models that are pre-trained on code) have emerged that can be fine-tuned specifically for test code repair purposes. While some studies have compared CLMs in different program generation tasks~\cite{jiang2023impact, zeng2022codelmstudy, lu2021codexglue, wang2021codet5}, to the best of our knowledge, this study is one of the first to examine their performance in the test repair context. Therefore, answering this research question will offer valuable practical insights and contribute to our understanding of test case repair using CLMs.

Additionally, we have explored multiple approaches to format input and output data for CLMs. This research question also seeks to evaluate how different input formats influence model performance in comparison to each other. Overall, this research question investigates the combinations of models and input formats that produce the best results in the context of test repair, which is essential for the effective utilization of CLMs.

\subsubsection{Selection of Models}
In general, CLMs can be categorized into three groups based on their architectures: encoder-only, decoder-only, and encoder-decoder. A systematic review conducted by Hou et al.~\cite{hou2023llm4se} explored the application of CLMs in software engineering tasks and identified more than 50 models that can be applied in various software engineering tasks. However, no study has applied these models specifically to test repair or has demonstrated that a single CLM outperforms others in software development tasks, particularly in program repair. Thus, since evaluating all available CLMs is computationally impossible, we have established the following filtering criteria to select CLMs used to answer RQ1.1.

\textbf{Direct Applicability.} Similar to the approach taken by Jiang et al.~\cite{jiang2023impact}, our focus is exclusively on models that can be seamlessly applied to repair source code without requiring additional modifications or architectural changes. Consequently, we exclude all models with an encoder-only architecture, as they necessitate the addition of an extra decoder component to generate repairs. Two notable examples of such models include CodeBERT~\cite{feng2020codebert} and GraphCodeBERT~\cite{guo2020graphcodebert}.
    
\textbf{Open Source Models.} We exclude all models that are not publicly available, as fine-tuning them with test code repair data is essential for our study. Examples of such models are Codex~\cite{codex-humaneval}, AlphaCode~\cite{li2022alphacode}, PaLM-Coder~\cite{chowdhery2022palm}, and GPT-4~\cite{openai2023gpt4}.

\textbf{Pre-training Data.} As our primary focus is on code generation, it is crucial that the models we employ have been pre-trained on large code datasets. Consequently, we do not consider models that have been pre-trained on general natural language datasets. For example, we exclude T5~\cite{raffel2023t5}, GPT-Neo~\cite{black2021gptneo}, and GPT-J~\cite{gpt-j} based on this criterion.

\textbf{Old Models.} We excluded models that have been superseded by newer versions. For instance, CodeT5~\cite{wang2021codet5} has been excluded because it is superseded by CodeT5+~\cite{wang2023codet5p}.

\textbf{Model Size.} To enhance practicality and broaden the applicability of fine-tuning, we excluded models exceeding one billion (1B) parameters in size. Utilizing larger models for fine-tuning often demands substantial computational resources, which may not be readily available in many development contexts. For instance, consider StarCoder~\cite{li2023starcoder}, a 15.5B decoder-only CLM, which underwent pre-training on 512 \textit{A100 80 GB GPUs} across 64 nodes. We fully realize that larger models will become easier to train over time and that therefore our repair performance results probably represent a lower bound in terms of what to expect in the future. 
    
\textbf{Model Performance.} While no study has identified top-performing models for the specific test repair context, we nevertheless reviewed related studies and excluded models that exhibited relatively lower performance compared to others. Our prioritization was based on the widely-recognized HumanEval~\cite{codex-humaneval} benchmark, which is designed for assessing code generation capabilities. This benchmark closely aligns with the challenges posed by the test case repair task. For example, we excluded PolyCoder~\cite{xu2022polycoder} due to its underperformance in HumanEval when compared to CodeGen~\cite{nijkamp2022codegen}.

By applying the exclusion criteria outlined above, we narrowed down our selection down to two CLMs: \textbf{CodeT5+ 770M~\cite{wang2023codet5p}} (CT5+), an encoder-decoder model with 770 million parameters, and \textbf{CodeGen Multi 350M~\cite{nijkamp2022codegen}} (CG), a decoder-only model with 350 million parameters. Furthermore, \textbf{PLBART Base 140M~\cite{ahmad2021unified}} (PLB), a relatively small encoder-decoder model that was not evaluated on the HumanEval benchmark, was selected as well to assess how a smaller model impacts test repair performance.

\TR{R3.9}{The beam size, which determines the number of repair candidates generated, is set to 200 for PLB and 40 for both CT5+ and CG. Our goal was to select the largest beam size feasible to maximize the exploration of the potential repair candidates' search space, thereby improving performance. The selected beam sizes represent the maximum we could utilize given our computational resources.}

\subsubsection{Input and Output Formatting}
Based on our problem definition in Section~\ref{sec:approach}, we perform test repair given the broken test code, the location of the breakage within the test code, and a set of code changes made to the SUT. With these inputs, we can construct an input sequence for a CLM in various ways. Additionally, the expected output that the model is trained to generate can have different formats. We defined four distinct input and output formats ($IO$s) in Section~\ref{sec:approach} and, in this research question, we explore the impact of these $IO$s on the performance of test case repair.

\subsubsection{RQ1.1 Results}
Table~\ref{tab:rq1_performance} presents the results of the three selected CLMs across various $IO$s, measured  using the four evaluation metrics. The CT5+ model achieves the highest EM, when utilizing Text-based Similarity Hunk Ordering ($IO_2$). Furthermore, in the majority of cases, CT5+ outperforms other models across different metrics and IOs. This superiority can be justified due to the larger number of parameters in CT5+.

Interestingly, we observed that PLB, despite having only 140M parameters, outperforms CG with 350M parameters. Furthermore, its performance (with $IO_2$) closely aligns with that of CT5+, with losses of 8.2, 0.8, 1.4, and 0.8 percentage points (pp) in terms of EM, PR, CodeBLEU, and BLEU, respectively. This observation suggests that the number of parameters is not the sole determinant of model performance. It implies that a high-performing CLM can be trained at a significantly lower cost without significantly compromising performance. Considering the substantial size difference between PLB and CT5+, sacrificing a few percentage points of performance for substantial cost reduction during training and inference seems reasonable. Thus, we recommend PLB over CG in the test repair context, and the choice between PLB and CT5+ is a trade-off between training cost and a marginal performance loss.

Moreover, concerning $IO$s, as depicted in Table~\ref{tab:rq1_performance}, $IO_2$ and $IO_3$ yield the best results across all models and metrics. Specifically, $IO_2$ provides the best results for CT5+, while $IO_3$ yields better results for PLB and CG. Overall, based on the results, $IO_3$ exhibits the highest performance in 7 out of 12 cases, outperforming $IO_2$, which achieves the highest performance in 3 out of 12 cases, with both being equal in the remaining 2 cases. As further discussed below, neither Base ($IO_1$) nor Edit Sequence Output ($IO_4$) achieves favorable results in any case, with $IO_4$ reaching the lowest performance in all instances. 

\TR{R2.1 R3.13}{By definition, the difference between $IO_2$ and $IO_3$ lies in their hunk representation: $IO_2$ uses line-level context, while $IO_3$ uses word-level context. The results indicate that the larger model, CT5+, performs better with line-level context, whereas the two smaller models, PLB and CG, demonstrate improved performance with word-level context. This suggests that larger models may be better at leveraging the additional information provided by line-level context, while smaller models might benefit more from the focused and concise nature of word-level information. However, it is not possible to draw a definitive conclusion based solely on this data, as the models vary in more than just size; they also differ in pre-training datasets and tasks. Further investigation is required to explore the relationship between model size and performance across different $IOs$. For instance, comparing CT5+ with 770 million parameters to versions with 220 million or 2 billion parameters could provide additional insights.}

\TR{R1.3}{We have explored a number of $IO$ formats and the fact that some of them perform worse should be reported to inform researchers and practitioners. Further, while we recommend against the use of $IO_4$ and $IO_1$, they remain viable options. For instance, $IO_4$ remains viable as the differences in EM and PR between $IO_4$ and $IO_2$ (the best-performing $IO$) are only  10.3\% and 13.8\%, respectively. The choice between the other two techniques, $IO_2$ and $IO_3$, is context-dependent and may require prior analysis and experimentation. For instance, $IO_2$ is recommended for CT5+, whereas $IO_3$ is preferred for PLB.}

\TR{R2.3}{We acknowledge that more advanced prompt engineering techniques, such as Chain-of-Thought prompting, Few-shot learning, or iterative refinement, as well as the larger context windows available in commercial LLMs, could further enhance the test case repair task.}

\TR{R1.13}{To assess the impact of trivial repairs on model performance, as discussed in Section~\ref{sec:tarbench_quality}, we conducted an additional experiment using our best-performing strategy: the CT5+ model with $IO_2$. In this experiment, we removed 7,196 trivial repairs from the training set and repeated the fine-tuning and evaluation processes. The results indicated a performance decline of 5.6 percentage points in EM and 2.8 percentage points in CodeBLEU. These findings highlight the positive contribution of trivial repairs to the overall model performance, reinforcing the value of including them in the training set.}

\begin{tcolorbox}[title=RQ1.1 Summary, breakable]\TRC{
The CT5+ model with $IO_2$ delivers the best performance for the test repair task, with 66.1\% exact match accuracy and 80.0\% plausible repair accuracy, making it our top recommendation. For resource-limited environments, the PLB model with $IO_3$ offers a good balance between performance and efficiency, with faster fine-tuning and inference speeds.}
\end{tcolorbox}

\subsection{\TRC{Test Repair Performance Against Baselines (RQ1.2)}}
\label{sec:rq12}

\TR{R1.12}{
In this research question, we aim to compare \approachName{} against a recent state-of-the-art method and two simple baselines. Specifically, we compare the best configuration of our approach (CT5+ with $IO_2$) with \ceprot{}~\cite{hu2023ceprot}, which addresses the automatic detection and updating of obsolete tests and focuses solely on the SUT method that the test code targets. Further, we also compare with two baselines that implement basic solutions: \sutcopy{} and \nocontext{}. This comparison aims to provide a comprehensive evaluation of the effectiveness and practicality of our approach against both advanced and basic baselines. In the following subsections, we detail the comparison between \approachName{} and \ceprot{}, define the \sutcopy{} and \nocontext{} baselines, and report the results of \approachName{}'s performance against these three baselines. We also provide a technical comparison between \approachName{} and \ceprot{} in Section~\ref{sec:relw_atr}.}

\subsubsection{\TRC{Collecting \ceprot{}'s Data}}
\TRC{
\ceprot{} is an approach that addresses both the automatic detection and updating of obsolete tests. This method fine-tunes the CodeT5 language model to handle both tasks efficiently. The authors have also constructed a dataset for the update task, containing 5,196 instances, with 520 instances selected as the evaluation (test) set.
In their study, the authors define an obsolete test case as one that is modified within a commit, given that the production method from the SUT, which is covered by the test, also undergoes changes. Although this definition is similar to our benchmark’s definition, \benchmarkName{} encompasses a broader spectrum of scenarios and does not require a change in the production method. \benchmarkName{} defines broken test cases based on the test execution results before and after changes in the test code, as detailed in Section~\ref{benchmark}.}

\TRC{
\ceprot{} achieves a 12.3\% EM and a 63.1\% CodeBLEU score in the test update generation task. To assess the generalizability of \approachName{} and ensure a fair comparison with \ceprot{}, we applied \approachName{} to \ceprot{}’s evaluation set. \approachName{} utilizes data from all code changes in a commit, and therefore, the raw data provided by \ceprot{} was insufficient for a direct application of \approachName{}. However, \ceprot{}  provided the GitHub repository name and the commit hash for each instance. Using this information, we collected all necessary data to enable \approachName{} to function correctly. We could not apply \ceprot{} to our benchmark as \ceprot{} did not provide the code needed to collect and construct the required data for their approach.}

\TRC{
After our data collection, we determined that \approachName{} could only be applied to a subset of 214 instances (41\%) from \ceprot{}’s evaluation set, which we refer to as compatible instances. This decision was based on several key factors described below.}

\TRC{
\textbf{Invalid Cases.} We found 30 instances to be invalid or impossible to collect or repair: 10 instances were excluded because the specified commit was not available on GitHub; 2 instances were excluded as the repository did not exist on GitHub; 12 instances were excluded because the source code of the test and the target code were identical, indicating no changes in the test; 2 instances were excluded as the identified test case was actually production code; 2 instances were excluded because the commits involved only test code changes without production code changes, likely indicating test code refactoring rather than repair; and 2 instances were excluded due to our data collection tool's inability to detect repairs for unknown reasons. These invalid cases were removed from our comparison.}

\TRC{
\textbf{Test Method Name Changes.} During \benchmarkName{}’s data collection, we matched test methods between versions based on their full class and method names. Consequently, we did not identify test method name change scenarios as repair instances. However, 38 instances in \ceprot{}’s evaluation set involved repairs that included changes to the test method name. \ceprot{} did not clarify how these methods were matched initially to ensure they pointed to the same test case. Therefore, we did not include these cases in our comparison.}

\TRC{
\textbf{No Source Changes.} In 24 instances of \ceprot{}’s evaluation set, the repairs involved only added lines without changing any existing lines. As mentioned in Section~\ref{benchmark_preprocessing}, \benchmarkName{} excludes these cases due to the absence of a test breakage location. \approachName{} requires the test breakage location to function, and without source changes, there is no breakage location. For this reason, we excluded these cases from our comparison.}

\TRC{
\textbf{Multi-hunk Repairs.} We found that 214 test instances involved multi-hunk test repairs, i.e., repairs with code changes in multiple chunks of the test method. As described in Section~\ref{benchmark_subjects}, our work focuses on single-hunk test repairs, which is a common practice in many program repair studies and provides clarity and focus in the model’s learning process. To evaluate \ceprot{}’s performance on multi-hunk instances, we replicated \ceprot{} and computed its performance on these cases, finding an EM of 7.9\% and a CodeBLEU score of 42.6\%. This indicates poor performance by \ceprot{} on multi-hunk instances, showing ineffective support for these cases, which clearly should be addressed by future research through iterative approaches. Consequently, we excluded multi-hunk instances from our comparison.}

\subsubsection{\TRC{\sutcopy{} and \nocontext{}}}
\TRC{
In \sutcopy{}, we examine the SUT changes in an arbitrary order and update the test code when we encounter an SUT changed part that is uniquely identifiable in the broken part of the test code. For example, if the string ``\textit{value1}'' is changed to ``\textit{value2}'' in the SUT, and ``\textit{value1}'' exists in the broken test code, \sutcopy{} replaces ``\textit{value1}'' with ``\textit{value2}'' to repair the test case. In \nocontext{}, we fine-tune the best performing CLM by excluding the repair context from the input, meaning we include only the broken test as input and expect the repaired test as output.}

\subsubsection{\TRC{RQ1.2 Results}}
\TRC{
We compute and report both exact match accuracy (EM) and CodeBLEU, as used by \ceprot{} for evaluating the test update generation task. However, there are two key differences in how we computed these metrics for \approachName{} when compared to \ceprot{}.}

\TRC{
First, we used beam search with a beam size of 40 for our best model, CT5+. Consequently, \approachName{} generates 40 repair candidates for each test repair instance, whereas \ceprot{} does not use beam search and generates only one repair candidate per instance. This affects the computation of EM. \approachName{} considers there is an exact match if at least one of the 40 candidates matches the ground truth, while \ceprot{} only compares the ground truth with a single repair candidate.}

\TRC{
Second, \ceprot{} generates the full test code as output, whereas \approachName{} generates only the repaired part of the test code. For computing CodeBLEU, \ceprot{} compares the full test code between the prediction and the ground truth, while \approachName{} compares only the repaired part. This difference affects the CodeBLEU scores because including the full test code inflates the CodeBLEU values. Indeed, since most of the test code remains unchanged in a repair, this leads to higher scores.}

\TRC{
To ensure a fair comparison, we applied \approachName{} to \ceprot{}’s test set and computed EM and CodeBLEU in the same manner as \ceprot{}. Specifically, we used only the first repair candidate out of our 40 candidates to compute EM and placed the repaired part within the full test code for computing CodeBLEU. Although \ceprot{} stated to have provided a script for CodeBLEU in their replication package, we were unable to locate it in their repository. Therefore, we used our own CodeBLEU implementation, which is included in our replication package.}

\begin{table}[tbp]
\centering
\resizebox{.80\linewidth}{!}{%
\begin{tabular}{lrr}
\toprule
Baseline & EM & CodeBLEU \\
\midrule
\ceprot{}~\cite{hu2023ceprot} & 21 & 60.4 \\
\cmidrule(lr){1-3}
\approachName{} (CT5+ \& $IO_2$) & 40.6 & 91.1 \\
\bottomrule
\end{tabular}
}
\caption{\TRC{Comparison of \ceprot{} with \approachName{}'s best model, CodeT5+ using $IO_2$, on \ceprot{}'s benchmark.}}
\label{tab:rq1_ceprot}
\end{table}

Table~\ref{tab:rq1_ceprot} presents the results of two techniques on the 214 compatible test instances. According to the results, \approachName{} achieved a CodeBLEU score of 91.1 and an EM score of 40.6, significantly outperforming \ceprot{}, which scored 60.4 in CodeBLEU and 21 in EM.

\TRR{C1.3}{To investigate the reasons behind \approachName{}'s superior performance, we performed a comparative analysis, manually reviewing instances where \approachName{} successfully repaired the test cases while \ceprot{} failed. Details of five handpicked comparisons are provided in the Appendix. Our analysis identified several recurring failure patterns in \ceprot{}'s repair attempts: failing to make any modifications to the test case (Comparison 1 in the Appendix), deleting significant portions of the test case (Comparisons 2 and 3), making superficial changes, such as modifying only whitespace characters (Comparison 5), and making irrelevant or minor adjustments that did not repair the test case (Comparison 4).}

\TRRC{We attribute \approachName{}'s superior performance to three key factors.
First, \approachName{} effectively utilizes SUT changes by prioritizing and selecting those most relevant to the test code. In contrast, \ceprot{} focuses exclusively on changes to the production method under test. Though this information is beneficial for repair, it may fail to capture the broader context needed for accurate test repair. This limitation is evident in Comparisons 3 and 5 in the Appendix.
Second, \ceprot{} represents code changes by including the two versions of the full production method along with the change's edit sequence. In contrast, \approachName{} includes only the changed lines, significantly reducing the token count and allowing more test code and SUT changes to fit within the input. Comparison 2 illustrates how this difference enables \approachName{} to better utilize contextual information.
Third, \approachName{} benefits from being fine-tuned on a substantially larger dataset of test repair instances (36.6k) compared to \ceprot{} (4.6k). Also, \approachName{} leverages a larger and more recent language model, CodeT5+ (770M parameters), whereas \ceprot{} uses the smaller and older CodeT5-base (220M parameters). These advantages are highlighted in Comparisons 1 and 4, where both approaches had sufficient repair contexts.}

\begin{table}[tbp]
\centering
\resizebox{\linewidth}{!}{%
\begin{tabular}{lrrrr}
\toprule
Baseline & EM & PR & CodeBLEU & BLEU \\
\midrule
\sutcopy{} & 10.8 & 13.8 & 61.9 & 52.1 \\
CT5 with \nocontext{} & 28.7 & 55.9 & 65.5 & 66.2 \\
\cmidrule(lr){1-5}
\approachName{} (CT5+ \& $IO_2$) & 66.1 & 80.0 & 79.3 & 80.0 \\
\bottomrule
\end{tabular}
}
\caption{Comparison of baseline models with \approachName{}'s best model, CodeT5+ using $IO_2$, on \benchmarkName{}.}
\label{tab:rq1_baselines}
\end{table}

\TRC{
Further, Table~\ref{tab:rq1_baselines} shows the result of baselines. The results indicate that CT5+ with \nocontext{} consistently outperforms \sutcopy{} across all metrics, with improvements of 17.9, 42.1, 3.6, and 14.1 pp in terms of EM, PR, CodeBLEU, and BLEU, respectively. This outcome aligns with expectations and clearly indicates that CLMs are indeed effective at automating test repair.}

\TRC{
Finally, \approachName{}'s best configuration (CT5+ with $IO_2$) consistently outperforms CT5+ with \nocontext{} across all metrics with improvements of 37.4, 24.1, 13.8, and 13.8 pp in terms of EM, PR, CodeBLEU, and BLEU, respectively. This confirms that using the repair context proposed by our approach significantly enhances CLMs' performance. Also, this implies that depending only on a generic CLM for test repair may not produce effective results. It emphasizes the critical need for context-dependent fine-tuning and customization of CLMs.}

\begin{tcolorbox}[title=RQ1.2 Summary, breakable]\TRC{
The best configuration of \approachName{} (CT5+ with $IO_2$) consistently and significantly outperforms the baselines---\ceprot{}, \nocontext{}, and \sutcopy{}---across all evaluated metrics. This finding highlights the effectiveness of using CLMs for test repair, provided that the model's input and output are carefully engineered and fine-tuned. Incorporating comprehensive and relevant information, as done in our approach by integrating the repair context, is crucial for achieving optimal performance.}
\end{tcolorbox}

\subsection{Test Repair Performance on Test Repair Characteristics (RQ2.1)}
\label{rq2.1}
Test cases can break for various reasons, each necessitating a specific category of repair. This research question investigates whether \approachName{}'s performance varies across repair categories. By addressing this question, we can gain a more detailed understanding of \approachName{}'s performance in test repair tasks, allowing us to determine their suitability for specific repair categories. Also, the results can highlight areas where future improvements are needed to effectively apply CLMs in the context of test repair.

\subsubsection{Test Repair Categories}
To answer RQ2.1, we examined the code changes involved in repairing test cases for all instances in \benchmarkName{}, and (1) categorized these changes into three main categories as discussed in the following, and (2) measured their complexity based on their number of abstract syntax tree (AST)-level edits. 

To extract the AST-Level edits, we parsed the broken and repaired test code into an AST. Similar to Tufano et al.~\cite{tufano2019empirical}, we rely on the GumTree Spoon AST Diff tool~\cite{gumtree} to compute the AST difference between the broken and repaired test code. This tool computes a sequence of AST-level edit actions that transforms the broken test code into the repaired version. Based on the extracted edits, we record the number of AST-level edits.

We then categorized the changes based on the features of the edit actions into: 
\begin{enumerate}
    \item \textbf{Invocation argument or return type change (ARG).} Adding, removing, or modifying arguments that are passed to an invocation, i.e., a method call or a constructor call, in the test code. This category also includes changing the expected return type of an invocation in the test code.
    \item \textbf{Invocation change (INV).} Adding or deleting an invocation, such as adding a new method call to the test code, falls into this category. It also includes the replacement of an existing invocation. For instance, the repair might replace the constructor \textit{Foo()} with \textit{Bar()}. The key distinction between this category and the previous one is that the former involves changes in the arguments and the return types for the same method or constructor. However, in this category, the invocation itself is changed, regardless of the arguments and return type.
    \item \textbf{Test oracle change (ORC).} Modifications to sections where test oracles are compared to the SUT's output, which include changes to assertions or expected exceptions. \TR{R3.14}{Modifications including changing the type of an expected exception or altering a line containing an assert statement—either changes to the type of assertion or its parameters—are counted in this category.}
\end{enumerate}

Note that a test repair may include more than one change category. In such cases, we assign it to a new category that combines multiple categories. Also, if a test repair does not belong to any of the categories above, we categorize it as Other (OTH). Finally, we computed performance metrics obtained from the best-performing model identified in RQ1.1 for each repair category and compared the results.

\begin{table}[tbp]
\centering
\resizebox{0.8\linewidth}{!}{%
\begin{tabular}{lcc}
\toprule
\textbf{Category} & \textbf{Runtime (\%)} & \textbf{Compile (\%)} \\ 
\midrule
ARG              & 8.3 (3781)             & 38.6 (17504)               \\ 
ORC              & 6.2 (2798)             & 10.7 (4857)                \\ 
INV              & 0.9 (408)              & 13.8 (6251)                \\ 
ARG+INV          & 0.8 (354)              & 11.0 (4972)                \\ 
ARG+ORC          & 0.7 (333)              & 3.0 (1349)                 \\ 
INV+ORC          & 0.3 (140)              & 1.7 (788)                  \\ 
ARG+INV+ORC      & 0.3 (130)              & 1.3 (603)                  \\
OTH              & 0.4 (184)              & 2.0 (921)                  \\
\bottomrule
\end{tabular}
}
\caption{\TRC{Distribution of runtime and compilation failures across different repair categories}}
\label{tab:failure_categories}
\end{table}

\TR{R1.14}{Across categories, repairs may stem from runtime or compilation errors. Table~\ref{tab:failure_categories} presents a detailed breakdown of failures by category. Take the ARG category, for example. Depending on the code change in the SUT, ARG repairs can result from either a runtime or compilation failure. If an argument is added or removed from a method, it leads to a compilation failure in the test code that calls this method. However, if the method's logic changes to expect a specific format for an existing argument or return value, it could trigger a runtime exception from within the SUT method or an assertion failure from the test code.}

\subsubsection{RQ2.1 Results}
\begin{figure}[tbp]
   \centering
   \includegraphics[width=\linewidth]{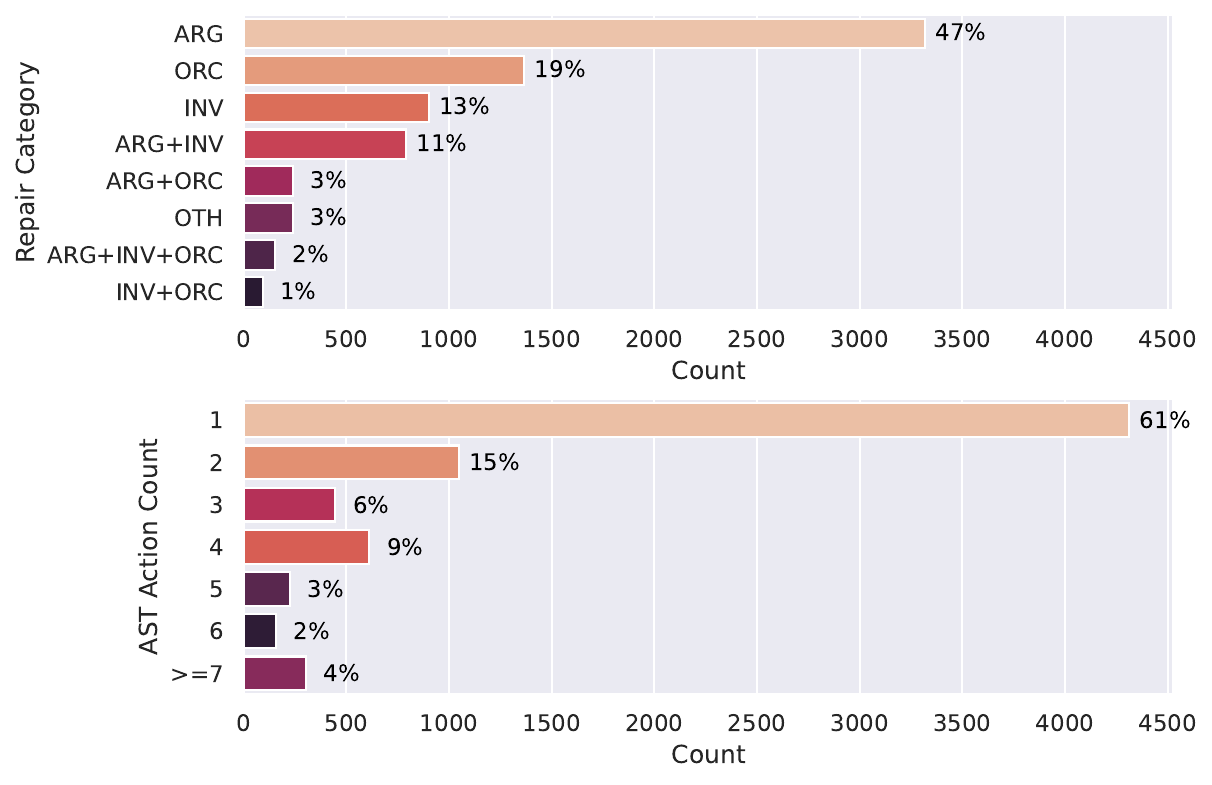}
   \caption{The distribution of test repair categories and AST-level edit action count within \benchmarkName{}'s test set.}
   \label{fig:rq21-char-dist}
\end{figure}

\begin{figure*}[tbp]
   \begin{subfigure}{0.49\textwidth}
       \centering
       \includegraphics[width=\textwidth]{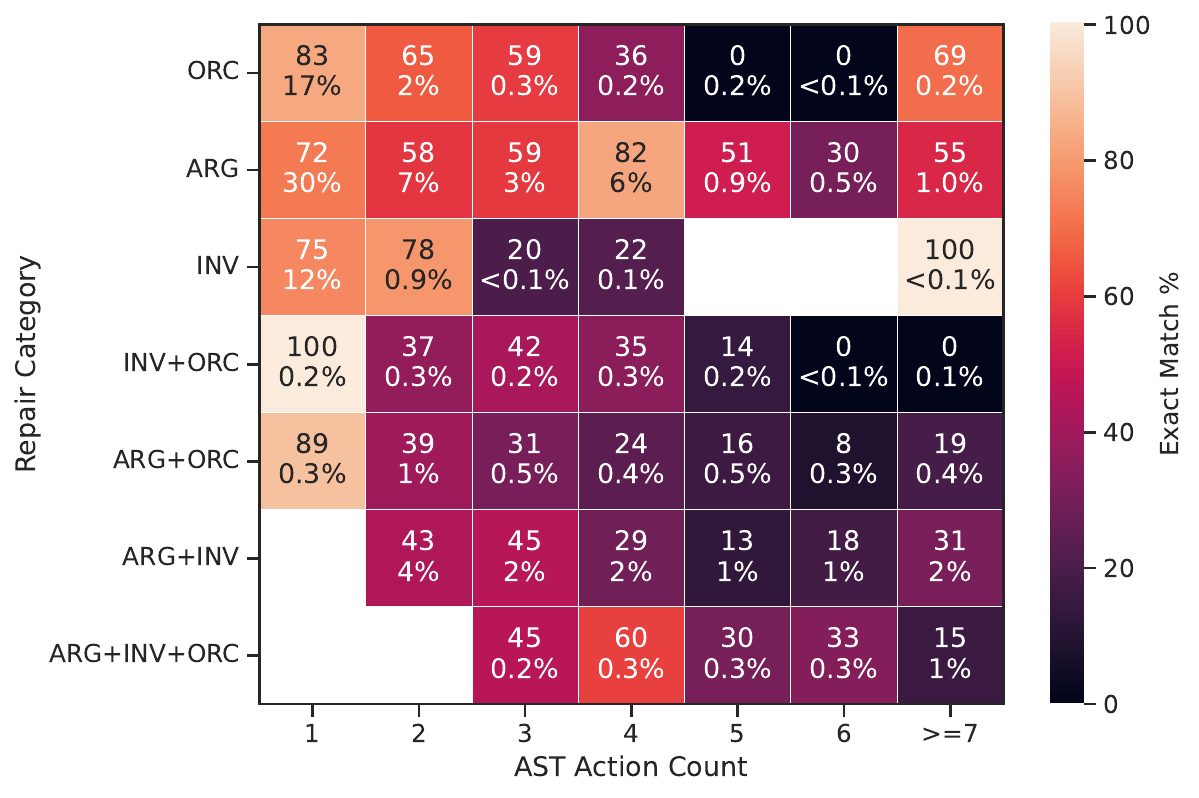}
       \caption{}
       \label{fig:rq21-cat-em}
   \end{subfigure}
   \begin{subfigure}{0.49\textwidth}
       \centering
       \includegraphics[width=\textwidth]{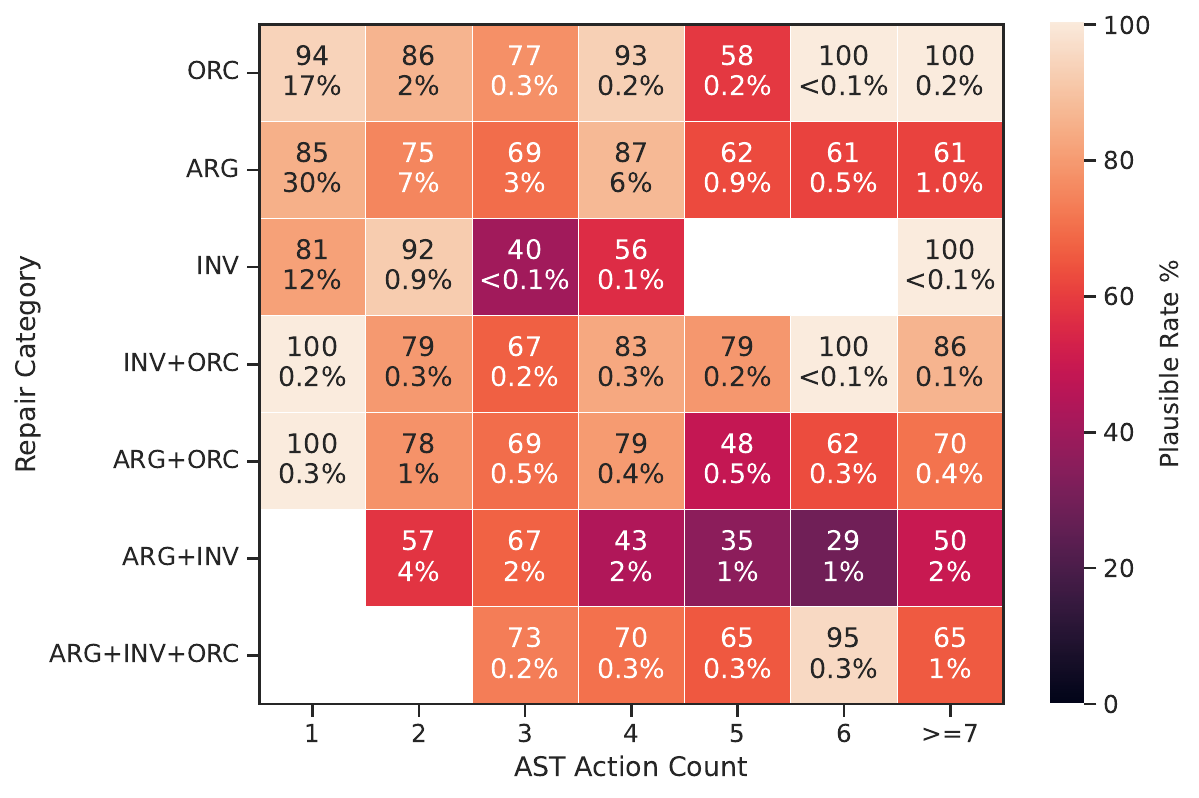}
       \caption{}
       \label{fig:rq21-cat-pr}
   \end{subfigure}
   \caption{Performance analysis of CodeT5+ using $IO_2$ (the best model) across repair categories, correlating with AST-level edit actions. Each cell displays the performance metric value (top) with the respective proportion of instances (bottom).}
   \label{fig:rq21-cat-eff}
\end{figure*}

Figure~\ref{fig:rq21-char-dist} shows the distribution of repair categories and the count of AST-level edit actions for 7,103 test case repairs in  the \benchmarkName{} benchmark (test set). The figure shows that that the majority (80\%) of repairs belong to a single category. Additionally, about 75\% of repairs entail either a single or double AST-level edit, implying that the vast majority test repairs do not require extensive structural changes. This may be explained by test methods being inherently simple, in terms of logic and size.

Figure~\ref{fig:rq21-cat-eff} presents two heatmaps showing the performance of the best model (CT5+ with $IO_2$) across repair categories and AST actions in terms of EM and PR metrics. The x-axis denotes the AST-level edit action count, and the y-axis denotes the repair category. Each cell in the heatmaps has three attributes: (1) the performance metric value, shown as the top number, (2) the color as determined by the performance metric value, with darker color indicating lower performance, and (3) the proportion of test repair instances (out of the 7,103 in the test set), shown as the bottom number.

Repairing a test involving multiple repair categories or AST actions is inherently more complex than repairing one focused on a single category or action. This complexity is confirmed by the results shown in Figure~\ref{fig:rq21-cat-eff}, which reveals a reduction in the model's performance as the number of repair categories and actions increases. Also, in Figures~\ref{fig:rq21-cat-em} and \ref{fig:rq21-cat-pr}, darker colors are notably concentrated towards the bottom-right of the heatmaps, where the number of repair categories and AST actions are higher. Thus it is safe to conclude that the model's performance degrades with more complex test case repairs, aligning with our expectations.

When comparing Figures~\ref{fig:rq21-cat-em} and \ref{fig:rq21-cat-pr}, we see a noticeable distinction: the bottom-right areas in Figure~\ref{fig:rq21-cat-em} are around the 20\% spectrum and appear considerably darker than their counterparts in Figure~\ref{fig:rq21-cat-pr}, which fall within the 60\% spectrum. This suggests that in more complex repair scenarios, although the model shows a lower success rate in generating exact match repairs (approximately 20\% of the time), it manages to generate repairs that execute successfully about 60\% of the time. Hence, the model remains applicable to more complex repairs, though its accuracy decreases compared to simpler repair tasks. Arguably, certain plausible repairs might hold semantic equivalence to the exact match repairs, despite differences in syntax. However, performing a semantic comparison on these repairs is a costly task. Identifying an applicable and scalable solution within this context should be addressed by future research.

Analyzing the proportions of repair instances shows that nearly 70\% of these instances are concentrated within the six top-left cells, identified by the presence of one or two AST actions alongside a single repair category. Focusing on this majority, the results highlight the model's effectiveness in repairing test oracles (ORC), achieving higher scores compared to the ARG category. Regarding EM, the model achieves 83\% and 65\% for the ORC category with one and two AST actions, respectively, surpassing the results for the ARG category with EM values of 72\% and 58\%. 
Similar results are achieved in terms of PR, with the model achieving 94\% and 86\% for the ORC category, surpassing those for the ARG category with 85\% and 75\%, respectively. Our observations reveal that ORC test repairs primarily involve changing literal values or the test's expected exceptions. Conversely, in ARG repairs, the changes show a larger variety, including changes, additions, or deletions of arguments, which could range from constructor calls, method calls, to literal values, among others. Consequently, the scope of potential repairs within the ARG category is broader than that within the ORC category, rendering it a more challenging repair category.

\TR{R2.4 R2.5}{
We believe that an iterative repair approach, as a future direction, could enhance the performance of \approachName{}, particularly in managing more complex test repairs. For instance, although ITER~\cite{ye2024iter} addresses program repair rather than test repair and is not a direct alternative to our approach, it is an influential technique that utilizes a specialized training method for iterative repair. Moreover, ITER supports multi-hunk repairs (also known as multi-location repairs~\cite{ye2024iter}), allowing it to address complex scenarios involving modifications across multiple locations. In addition, generating synthetic training instances could further improve \approachName{}'s performance. As mentioned earlier, complex repair types are underrepresented in our dataset. By generating synthetic data, we could introduce a greater variety of repair complexities in our training set, potentially enhancing the model's effectiveness in handling complex repairs. For example, a previous study~\cite{ye2024selfapr} generated one million synthetic training samples for the program repair task, which demonstrated superior performance compared to traditional learning methods.}

\subsubsection{\TRC{Qualitative Analysis}}

\TR{R1.17 R2.1}{To complement the quantitative results and gain deeper insight into \approachName{}'s performance in real-world scenarios, we conducted a qualitative analysis. By examining both successful and failed cases, we aimed to identify factors that contribute to \approachName{}'s repair effectiveness and highlight areas for improvement. This analysis not only strengthens the validation of our approach but also informs future advancements in test case repair. We manually analyzed many examples from different repair categories and projects, and selected eight representative and insightful cases that illustrate successes and common failures. To keep the main text concise, details regarding these examples are provided in the Appendix.}

\TRC{
We present four successful examples that demonstrate both exact-match and plausible repairs, underscoring the effectiveness of \approachName{} and its intelligent repairing of test cases. In some examples (successful examples 1 and 2 in the Appendix), \approachName{} generated exact-match repairs by accurately capturing the relevant context of the SUT and identifying key code change patterns. In other examples (successful examples 3 and 4), \approachName{} proposed a functionally equivalent repair, even though it did not exactly match the ground truth, maintaining the intent of the original test case. These examples highlight the model's ability to prioritize crucial information and apply learned patterns, even in challenging scenarios.}

\TRC{
Our analysis of failed cases identified two main factors contributing to unsuccessful repairs: (1) when multiple distinct changes are required (failure examples 2 and 3), and (2) when deeper contextual understanding of the SUT is necessary (failure examples 1 and 4), especially from both unchanged and changed code. In one failure for instance (failure example 1), the model was unable to repair an assertion due to a lack of contextual information about the underlying unchanged logic and API of the SUT. In another case (failure example 3), the model partially predicted the correct repair but introduced syntax errors due to the high volume of the required changes.}

\TRC{
Overall, the qualitative analysis highlights the strengths of \approachName{} in selecting, representing, and learning the repair context while revealing areas for future improvement, such as enhancing its capacity to process broader contextual information and manage multi-step repairs more effectively. These insights can guide future research to refine test case repair methodologies and address unresolved challenges.}

\begin{tcolorbox}[title=RQ2.1 Summary, breakable]\TRC{
\approachName{}'s performance declines as the complexity of repairs increases, particularly when the repairs involve a higher number or variety of code edits. While the model achieves a exact match accuracy of around 20\% even for more complex tasks, its accuracy is notably higher for simpler repairs. We also observed that the model's performance in the ARG category is relatively lower compared to that in other simpler repair instances, for reasons explained above.} 

\TRC{
These limitations highlight areas of improvement in future work. A multi-step and iterative approach is a reasonable avenue for exploration. Also, diversifying the training set with synthetic data to better balance the frequency of repair categories and complexity is a promising direction.}
\end{tcolorbox}

\subsection{Test Repair Trustworthiness Prediction (RQ2.2)}
\label{rq2.2}
Being able to determine, in practice, whether to trust a proposed repair would make \approachName{} much more practical as practitioners would not waste time considering low quality repairs. The question is then whether we can accurately predict the performance of \approachName{} for a given broken test case.

\subsubsection{Creating the Prediction Model}
To address this research question, we first propose the following features, which we hypothesize may be good indicators of repair correctness to be used by the prediction model.

\begin{itemize}
    \item \textbf{Similarity between the input's test context and the repair context.} We compute the similarity between the broken lines of the test code and the SUT's changed lines before the change, that are included in the input. We extract four features that relate to this similarity. The two initial features include the maximum and average cosine similarities derived from TF-IDF vectors. The third and fourth features involve the count of common AST nodes at hunk and node levels, respectively.
    \item \textbf{Overall complexity of changes in SUT.} We calculate the total number of changed files in the SUT as well as the total number of added and deleted lines.
    \item \textbf{Complexity of the test code.} We compute the number of lines (LOC) of the broken test method.
\end{itemize}

Secondly, we constructed a dataset using the above-mentioned features to train models for predicting whether a test is likely to be properly repaired. To achieve this, we used two distinct labels as the ground truth: (1) exact match repair and (2) plausible repair, hence creating two distinct prediction models. The labels indicating the model's success in generating exact match or plausible repairs were extracted from the results of the best-performing model. For instance, if the model generates an exact match repair for a particular broken test case, we identify it as a positive exact match label for that test. Since we have an imbalanced dataset (66\% exact match repairs and 80\% plausible repairs), we used random oversampling to improve performance for the minority class.

Thirdly, we conducted a 5-fold cross validation on the test set of \benchmarkName{}. For each of the five folds, we trained a Random Forest (RF) model. We chose RF due to its explainability and ability to generate accurate models while being robust to overfitting~\cite{delgado2014mlclf}. RF enables us to measure the importance of each feature on the decision-making process. Our RF model incorporates all available features and predicts the probability that our model correctly repairs a given test. Finally, we report and analyze the average performance across folds, including average precision, recall, and F1 score.

\subsubsection{RQ2.2 Results}
\begin{table}[tbp]
\centering
\resizebox{0.75\linewidth}{!}{%
\begin{tabular}{lrrr}
\toprule
Label & Precision & Recall & F1 \\
\midrule
Exact Match & 87 & 88 & 88 \\
Plausible & 90 & 94 & 92 \\
\bottomrule
\end{tabular}
}
\caption{Test repair effectiveness prediction results using the Random Forest classifier.}
\label{tab:rq22_confidence}
\end{table}

Table~\ref{tab:rq22_confidence} presents the performance of the RF model, achieving precision, recall, and F1 scores of 87\%, 88\%, and 88\% for the exact match label, and 90\%, 94\%, and 92\% for the plausible label. The RF models are therefore highly accurate in predicting whether our best test repair CLM (CT5+ with $IO_2$) can effectively repair a given broken test case. Notably, the RF model operates exclusively on input information, without utilizing any data related to the output of CLMs. This model characteristic enhances its applicability since there is no requirement for additional prompting of the CLM to get extra information. Furthermore, enhancing the accuracy of the RF model should be possible with richer training data including a greater number of repair instances and a wider variety of repairs. Indeed, as developers use our CLM and offer feedback on the correctness of the generated test repairs in various test repair scenarios, there is significant potential for refining the RF model.

Let us discuss the practical implications of the RF model's results. Regarding precision, when the RF model predicts that our CLM's repair is trustworthy, developers would benefit from correct test repairs in 87\% and 90\% of the cases in terms of exact match and plausible repair, respectively. In other words, by relying on the RF model, in only 13\% and 10\% of the cases would developers need to further repair incorrectly generated CLM repairs. Therefore, such predictions provide useful guidance to developers.

Examining the recall values allows us to understand the number of missed opportunities in terms of generated correct repairs. In 12\% of cases for exact match repairs and 6\% for plausible repairs, the RF model misleads  developers in suggesting that correct automatic repairs should not be trusted. Such cases do not entail extra effort for developers, as they would anyway have to repair such test cases manually without the CLM.

Last, we employed the \textit{Permutation Feature Importance} technique to measure the impact of the various features on the RF model's performance. This method assesses the impact of individual features by shuffling each feature's values while keeping others constant and observing the resulting changes in the model's predictive accuracy. By averaging the feature importance scores obtained across 5 cross-validation folds, we identified that, the most important features, in order of importance, are (1) the average similarity between TF-IDF vectors of test broken lines and the pre-change SUT lines, (2) the count of common AST nodes between the two, and (3) the LOC of the broken test method.
This suggests that the most important features in predicting the success of our CLM's test repair are related to the relationship between the broken lines of the test case and the pre-change SUT lines. Therefore, it is crucial to provide the most relevant SUT changes to the model's input while filtering out irrelevant changes. Further exploration focusing on this aspect, as well as incorporating data beyond the scope of SUT changes, remains a potential area for future research.

\begin{tcolorbox}[title=RQ2.2 Summary, breakable]\TRC{
We introduced a simple machine learning technique that accurately predicts whether \approachName{} will successfully repair a given broken test case. This capability enhances the application of \approachName{} by identifying instances where developer intervention is necessary. Additionally, our feature importance analysis highlights the critical role of the repair context, suggesting that refining this aspect could lead to a higher success rate in automated repairs.}
\end{tcolorbox}

\subsection{Impact of Fine-tuning Data Size on Test Repair Performance (RQ3.1)}
\label{rq3.1}
Often, CLMs are initially trained for general tasks and then fine-tuned for specific tasks such as test case repair. However, fine-tuning is a time-consuming process that requires substantial amounts of data, especially in the context of test case repair where test breakages are not common~\cite{kahiwa2021refactoringbreak, nagy2022cooccurrence}. Moreover, extensive fine-tuning datasets may be necessary to enhance the performance of a large CLM with hundreds of millions of parameters when compared to its pre-trained version. This RQ aims to analyze how the performance of test repair in CLMs is affected by the amount of fine-tuning data. By answering this question, we can gain valuable insights into the data requirements for fine-tuning to achieve effective test case repair.

\subsubsection{Downsizing Fine-tuning Data}
\label{sec:rq31_experiments}
To address this research question, we utilized the training set from \benchmarkName{}. We created four distinct fine-tuning data subsets, each containing 20\%, 40\%, 60\%, and 80\% of the original training data, achieved by excluding older data, for each project, based on commit dates. This process ensured a reduction in data size while preserving project diversity. Using these subsets, we fine-tuned four models using the best-performing settings identified in RQ1.1. Subsequently, we conducted an analysis to examine how varying sizes of the fine-tuning data impact the performance of test case repair.

\subsubsection{RQ3.1 Results}
\begin{figure}[tbp]
   \centering
   \includegraphics[width=\linewidth]{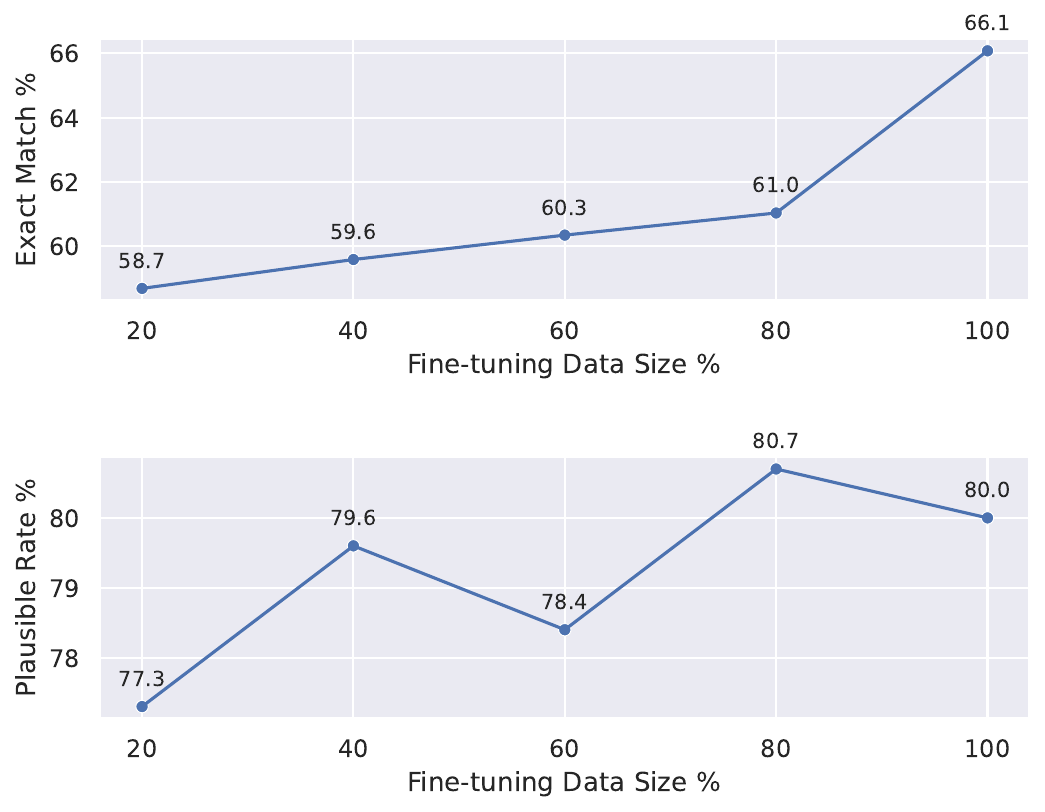}
   \caption{Test case repair performance on varying fine-tuning data sizes.}
   \label{fig:rq31}
\end{figure}

Figure~\ref{fig:rq31} depicts the impact of fine-tuning data sizes on test case repair performance. Initially, as the fine-tuning data size increases, we observe rising trends in both the EM and PR metrics, as expected. However, the effect is more noticeable on the EM than on the PR. For instance, increasing the fine-tuning data size from 20\% to 100\% results in an improvement of 7.4 pp in the EM, compared to a 2.7 pp improvement in the PR.

As a result, if we consider the PR as our criterion, our findings suggest that using approximately 7,000 fine-tuning instances yields a test repair performance similar to using around 36,000 instances. On the contrary, when considering the more conservative metric, EM, as our criterion, a trade-off between performance and data size becomes evident. Moving from 7,000 fine-tuning instances to 36,000 and starting from an initial 58.7\% EM, each addition of 7,000 instances leads to an average 1.85 pp improvement in EM.

\begin{tcolorbox}[title=RQ3.1 Summary, breakable]\TRC{
Our results show that having a larger fine-tuning dataset improves \approachName{}'s test repair performance, particularly in terms of exact match accuracy, which is the most conservative evaluation metric. While collecting high-quality training data can be expensive, generating synthetic training data may be a beneficial and cost-effective approach worth exploring in future research.}
\end{tcolorbox}

\subsection{Generalization of Fine-tuned Test Repair CLMs (RQ3.2)}
\label{rq3.2}
To complement RQ3.1, this RQ aims to determine whether models that are fine-tuned on a number of projects can be applied to other unseen projects without further fine-tuning. The results can reveal whether a more generic approach can be employed to address test code repair, eliminating or minimizing the need for project-specific fine-tuning.

\subsubsection{Excluding Projects from Fine-tuning Data}
\label{sec:rq32_experiments}
To address this research question, our approach involved creating random subsets of projects. Each subset was chosen to be excluded from the training set while keeping the test set and validation set unchanged. More precisely, we formed 10 folds by using a stratified sampling method, each fold comprising a specific number of projects. Six projects were sampled for all folds, except for the 10th fold, which included five projects due the dataset containing a total of 59 projects. The stratified sampling approach ensured that each fold represented the overall project population, categorized into three segments based on the number of test repair instances: small, medium, and large. The data in the last two rows of Table~\ref{tab:rq32_generalization} displays the count of test instances from the chosen projects in each fold, along with the number of train instances that were excluded from the training set.

Subsequently, for each fold, we fine-tuned a new model based on the best-performing settings (CT5+ \& IO2) identified in RQ1.1. The newly fine-tuned models were trained on the modified dataset, excluding the subset of projects corresponding to their respective fold. Finally, to assess the effectiveness and generalizability of these fine-tuned models, we conducted evaluations on the excluded projects. This evaluation process provided a measure of the models' performance on unseen project data, offering insights into their ability for generalization.

\subsubsection{RQ3.2 Results}
\begin{table*}[tbp]
\centering
\resizebox{\linewidth}{!}{%
\begin{tabular}{lrrrrrrrrrr}
\toprule
 & Fold 1 & Fold 2 & Fold 3 & Fold 4 & Fold 5 & Fold 6 & Fold 7 & Fold 8 & Fold 9 & Fold 10 \\
\midrule
Project-agnostic EM & 55.4 & 70.3 & 60.7 & 69.7 & 62.9 & 67.6 & 55.7 & 64.8 & 34.3 & 49.2 \\
Best (CT5+ \& $IO_2$) EM & 56.3 & 72.2 & 65.1 & 69.9 & 65.5 & 71.8 & 60.7 & 70.9 & 55.8 & 51.0 \\
\midrule
Evaluation Test Instances & 453 & 327 & 476 & 1169 & 437 & 632 & 458 & 2135 & 534 & 482 \\
Excluded Train Instances & 2583 & 1819 & 2613 & 3445 & 2023 & 3915 & 2681 & 13049 & 1919 & 2592 \\
\bottomrule
\end{tabular}
}
\caption{Generalization analysis results of fine-tuned test case repair models: The table shows the exact match accuracy (EM) and the counts of training and test instances of the selected projects across 10 folds. In each fold, six random projects were selected and excluded from the training data. EM values are specifically computed on the testing data solely from the excluded projects in each fold.}
\label{tab:rq32_generalization}
\end{table*}

The results presented in Table~\ref{tab:rq32_generalization} report the EM values of two models across 10 folds, as outlined earlier. The \textit{Project-agnostic} model refers to the model that was not exposed to the selected projects during fine-tuning in each fold, whereas the \textit{Best} model denotes the top-performing model identified in RQ1.1, fine-tuned on the entire training dataset. As shown in Table~\ref{tab:rq32_generalization} , the difference between the two models in terms of EM is relatively small, with an average EM difference of 4.87 pp.

To assess the statistical significance of the difference between the Project-agnostic model and the Best model across folds, we conducted a paired t-test. Our dataset satisfied the normality assumption, confirmed by the Shapiro-Wilk normality test, with p-values of 0.16 and 0.25 (p > 0.05) for the Project-agnostic EM data and the Best EM data, respectively. The paired t-test yielded a statistically significant difference (p-value = 0.034 < 0.05) between the two models. Moreover, to measure the observed difference's magnitude, we computed Cohen's d as the effect size, resulting in a value of 0.513. According to Cohen's guidelines~\cite{Cohen1988}, this effect size falls within the moderate range, leaning closer to a small effect size, given that values below 0.5 are categorized as indicative of a small effect.

The EM values of the Project-agnostic models, except for two of the folds, are above 50\%, averaging 59.1\% across all folds. Also, despite the statistical significance in the differences observed across folds, we note a small to moderate effect size and an average difference of 4.87 EM pp. These findings suggest that Project-agnostic models represent an acceptable alternative. Although Project-agnostic models do not fare quite like models fine-tuned on the entire training data, their results imply that \approachName{} is not restricted solely to project-specific fine-tuning and can generalize to unseen projects.

The above results have significant practical implications regarding  applicability. Currently, the most effective test repair methods~\cite{li2019intent} rely on dynamic symbolic execution, requiring project-specific analysis and execution, which poses a considerable challenge in practice. In contrast, as suggested by the above results, fine-tuned CLMs can produce satisfactory outcomes for unseen projects without the need for any dedicated project analysis.

\TR{R1.16}{The primary goal of this RQ is to evaluate the impact of cross-project training by comparing the performance of our best model against project-agnostic models, regardless of the specifically performance level observed of each individual fold. Notably, our best model achieves an EM of around 51\% and 56\% on folds 1 and 10, respectively, while achieving around 72\% and 71\% on folds 6 and 8.}

\TRC{To further investigate performance variability across different folds, we employed the same metrics used in RQ2.1, specifically the number of AST-level edit actions and repair categories (ARG, INV, ORC). Both the AST actions metric, outlined in Section~\ref{rq2.1}, and the number of repair categories, ranging from 1 to 3, measure the complexity of the repair.
For each fold, we calculated the average value of these metrics across repair instances and computed the Pearson correlation coefficient ($r$) between each metric and the EM. For our best-performing model, we found negative correlations of $r=-0.57$ for AST actions and $r=-0.59$ for repair categories. The project-agnostic models showed $r=-0.21$ and $r=-0.51$, respectively. These results indicate a moderate but significant negative correlation, suggesting that as AST actions or repair categories increase, model performance tends to decrease. This trend explains the lower EM values in certain folds, a pattern consistent in both project-agnostic and our best-performing model. For instance, in fold 10, where the EM is 51\%, the average number of AST actions are 3.28 with a standard deviation of 3.22, whereas in fold 8, where the EM is 71\%, this average is 1.90 with a standard deviation of 2.07. This analysis aligns with the findings of RQ2.1, reinforcing the observed relationship between repair complexity and performance without altering our conclusion in this RQ regarding the limited impact of project-agnostic models on performance.}

\begin{tcolorbox}[title=RQ3.2 Summary, breakable]\TRC{
Our findings show that fine-tuning on project-specific data has a relatively limited impact on performance. Despite the challenges introduced by complex test repair tasks in certain projects, this suggests that \approachName{} provides a relatively generalizable solution that is not project-specific.}
\end{tcolorbox}

\subsection{Implementation and Resources}
\label{sec:implementation}

\textbf{Implementation.} We implemented CLMs using the Hugging Face Transformers~\cite{wolf2020transformers} library and the Accelerate~\cite{gugger2022accelerate} library. In each fine-tuning setting, we trained the model for 4 epochs and employed an early-stopping strategy, which involved monitoring the validation loss after each epoch. If the validation loss did not decrease after 1 epoch, we stopped fine-tuning and selected the model with the smallest validation loss. We configured a beam size of 200 for PLB and 40 for CT5+ and CG, along with batch sizes of 8 for PLB, 1 for CT5+, and 2 for CG. We used the Adam optimizer with weight decay (AdamW) and set a learning rate of $1 \times 10^{-5}$, in combination with a cosine learning rate scheduler.

\TR{R2.3}{We anticipate that the primary contributions of our work, including relevant hunk selection, representation, and fine-tuning of CLMs, are not dependent on any specific programming language and can therefore be widely adopted. The CLMs we utilize are pre-trained on multiple languages, such as Java and Python, and our most effective hunk selection strategy is language-agnostic, as it relies on the textual content of code changes. However, the current implementation of our approach focuses specifically on Java, particularly for repairing JUnit tests. To achieve optimal results for other programming languages, we recommend additional efforts in gathering language-specific test repair data and selecting a CLM with pre-training that is more tailored to the target language.}

\textbf{Computation Resources.} We conducted our fine-tuning experiments on a machine with two Nvidia Quadro RTX 6000 GPUs (each with 24GB GPU memory), an Intel Xeon Gold 6234 16-Core CPU, and 187GB of RAM. We used the Digital Research Alliance of Canada~\cite{computecanada} servers for concurrently executing test cases at a large scale.

\subsection{Threats to Validity}
\TRC{\textbf{External Validity.}}
\TR{R1.2}{Two threats to the external validity of our evaluation are the use of possibly simple test repairs and the applicability of \approachName{} to real-world development scenarios. To mitigate the former, we have taken steps to ensure our evaluation set is fair and realistic by excluding simple test repairs that can be addressed by an IDE or CI automation tools. Specifically, we identified and excluded trivial repairs (discussed in Section~\ref{sec:tarbench_quality}), such as renaming class or method names, as their inclusion would have inflated our results. The number of these trivial repairs was not negligible, reinforcing the importance of their exclusion for an accurate assessment.}

\TRC{The applicability of \approachName{} to real-world development environments was carefully considered in our data collection process. We aimed to replicate real-world conditions by sourcing high-quality data samples from over 1,200 representative Java open-source projects on GitHub, focusing on test case repairs extracted from actual commits. We also executed the test cases to further ensure the authenticity and reliability of the repairs. No synthesized elements (e.g., faults and repairs) were introduced; all data were derived from real-world development environments.}

\TRC{\textbf{Internal Validity.}}
\TR{R1.15 R3.15}{We acknowledge that the PR metric has a limitation in not always accounting for the preservation of the tests' semantics and intent. For instance, the removal of assertions could be considered a plausible repair, even if the ground truth repair does not involve such a change (though there are cases, such as when a feature is removed from the SUT, where this is necessary). In the following paragraphs, we discuss the strategies employed to mitigate this issue.}

\TRC{Our approach focuses on repairing only the breakage lines, which is discussed Section~\ref{sec:approach}. This limits the capability of the model to change the semantics of the test code as it can only change a few lines. Also, the PR metric is not the only metric we used for evaluation. We also employed three other metrics: exact match accuracy (EM), CodeBLEU, and BLEU. These three metrics are highly sensitive to unwanted changes, such as removing assertions, because they are text-based and not execution-based. Therefore, the diversity of the metrics we use mitigates the limitation of the PR metric.}

\TRC{Further, to assess the impact of the PR metric limitation on our best-performing model (CT5+ with $IO_2$), we conducted a detailed analysis. We focused on predictions where the model generated at least one plausible repair candidate without producing any exact matches, as exact matches are definitely correct and irrelevant to this particular analysis. From these non-exact-match and plausible instances, we identified a subset of 377 cases where the keyword "assert" appeared in either the ground truth repair or the model’s plausible repair candidate.
In the most conservative scenario, we considered the possibility that all 377 of these instances might be misclassified as plausible when they are actually non-plausible. For instance, 123 of these cases involved the deletion of assertions, while the ground truth did not involve such deletions. Even under this worst-case assumption, where all 377 instances are reclassified as non-plausible, the PR value for our model would only decrease by 5.3 percentage points, from 80.0\% to 74.7\%. This small reduction demonstrates that the overall impact on the model's performance is limited.}

\section{Related work}
\label{sec:related-work}
In this section, we discuss existing research related to the automated repair of broken tests, along with an analysis of their challenges. Additionally, we provide a concise overview of automated program repair studies that use language models. These two areas of research share similarities in their use of language models for repairing source code, even though their purposes differ. 

\subsection{Automatic Repair of Broken Test Cases}
\label{sec:relw_atr}

Despite the importance of repairing broken tests, there are only a few studies that focus on this problem, as discussed below. The first work in this context is a tool called \textit{ReAssert}~\cite{daniel2009reassert} that repairs broken JUnit test cases using a set of heuristic repair strategies (e.g., replacing the expected value in an assertion statement). Daniel et al. ~\cite{daniel2010test} extended \textit{ReAssert} by using symbolic execution to repair broken tests. In order to repair the tests this way, the expected value of the assertion is modified using a solution to symbolic constraints. These constraints are built based on the literals which contribute only to the assertion's expected value. In this way, the actual value of the assertion remains unmodified. The solution to the symbolic constraints are used to update literal values in the test case, and if no solution exists, the test is deemed unrepairable~\cite{daniel2010test}. 

Mirzaaghaei et al.~\cite{mirzaaghaei2014automatic} developed a framework named \textit{TestCareAssistant} (TCA) to repair broken tests or generate new tests based on existing ones. TCA repair algorithms focus on repairing tests whose covering code undergoes specific types of changes, such as adding parameters to a method being covered by a test. The repair algorithms work by replacing variables and values within the test to identify potential candidate repairs~\cite{mirzaaghaei2014automatic}. 

Xu et al.~\cite{xu2014using} proposed \textit{TestFix} that uses a genetic algorithm to repair broken test cases by finding a sequence of adding or deleting method calls which results in a repaired test. \textit{TestFix} can only repair tests with a single assert statement~\cite{xu2014using}.

Li et al.~\cite{li2019intent} developed \textit{TRIP} that focuses on preserving the intent of the test during its automatic repair~\cite{li2019intent}. \textit{TRIP} generates candidate repairs by modifying the test using a search-based technique. The test case is modified while considering the test-accessible elements (e.g., a public method). The search algorithm replaces broken calls to elements that are no longer test-accessible with one or more calls to elements that are accessible in the updated SUT~\cite{li2019intent}. \textit{TRIP} uses dynamic symbolic execution of the original test and repair candidates to determine the similarity between their intent, which is used to rank candidates~\cite{li2019intent}.

\begin{table*}[tbp]
    \resizebox{\linewidth}{!}{%
    \begin{tabular}{lp{0.45\linewidth}rrll}
        \toprule
        \multirow{2}{*}{Project} & \multirow{2}{*}{Repair Categories} & \multicolumn{2}{c}{Evaluation Benchmark} & \multicolumn{2}{c}{Reproducibility} \\
        \cmidrule(lr){3-4} \cmidrule(lr){5-6}
        & & \# Projects & \# Tests & Repeatable & Reusable \\
        \midrule
        \textit{ReAssert}~\cite{daniel2009reassert} & Test oracles  & 6 & 170 & No & No \\
        \textit{Symbolic Test Repair}~\cite{daniel2010test} & Repairs that require modifying literal values & 14 & 235 & No & No \\
        \textit{TestCareAssistant}~\cite{mirzaaghaei2014automatic} & Method parameters and return types & 5 & 138 & No & No  \\
        \textit{TestFix}~\cite{xu2014using} & Single assertion statements & 1 & 6 & No & No  \\
        \textit{TRIP}~\cite{li2019intent} & All except repairs requiring long sequences of functional calls or repairs involving generic types & 4 & 91  & Yes & No  \\
        \approachName{} & Any repair category & 59 & 45,373 & Yes~\cite{targetGitHub} & Yes  \\
        \bottomrule
    \end{tabular}
    }
    \caption{Criteria for comparing our work with existing work}
    \label{table:criteria}
\end{table*}

Our analysis of the aforementioned existing work reveals the following three main issues that our work (\approachName{}) aims to address. 

\textbf{Repair Categories:} Overall, as discussed in Section~\ref{rq2.1}, a repair category may cover one or more change categories including: (1) Invocation argument or return type change (ARG), (2) Invocation change (INV), and (3) Test oracle change (ORC).

Unlike existing work, thanks to our reliance on language models, \approachName{} is not limited to handling specific repair categories, with the caveat that it will of course perform better on repair categories that are more common in the training dataset. As shown in column~\textit{Repair Categories} of Table~\ref{table:criteria}, \approachName{} is capable of repairing failing test oracles, in contrast to \textit{TestFix}~\cite{xu2014using}. \textit{ReAssert}~\cite{daniel2009reassert} and \textit{Symbolic Test Repair}~\cite{daniel2010test} are unable to modify method calls in the broken test, which is something that \approachName{} can accomplish. Finally, unlike \textit{TestCareAssistant}~\cite{mirzaaghaei2014automatic}, \approachName{} is able to modify the literal values of variables. Thus, \approachName{} is a more generic approach compared to those discussed above. \approachName{} is also a comprehensive test repair tool capable of addressing a combination of repair types covered by existing studies, while the other described approaches would need to be combined with one another to achieve similar comprehensiveness.

\textbf{Evaluation Benchmark:}
As shown in the \textit{Evaluation Benchmark} column of Table~\ref{table:criteria}, existing studies utilize different benchmarks to evaluate their proposed approaches, with none of theses studies using common benchmarks to enable comparisons. Moreover, reported accuracy percentages range from 44.7\% to 100\% in terms of plausible repair accuracy (proportion of tests that run without a failure after being repaired), using benchmarks containing less than 250 broken tests. Such limited benchmark size raises concerns about the validity of their evaluation regarding the effectiveness and applicability of these approaches across various software projects and repair scenarios.

To address this limitation, we evaluated \approachName{} using a significantly larger and more comprehensive benchmark, \benchmarkName{}, which is constructed from 59 open-source Java projects, comprising a total of 45,373 broken tests. In this evaluation, \approachName{} generated plausible repairs for 80\% of the broken tests, demonstrating its effectiveness and scalability on a much larger and more diverse set of real-world scenarios.

\textbf{Reproducibility and Future Research:} 
As shown in the \textit{Reproducibility} column of Table~\ref{table:criteria}, all studies, except one, do not provide their benchmarks and the source code of their implementations and experiments, thus making their replication impossible. Also, in the case of \textit{TRIP}~\cite{li2019intent}, only a dataset comprising 91 broken tests is made available. However, \textit{TRIP}'s source code is unavailable and that makes it unusable and inapplicable to other projects. Consequently, none of the existing studies are deemed reproducible. To address this issue and provide a foundation for future research in the context of broken test repairs, we provide the implementation of \approachName{}, \benchmarkName{}, and the script for reproducing all experiments~\cite{targetGitHub,tarbenchFigshare}.

Additionally, extending existing solutions may prove challenging due to limited access to their source code and their primary focus being Java and JUnit, with the exception of \textit{Symbolic Test Repair} which was also evaluated on .NET tests~\cite{daniel2010test}. Similarly, \approachName{} was fine-tuned specifically to repair tests written in Java with the JUnit framework. Nonetheless, it is worth noting that in contrast with existing solutions, fine-tuning \approachName{} for repairing tests in other languages and testing frameworks can be easily achieved. 
This is because the underlying language models are pre-trained on multiple programming languages. For example, \textit{PLBART} underwent pre-training on both Python and Java~\cite{ahmad2021unified}.

\TR{R1.12}{
We highlighted above reproducibility issues, noting that most works do not provide their benchmarks and implementation details, making it impossible to compare them with our approach. Though \textit{TRIP}~\cite{li2019intent} made its dataset and repair tool available, we encountered significant challenges in comparing \approachName{} with \textit{TRIP}. Firstly, the \textit{TRIP} repair tool failed to execute on our dataset. Since the authors only provided executables without the accompanying source code, we were unable to troubleshoot and resolve the failure. Also, we were unable to get assistance from the authors. Secondly, our methodology requires access to  the Git repositories of the projects to collect test repair data. However, the TRIP dataset does not include links to Git repositories, commits, or tags. Consequently, we were unable to perform a meaningful comparison between \approachName{} and TRIP.
}

\TRC{
Additionally, being the most recent and closest work to ours, we empirically compared \ceprot{}~\cite{hu2023ceprot} with \approachName{} in Section~\ref{sec:rq12} to evaluate their respective performance. In the following, we analyze in depth the issues and limitations of \ceprot{} and its empirical validation in comparison to \approachName{} and \benchmarkName{}. This analysis includes the examination of the dataset, approach, and experimental design.}

\TRC{
\textbf{Dataset Diversity and Quality:} \ceprot{} aligns test methods with their corresponding production methods by matching file paths and method names. It then defines a test repair as a co-evolving production-test pair within the same commit. When both a test method and its associated production method are modified in a commit, \ceprot{} labels the pair as an obsolete test case repair. However, this approach limits the diversity of the dataset. A test method might interact with multiple classes or methods in the production code for purposes such as initialization, which are not directly tested in the specific test case. These additional classes or methods may also need to evolve alongside the production code, but \ceprot{} fails to identify these cases.}

\TRC{
Additionally, this approach can potentially collect instances that are not genuine test repairs. For example, \ceprot{} might mistake test code refactoring cases for obsolete test repairs. If a variable is renamed in both the test and production code, \ceprot{} incorrectly identifies it as a repair instance, even though it is not a true repair. \ceprot{} can also capture unsuccessful or partial repairs because it does not verify the success of the repair.}

\TRC{
In contrast, \benchmarkName{} addresses these limitations by ensuring greater dataset diversity and quality through a three-step test execution process. First, \benchmarkName{} verifies that the test case works correctly in its original state by executing it on the original production code and confirming its success. Second, it confirms that the test case has become obsolete by executing the original test code on the updated production code and expecting it to fail. Finally, it ensures that the updated test code executes successfully on the updated production code. This method effectively overcomes the limitations of \ceprot{} by providing a dataset with higher quality and more diversity.}

\TRC{
\textbf{Trivial Repairs:} \benchmarkName{} identifies and excludes trivial repairs, i.e.,  renaming class or method names, which can be automatically accomplished by code editors and IDEs. This ensures that its evaluation results are realistic. In contrast, \ceprot{} does not follow this practice. In our application of \approachName{} to \ceprot{}’s subset of 214 test instances (Section~\ref{sec:rq12}), we identified 37 (17\%) trivial repairs. Including these trivial repairs produces overly optimistic results.
}

\TRC{
\textbf{Dataset Size:} \approachName{} is fine-tuned and evaluated on a significantly larger dataset compared to \ceprot{}. Specifically, \ceprot{} includes 4,676 training instances, whereas \benchmarkName{} comprises 36,639 training instances. This substantial difference enhances the effectiveness of the fine-tuning process for \approachName{}. Additionally, \ceprot{} has only 520 evaluation instances, in contrast to the 7,103 evaluation instances in \benchmarkName{}. This larger and more varied evaluation set provides a more diverse and realistic assessment of the model's performance.}

\TRC{
\textbf{Dataset Issues:} In addition to the data limitations previously mentioned, our analysis of \ceprot{}'s data revealed other issues.
First, we observed discrepancies in the commit attribution between test cases and focal (production) methods. Specifically, in 511 instances (11\%) within the training set and 60 instances (11\%) within the test set, the commit attributed to the test case differed from that of the focal method. This contradicts the authors' assertion that the test and focal methods are changed in the same commit. Moreover, 33 instances from the training set and 2 instances from the test set even differed in the project associated with the commits.}

\TRC{
Second, we identified instances where the "test\_src" and "test\_tgt" values, which are preprocessed code and are utilized for training and testing, were completely identical. This occurred in 234 instances (5\%) of the training set and 22 instances (4\%) of the testing set, indicating no code change in the test repair, potentially leading the model to learn to generate outputs that exactly match the inputs.}

\TRC{
Third, we found that in 355 training instances (7\%) and 36 testing instances (7\%), the method names of the source and target test code were different. Since the authors did not provide clarification on how they matched these cases, they are potentially mismatches.}

\TRC{
These findings highlight inconsistencies in the data, which could impact the validity of the authors' results and the overall reliability of their model.
}

\TRC{\textbf{Limitations:} We identified two limitations in \ceprot{}’s approach and experiments that \approachName{} effectively addresses. First, \ceprot{}'s authors state that their train-test split is done randomly. However, for a realistic evaluation, it is crucial that more recent test case repairs are not included in the training data. This ensures that the model is not exposed to future information, which could contain the solutions to older test repairs. To address this, in each project within \benchmarkName{}, we ensured that older commits were included in the training set while newer commits were reserved for the test set.}

\TRC{Second, \ceprot{}'s implementation has a restrictive token limit of 150 for the test method, focal method, and edit sequence, truncating any instances that exceed this limit. Additionally, \ceprot{} imposes a 150-token limit for output generation. In contrast, \approachName{} supports a much larger token limit of 512 for both input and output. This allows \approachName{} to process the full test code and all relevant SUT (System Under Test) code changes, ensuring a more comprehensive and accurate analysis.}

\TRC{To conclude, in addition to outperforming (Section~\ref{sec:rq12}) the most recent and relevant approach (\ceprot{}), \approachName{} offers a more comprehensive and adaptable solution for repairing broken test cases. \approachName{} leverages the flexibility of language models to address a wider range of repair categories. Additionally, in contrast to \ceprot{}, \approachName{} is fine-tuned and evaluated using an extensive dataset (\benchmarkName{}), which provides a large-scale, realistic assessment of repair effectiveness. This also addresses the scalability and validity concerns present in other studies. Finally, \approachName{} distinguishes itself by being fully reproducible and extensible, with its codebase and dataset openly available, thereby setting a new baseline for future research in the field.}

\subsection{Automated Program Repair}
Automated program repair (APR) is a closely related research field to automated test repair, as both aim to automatically repair source code. However, APR focuses on repairing faulty code by taking a defective program along with a test suite that exposes the program's faults. The goal of APR is to generate a repair that corrects the program's faults to ensure that the provided test suite passes comprehensively.

While APR techniques are effective at repairing faulty code, they are not applicable to repairing broken test cases due to (1) APR's correctness and training relying on a set of tests, whereas in test case repair the tests themselves are flawed, (2) APR addressing problems arising from flawed behavior in the SUT code. In contrast, broken tests result from changes in the SUT. This necessitates the provision of SUT changes along with the broken tests to the model to learn repairs as we will discuss in Section~\ref{sec:approach}. Finding and providing SUT changes is challenging given the limited input size of language models and large sets of SUT code changes. 

Over the last few years, there has been significant work put into the application of language models to repair program faults. Most applications tend to achieve a 15\% to 20\% repair rate on program faults, with the best results reach repair rates as high as 23\%. Fortunately, there are a couple of widely used benchmark datasets designed for APR, meaning that different studies are easily comparable. One of the more commonly used datasets is \textit{Defects4J}~\cite{just2014defects4j}, which contains hundreds of bugs from Java programs. There are two widely used versions of Defects4J, v1.2 and v2.0 which contain 393 bugs and 837 bugs, respectively~\cite{defects4jGit}. The other commonly used dataset is \textit{QuixBugs}~\cite{lin2017quixbugs}, which contains both Java and Python bugs. \textit{QuixBugs} is a much smaller dataset, containing only 40 bugs.

Liu et al.~\cite{liu2019tbar} built a tool called \textit{TBar}, which uses templates to repair bugs in Java programs. These templates are patterns of common bugs which specify a strategy to use for repair (e.g. inserting a null check before a buggy statement). Liu et al. found that \textit{TBar} could repair 18.7\% of the considered bugs~\cite{liu2019tbar}. \textit{TBar} is commonly used as a point of comparison for other subsequent works in this area.

Jiang et al.~\cite{jiang2021cure} built a neural machine translation model for APR called \textit{CURE}, which uses novel beam search strategies to select repair candidates. \textit{CURE} was able to repair 19.2\% of analyzed bugs, and the authors noted that some of those repairs were cases where state-of-the-art techniques failed due to having no applicable repair patterns~\cite{jiang2021cure}.

Drain et al.~\cite{drain2021generating} presented \textit{DeepDebug}, a model which uses pre-trained transformers to generate fixes for faulty functions written in Java. \textit{DeepDebug} achieves a 14.9\% exact match accuracy overall. Drain et al. also note that datasets for fault-related changes are collected via heuristics (e.g. all changes in a project repository which contain specific words such as "bug"), resulting in noisy datasets that impact their accuracy~\cite{drain2021generating}. Zhu et al.~\cite{zhu2021syntax} developed \textit{Recoder}, which uses a tree-based Transformer~\cite{vaswani2017attention} in order to generate program repairs. \textit{Recoder} achieved a 16.7\% repair rate, which outperformed state-of-the-art models on the same dataset~\cite{zhu2021syntax}.

Ye et al.~\cite{ye2022neural} produced a neural translation model, \textit{RewardRepair}, that was designed to produce compilable patches, as they noted that previous models tend to generate patches which do not compile. They do this by factoring compilation results in the training loss function. \textit{RewardRepair} was able to repair 23\% of the assessed bugs, and had a 45.3\% compilation rate in the top 30 candidate patches~\cite{ye2022neural}. Xia and Zhang~\cite{xia2022less} proposed \textit{AlphaRepair}, which was designed to produce patches without fine-tuning the underlying pre-trained model on repair datasets. This was done with the intent of avoiding common issues regarding the quality and quantity of program repair datasets~\cite{xia2022less} required for fine-tuning. The buggy line is completely masked before being fed into \textit{AlphaRepair} along with surrounding code context, so that the APR problem is framed as a code generation problem. \textit{AlphaRepair} was able to repair 18.9\% of the utilized buggy programs~\cite{xia2022less}. Jiang et al.~\cite{jiang2023knod} developed an APR approach called \textit{KNOD}, which aims to use domain knowledge to inform candidate patch generation. Domain knowledge is used to train \textit{KNOD} by using grammar logic formulae as part of the loss function. \textit{KNOD} achieved a 16.8\% repair rate on the analyzed benchmarks~\cite{jiang2023knod}.

Jiang et al.~\cite{jiang2023impact} recently studied the performance of language models and DL-based APR techniques in order to compare the performance of program repair techniques. They applied four different language models to the APR problem: \textit{PLBART}~\cite{ahmad2021unified}, \textit{CodeT5}~\cite{wang2021codet5}, \textit{CodeGen}~\cite{nijkamp2022conversational}, and \textit{InCoder}~\cite{fried2022incoder}. These language model-based techniques were compared to four DL-based APR techniques: \textit{CURE}~\cite{jiang2021cure}, \textit{RewardRepair}~\cite{ye2022neural}, \textit{Recoder}~\cite{zhu2021syntax}, and \textit{KNOD}~\cite{jiang2023knod}.
Results showed that language models are able to be competitive with and surpass the DL-based techniques~\cite{jiang2023impact}. The latter generated 11.6\% correct repairs on average, while the former were able to generate 10.7\% correct repairs on average before fine-tuning, and 22.3\% correct repairs on average after fine-tuning. Jiang et al. specify that data leakage is a point of concern for language model-based techniques, as there is a risk that code from open-source code repositories is used in the training dataset, and later contained in the test dataset when the models are evaluated~\cite{jiang2023impact}.

We can perform a comparison of our results to the results of APR techniques by comparing their rate of plausible repairs with ours. For APR, a plausible repair is defined as a repair which passes the test suite revealing the fault~\cite{liu2019tbar}, \cite{jiang2021cure}, \cite{ye2022neural}, \cite{zhu2021syntax}, \cite{xia2022less}. Although we do not have a test suite to determine plausible repairs in the context of test repair, this is similar to how we define plausible test repairs. A plausible repaired test passes on the new SUT version. The APR methodologies achieve a PR that ranges from 21.1\% to 32.1\%, with an average of 26.1\%~\cite{liu2019tbar}, \cite{jiang2021cure}, \cite{zhu2021syntax}, \cite{xia2022less}, \cite{jiang2023knod}. In contrast, we were able to achieve a PR of 80\%. Though we cannot use the same benchmarks and there are significant differences in the problems being addressed between APR and test repair, this comparison provides a rough indication of how comparatively good \approachName{}'s results are.

\section{Conclusion}
\label{sec:conclusion}
This paper introduces \approachName{} (\approachNameAbbr{}), a new method that utilizes language models to automatically repair broken test cases. \approachName{} formulates test repair as a language translation task and fine-tunes pre-trained CLMs. The best-performing CLM (CodeT5+ 770M) achieves a 66.1\% exact match accuracy (EM) and an 80\% plausible repair accuracy (PR) on an extensive benchmark.

Beyond \approachName{}, we developed \benchmarkName{}, a benchmark comprising 45,373 broken test repairs across 59 distinct projects, making it by far the most comprehensive benchmark publicly available. Additionally, we address crucial research questions through a large-scale empirical study to explore different configurations of \approachName{}, analyze its effectiveness across repairs, introduce a model for predicting the repair reliability of \approachName{}, and assess its generalizability across diverse software projects.

We show that leveraging the test repair context in our best configuration outperforms the performance obtained without utilizing this context. We observe notable improvements of 37.4, 24.1, 13.8, and 13.8 percentage points in terms of EM, PR, CodeBLEU, and BLEU, respectively.
Additionally, we noted a small difference between the performance of PLBART 140M and CodeT5+ 770M, the smallest and largest CLMs in our study, respectively. Thus, a high-performing CLM can be fine-tuned and used at a significantly reduced cost without  significantly sacrificing effectiveness.

Our findings also show that the effectiveness of \approachName{} degrades when handling more complex test repairs, particularly those involving multiple repair categories or AST edit actions. Furthermore, our approach seems more effective at repairing test oracles when compared to other repair categories.
Moreover, our repair reliability predictive model helps make \approachName{} more applicable in practice, offering practical guidance to developers by filtering out low-quality generated repairs, thus saving valuable time and effort.
As it may be intuitively expected, we found that a decrease in fine-tuning data size negatively impacts test repair effectiveness in terms of EM. Finally, our results demonstrate that \approachName{} is not limited to project-specific fine-tuning and has the capability to generalize across previously unseen projects.

In conclusion, the findings detailed above strongly suggest that \approachName{} can be an effective and practical solution to support test repair. 
Our study is the first that applies CLMs for repairing broken tests. With the promising results obtained, we anticipate several avenues of research for refining this approach. Potential improvements include, but are not limited to: (1) an investigation of plausible repairs that are not an exact match for their semantic equivalence with the ground truth, measuring their quality and reliability, (2) exploring ways to further improve the repair context by expanding the search scope beyond SUT changes and by eliminating irrelevant context, and (3) extending the diversity of programming languages in the benchmark.

\section*{Acknowledgement}
This work was supported by a research grant from Huawei Technologies Canada, Mitacs Canada, as well as the Canada Research Chair and Discovery Grant programs of the Natural Sciences and Engineering Research Council of Canada (NSERC) and the Research Ireland grant 13/RC/2094-2. This research was also enabled in part by the computation support provided by Digital Research Alliance of Canada~\cite{computecanada}.

\bibliographystyle{IEEEtran}
\bibliography{main.bib}

\clearpage

\onecolumn
\begin{appendices}

\section{Examples of Test Case Repairs Generated by \approachName{} on \benchmarkName{}: Successes and Failures}
In this section, we present eight detailed examples---four successful repairs and four failed repairs---to complement the qualitative analysis of \approachName{} discussed in the paper.

\TR{C1.4}{Successful and failed repairs are determined using two well-established evaluation metrics: Exact Match Accuracy (EM) and Plausible Repair Accuracy (PR). These metrics are clearly defined, justified, and consistently applied throughout our study. To reiterate, a repair is a successful exact match repair if it is identical to the ground truth repair.
A repair is a successful plausible repair if it compiles and runs to completion without any failures. In selecting the examples, we followed these criteria. Additionally, we ensured diversity in the repair characteristics and avoided plausible examples that did not uphold the semantics and intent of the test case (see limitations of PR outlined in our paper).}

\subsection{Failure Examples}
\subsubsection{Failure Example 1}
\label{sec:fe1}

The analysis of the failure case in Figure~\ref{fig:fail1} reveals several critical insights. According to the ground truth in Figure~\ref{fig:fail1_gt}, no changes were initially expected in the output based on the provided input. However, following the repair, the system expected the addition of a new annotation, \textit{@CanIgnoreReturnValue} (\textit{@CIRV}), which was expected to appear above a method declaration, along with its corresponding import.

The predictions generated by \approachName{}, as seen in Figure~\ref{fig:fail1_cr}, indicate that the model did not attempt to change the expected output lines and continued to expect no changes. This misunderstanding is evident in the method calls of the model's predictions, where the model, in one instance, even removed a method call. Although the relevant repair context, shown in Figure~\ref{fig:fail1_ric} in line 6, hinted at a behavioral change related to \textit{@CIRV}, this context alone was insufficient for the model to correctly perform the repair.

A deeper analysis shows that the model lacked critical contextual information. Specifically, it needed details about the system under test (SUT) and the test code that were not provided. Firstly, the model required awareness of the API associated with the \textit{helper} variable and the presence of the \textit{addOutputLines} method. Secondly, it needed a more comprehensive understanding of the SUT's logic, including how annotations like \textit{@CIRV} are applied.

In conclusion, this failure can be attributed to missing contextual information. To improve repair performance in such cases, a strategy that incorporates repository-level repair context-—considering both unchanged and changed parts of the codebase-—would likely perform better.

\begin{figure}[H]
    \caption{\textbf{Failure Example 1} -- An example of missing essential repository-level unchanged contextual information from the model's input.\\\textbf{\benchmarkName{} ID:} \textit{google/error-prone:25}\\\textbf{GitHub Commit Reference:} \href{https://github.com/google/error-prone/commit/428aa98972}{\textcolor{blue}{\textit{\underline{https://github.com/google/error-prone/commit/428aa98972}}}}}
    \label{fig:fail1}
    \centering
    \begin{subfigure}{\linewidth}
        \centering
        \begin{tcolorbox}[codebox]
        \begin{lstlisting}[language=Java,gobble=10,numbers=left,style=codeHighlighting,showstringspaces=false]
            @Test
            public void testBuilder_abstractClass() {
            <@\quad@>helper
            <@\quad@><@\quad@>.addInputLines(
            <@\quad@><@\quad@><@\quad@>"Builder.java",
            <@\quad@><@\quad@><@\quad@>"package com.google.frobber;",
            <@\quad@><@\quad@><@\quad@>"public abstract class Builder {",
            <@\quad@><@\quad@><@\quad@>"  public abstract Builder setName(String name);",
            <@\quad@><@\quad@><@\quad@>"  public abstract Builder clone();",
            <@\quad@><@\quad@><@\quad@>"  public abstract Builder copy();",
            <@\quad@><@\quad@><@\quad@>"}")
            <@\quad@><@\quad@>&- .&^expectUnchanged^&(&^)^&&
            <@\quad@><@\quad@>`+ .`@@addOutputLines@@`(`
            <@\quad@><@\quad@>@@+   "Builder.java",@@
            <@\quad@><@\quad@>@@+   "package com.google.frobber;",@@
            <@\quad@><@\quad@>@@+   "import com.google.errorprone.annotations.CanIgnoreReturnValue;",@@
            <@\quad@><@\quad@>@@+   "public abstract class Builder {",@@
            <@\quad@><@\quad@>@@+   "  @CanIgnoreReturnValue",@@
            <@\quad@><@\quad@>@@+   "  public abstract Builder setName(String name);",@@
            <@\quad@><@\quad@>@@+   "  public abstract Builder clone();",@@
            <@\quad@><@\quad@>@@+   "  public abstract Builder copy();",@@
            <@\quad@><@\quad@>@@+   "}")@@
                .doTest();
            }
        \end{lstlisting}
        \end{tcolorbox}
        \subcaption{\textbf{Ground Truth} Repair}
        \label{fig:fail1_gt}
    \end{subfigure}
\end{figure}
\begin{figure}[H]\ContinuedFloat    
    \begin{subfigure}{\linewidth}
        \centering
        \begin{tcolorbox}[codebox]
        \begin{lstlisting}[language=Java,gobble=10,numbers=left,style=codeHighlighting,showstringspaces=false]
            &-  .&^expectUnchanged^& ( )&
            `+  .`@@expectNoChanges@@` ( )`
            
            ----------------------------------
            
            &-  &^.expectUnchanged^& &^( )^
            `+  `@@/@@@@/@@` `@@Deleted@@
            
            ----------------------------------
            
            &-  .&^expectUnchanged^& ( )&
            `+  .`@@expectNoChange@@` ( )`
            
            ----------------------------------
            
            &-  .&^expectUnchanged^& ( )&
            `+  .`@@expectNoChanged@@` ( )`
        \end{lstlisting}
        \end{tcolorbox}
        \subcaption{Selected Top \textbf{Repair Candidates} Generated by \approachName{}}
        \label{fig:fail1_cr}
    \end{subfigure}
\end{figure}
\begin{figure}[H]\ContinuedFloat
    \begin{subfigure}{\linewidth}
        \centering
        \begin{tcolorbox}[codebox]
        \begin{lstlisting}[language=Java,gobble=10,numbers=left,style=codeHighlighting,showstringspaces=false]
            `+ /``/`` skip @AutoValue and @AutoBuilder methods`
            `+ if ( isAbstractAutoValueOrAutoBuilderMethod ( methodSymbol , state ) ) {`
            <@\quad@>`+ return Description.NO_MATCH ;`
            `+ }`
            &- /&&/&& &^OK^& &^,^& &^now^& &^the^& &^real^& &^implementation^& &^:^& &^For^& &^each^& &^possible^& &^return^& &^branch^& , &^does^& &^the^& &^expression^
            `+ /``/`` `@@if@@` `@@the@@` `@@method@@` `@@looks@@` `@@like@@` `@@a@@` `@@builder@@` `@@,@@` `@@or@@` `@@if@@` `@@it@@` `@@always@@` `@@returns `this` @@`, `@@then@@` `@@make@@` `@@it @CIRV@@``
            &- /&&/&& returned look like "this" or instance methods that are also @CanIgnoreReturnValue.&
            &- if ( methodReturnsIgnorableValues ( methodTree , state ) ) {&
            `+ if ( `@@methodLooksLikeBuilder ( methodSymbol ) || @@`methodReturnsIgnorableValues ( methodTree , state ) ) {`
        \end{lstlisting}
        \end{tcolorbox}
        \subcaption{Relevant \textbf{Repair Context} Selected by \approachName{} and Included in the Model's Input}
        \label{fig:fail1_ric}
    \end{subfigure}
\end{figure}


\subsubsection{Failure Example 2}
\label{sec:fe2}

Figure~\ref{fig:fail2} presents the second failure example. As seen in Figure~\ref{fig:fail2_gt}, the ground truth repair involves modifying two assertions in the test case. In the first assertion, the \textit{OpenTelemetryConstants} class is replaced with the \textit{AttributeConstants} class in two instances. In the second assertion, the method for accessing the \textit{span kind} has changed from direct access to accessing it via its \textit{attributes}.

As shown in Figure~\ref{fig:fail2_cr}, \approachName{} successfully repaired the first assertion in two of the five provided candidates (lines 9 and 21 of Figure~\ref{fig:fail2_cr}). However, it failed to repair the second assertion in any candidate. Notably, in two candidates (lines 15 and 20), the second assertion was not generated at all. Upon reviewing the repair context included in the input (Figure~\ref{fig:fail2_ric}), we observe that line 7 provides a hint about the \textit{span kind} becoming an attribute. Additionally, the repair context excluded due to token limits (line 4 of Figure~\ref{fig:fail2_rnic}) could have offered further insights, particularly about the \textit{SPAN\_KIND\_CLIENT}, if it had been included.

In conclusion, we believe this failure stems from the wide range of changes required for the repair, which likely reduced the model's focus on each individual change. While additional repair context offers valuable clues, employing a multi-step or iterative repair process, which concentrates the model's attention on each assertion separately, could enhance repair success in such cases.

\begin{figure}[H]
    \caption{\textbf{Failure Example 2} -- A repair involving a wide variety of code changes, which reduces the model's focus on repairing each assertion.\\\textbf{\benchmarkName{} ID:} \textit{apache/shardingsphere:10788}\\\textbf{GitHub Commit Reference:} \href{https://github.com/apache/shardingsphere/commit/b5230f5c681}{\textcolor{blue}{\textit{\underline{https://github.com/apache/shardingsphere/commit/b5230f5c681}}}}}
    \label{fig:fail2}
    \centering
    \begin{subfigure}{\linewidth}
        \centering
        \begin{tcolorbox}[codebox]
        \begin{lstlisting}[language=Java,gobble=10,numbers=left,style=codeHighlighting,showstringspaces=false,breaklines=true]
            @Test
            public void assertMethod() {
            <@\quad@>OpenTelemetryCommandExecutorTaskAdvice advice = new OpenTelemetryCommandExecutorTaskAdvice();
            <@\quad@>advice.beforeMethod(getTargetObject(), null, new Object[]{}, "OpenTelemetry");
            <@\quad@>advice.afterMethod(getTargetObject(), null, new Object[]{}, null, "OpenTelemetry");
            <@\quad@>List<SpanData> spanItems = COLLECTOR.getSpanItems();
            <@\quad@>assertThat(spanItems.size(), is(1));
            <@\quad@>SpanData spanData = spanItems.get(0);
            <@\quad@>assertThat(spanData.getName(), is("/ShardingSphere/rootInvoke/"));
            &-  assertThat(spanData.getAttributes().get(AttributeKey.stringKey(&^OpenTelemetryConstants^&.COMPONENT)),& &is(&^OpenTelemetryConstants^&.COMPONENT_NAME));&
            `+  assertThat(spanData.getAttributes().get(AttributeKey.stringKey(`@@AttributeConstants@@`.COMPONENT)),` `is(`@@AttributeConstants@@`.COMPONENT_NAME));`
            &-  assertThat(spanData.&^getKind^&(), is(&^SpanKind^&.&^CLIENT^&));&
            `+  assertThat(spanData.`@@getAttributes@@`()`@@.get(AttributeKey.stringKey(AttributeConstants.SPAN_KIND))@@`,` `is(`@@AttributeConstants@@`.`@@SPAN_KIND_CLIENT@@`));`
            }
        \end{lstlisting}
        \end{tcolorbox}
        \subcaption{\textbf{Ground Truth} Repair}
        \label{fig:fail2_gt}
    \end{subfigure}
\end{figure}
\begin{figure}[H]\ContinuedFloat
    \begin{subfigure}{\linewidth}
        \centering
        \begin{tcolorbox}[codebox]
        \begin{lstlisting}[language=Java,gobble=10,numbers=left,style=codeHighlighting,showstringspaces=false]
            &-  assertThat ( spanData.getAttributes ( ) .get ( &^AttributeKey^&.&^stringKey ( OpenTelemetryConstants.^&COMPONENT& &) ) , is ( &^OpenTelemetryConstants^&.COMPONENT_NAME ) ) ;&
            `+  assertThat ( spanData.getAttributes ( ) .get ( `@@AttributeConstants@@`.COMPONENT ) ) , is (` ``@@AttributeConstants@@`.COMPONENT_NAME ) ) ;`
            &-  assertThat ( spanData.getKind ( ) , is ( SpanKind.&^CLIENT^& ) ) ;&
            `+  assertThat ( spanData.getKind ( ) , is ( SpanKind.`@@INTERNAL@@` ) ) ;`
            
            ----------------------------------
            
            &-  assertThat ( spanData.getAttributes ( ) .get ( AttributeKey.stringKey ( &^OpenTelemetryConstants^&.COMPONENT& &) ) , is ( &^OpenTelemetryConstants^&.COMPONENT_NAME ) ) ;&
            `+  assertThat ( spanData.getAttributes ( ) .get ( AttributeKey.stringKey ( `@@AttributeConstants@@`.COMPONENT ) )` `, is ( `@@AttributeConstants@@`.COMPONENT_NAME ) ) ;`
            &-  assertThat ( spanData.getKind ( ) , is ( SpanKind.&^CLIENT^& ) ) ;&
            `+  assertThat ( spanData.getKind ( ) , is ( SpanKind.`@@INTERNAL@@` ) ) ;`
            
            ----------------------------------
            
            &-  assertThat ( spanData.getAttributes ( ) .get ( &^AttributeKey^&.&^stringKey ( OpenTelemetryConstants.^&COMPONENT& &) ) , is ( &^OpenTelemetryConstants^&.COMPONENT_NAME ) ) ;&
            `+  assertThat ( spanData.getAttributes ( ) .get ( `@@AttributeConstants@@`.COMPONENT ) ) , is (` ``@@AttributeConstants@@`.COMPONENT_NAME ) ) ;`
            
            ----------------------------------
            
            &-  assertThat ( spanData.getAttributes ( ) .get ( AttributeKey.stringKey ( &^OpenTelemetryConstants^&.COMPONENT& &) ) , is ( &^OpenTelemetryConstants^&.COMPONENT_NAME ) ) ;&
            `+  assertThat ( spanData.getAttributes ( ) .get ( AttributeKey.stringKey ( `@@AttributeConstants@@`.COMPONENT ) ) ,` `is ( `@@AttributeConstants@@`.COMPONENT_NAME ) ) ;`
            
            ----------------------------------
            
            &-  assertThat ( spanData.getAttributes ( ) .get ( &^AttributeKey^&.&^stringKey ( OpenTelemetryConstants.^&COMPONENT& &) ) , is ( &^OpenTelemetryConstants^&.COMPONENT_NAME ) ) ;&
            `+  assertThat ( spanData.getAttributes ( ) .get ( `@@AttributeConstants@@`.COMPONENT ) ) , is (` ``@@AttributeConstants@@`.COMPONENT_NAME ) ) ;`
            &-  assertThat ( spanData.getKind ( ) , is ( &^SpanKind^&.&^CLIENT^& ) ) ;&
            `+  assertThat ( spanData.getKind ( ) , is ( `@@AttributeConstants@@`.`@@SPAN_KIND_INTERNAL@@` ) ) ;`
        \end{lstlisting}
        \end{tcolorbox}
        \subcaption{Selected Top \textbf{Repair Candidates} Generated by \approachName{}}
        \label{fig:fail2_cr}
    \end{subfigure}
\end{figure}
\begin{figure}[H]\ContinuedFloat
    \begin{subfigure}{\linewidth}
        \centering
        \begin{tcolorbox}[codebox]
        \begin{lstlisting}[language=Java,gobble=10,numbers=left,style=codeHighlighting,showstringspaces=false]
            &- .setAttribute ( &^OpenTelemetryConstants^&.COMPONENT , &^OpenTelemetryConstants^&.COMPONENT_NAME )&
            `+ .setAttribute ( `@@AttributeConstants@@`.COMPONENT , `@@AttributeConstants@@`.COMPONENT_NAME )`
            &- .setAttribute ( OpenTelemetryConstants.DB_TYPE , OpenTelemetryConstants.DB_TYPE_VALUE )&
            &- .setAttribute ( &^OpenTelemetryConstants^&.DB_STATEMENT , sql )&^ ;^&&
            `+ .setAttribute ( `@@AttributeConstants@@`.DB_STATEMENT , sql )`
            &- &^if^& ( &^null^& &^!=^& &^rootSpan^& ) &^{^&&
            `+ `@@.setAttribute@@` ( `@@AttributeConstants.SPAN_KIND@@` `@@,@@` `@@AttributeConstants.SPAN_KIND_INTERNAL@@` ) `@@;@@``
            &- spanBuilder.setParent ( Context.current ( ) .with ( &^rootSpan^& ) ) ;&
            `+ spanBuilder.setParent ( Context.current ( ) .with ( `@@parentSpan@@` ) ) ;`
            &- }&
            &- &^return^& spanBuilder.startSpan ( ) ;&
            `+ `@@Span@@` `@@result = @@`spanBuilder.startSpan ( ) ;`
            `+ target.setAttachment ( result ) ;`
            `+ return result ;`
        \end{lstlisting}
        \end{tcolorbox}
        \subcaption{Relevant \textbf{Repair Context} Selected by \approachName{} and Included in the Model's Input}
        \label{fig:fail2_ric}
    \end{subfigure}
\end{figure}
\begin{figure}[H]\ContinuedFloat
    \begin{subfigure}{\linewidth}
        \centering
        \begin{tcolorbox}[codebox]
        \begin{lstlisting}[language=Java,gobble=10,numbers=left,style=codeHighlighting,showstringspaces=false]
            &- .setAttribute ( &^OpenTelemetryConstants^&.COMPONENT , &^OpenTelemetryConstants^&.COMPONENT_NAME )&
            `+ .setAttribute ( `@@AttributeConstants@@`.COMPONENT , `@@AttributeConstants@@`.COMPONENT_NAME )`
            &- .&^setSpanKind^& ( &^SpanKind^&.&^CLIENT^& ) ;&
            `+ .`@@setAttribute@@` ( `@@AttributeConstants@@`.`@@SPAN_KIND@@` `@@, AttributeConstants.SPAN_KIND_CLIENT @@`) ;`
            &- &^Span^& &^result = ^&spanBuilder.startSpan ( ) ;&
            `+ `@@return@@` spanBuilder.startSpan ( ) ;`
            ^- target.setAttachment ( result ) ;^
            ^- return result ;^
        \end{lstlisting}
        \end{tcolorbox}
        \subcaption{Relevant \textbf{Repair Context} Not Selected by \approachName{} and Excluded from the Model's Input}
        \label{fig:fail2_rnic}
    \end{subfigure}
\end{figure}


\subsubsection{Failure Example 3}

The third failure example, illustrated in Figure~\ref{fig:fail3}, highlights a repair that primarily involves changing the way fields are created and retrieved from the \textit{GeoShapeMapper} class. The ground truth repair, shown in Figure~\ref{fig:fail3_gt}, replaces the use of the \textit{Document} class and the \textit{mapper.addFields} method with the \textit{mapper.indexableFields} method. Additionally, the first \textit{assertEquals} statement in the test case is removed, as it becomes redundant following these changes, while the second \textit{assertEquals} is updated accordingly to reflect the new implementation.

In reviewing \approachName{}'s top five repair candidates (Figure~\ref{fig:fail3_cr}), we observe that all candidates correctly employ the new \textit{mapper.indexableFields} method, remove the obsolete field handling logic, remove the redundant assertion, and update the remaining assertion. Despite these successes, none of the repair candidates are fully correct, as they introduce issues that result in compilation errors. Specifically, the first, third, and fourth candidates (lines 7, 29, and 40 in Figure~\ref{fig:fail3_cr}) fail to close the double quotation marks for a long string. Meanwhile, the second and fifth candidates (lines 18 and 51) misuse the API of the \textit{Columns} class.

Examining the relevant repair context in Figure~\ref{fig:fail3_ric}, we see that the first three items (lines 2, 7, and 12) provide the model with information to understand the new \textit{indexableFields} method. Additionally, the last item (line 17) demonstrates an alternative approach for adding columns. Although adding columns is not the central focus of the required repair, this extra information about the API usage contributes to the failure of these candidates.

In conclusion, similar to the example in Section~\ref{sec:fe2}, this failure appears to result from the wide range of changes required. We hypothesize that a multi-step or iterative repair approach--where the model sequentially addresses each distinct aspect of the repair (e.g., the core field change and the column API change)—-could improve its success rate in cases involving multiple changes.

\begin{figure}[H]
    \caption{\textbf{Failure Example 3} -- A repair involving a wide variety of code changes changes.\\\textbf{\benchmarkName{} ID:} \textit{stratio/cassandra-lucene-index:571}\\\textbf{GitHub Commit Reference:} \href{https://github.com/Stratio/cassandra-lucene-index/commit/f9234609}{\textcolor{blue}{\textit{\underline{https://github.com/Stratio/cassandra-lucene-index/commit/f9234609}}}}}
    \label{fig:fail3}
    \centering
    \begin{subfigure}{\linewidth}
        \centering
        \begin{tcolorbox}[codebox]
        \begin{lstlisting}[language=Java,gobble=10,numbers=left,style=codeHighlighting,showstringspaces=false,breaklines=true,basicstyle=\fontsize{6}{6}\ttfamily, numberstyle=\fontsize{6}{6}]
            @Test
            public void testAddFieldsWithValidUnionMultiPolygon2() {
            <@\quad@>GeoShapeMapper mapper = geoShapeMapper().column("column").maxLevels(10).build("field");
            <@\quad@>&- Columns columns = new Columns()&^;^&&
            <@\quad@>`+ Columns columns = new Columns()`
            <@\quad@> &- &^columns^&.add("column",&
            <@\quad@><@\quad@>`+ .add("column",`
            <@\quad@><@\quad@><@\quad@>  "MULTIPOLYGON (((40 40, 20 45, 45 30, 40 40)),((20 35, 10 30, 10 10, 30 5, 45 20, 20 35),(30 20, 20 15, 20 25, 30 20)))");
            <@\quad@><@\quad@>&- Document document = new Document();&
            <@\quad@><@\quad@>&- mapper.addFields(document, columns);&
            <@\quad@><@\quad@>&- &^assertEquals("Fields are not properly created", 1, document^&.&^getFields^&(&^"field"^&)&^.length)^&;&
            <@\quad@><@\quad@>`+ `@@List<IndexableField>@@` `@@fields@@` `@@=@@` `@@mapper@@`.`@@indexableFields@@`(`@@columns@@`);`
            <@\quad@><@\quad@>&- assertEquals("Fields are not properly created", 1, &^document^&.&^getFields().^&size());&
            <@\quad@><@\quad@>`+ assertEquals("Fields are not properly created", 1, `@@fields@@`.size());`
            }
        \end{lstlisting}
        \end{tcolorbox}
        \subcaption{\textbf{Ground Truth} Repair}
        \label{fig:fail3_gt}
    \end{subfigure}
\end{figure}
\begin{figure}[H]\ContinuedFloat
    \begin{subfigure}{\linewidth}
        \centering
        \begin{tcolorbox}[codebox]
        \begin{lstlisting}[language=Java,gobble=10,numbers=left,style=codeHighlighting,showstringspaces=false,basicstyle=\tiny\ttfamily,numberstyle=\tiny]
            &- &^Columns^& &^columns^& = new Columns ( ) ;&
            &- &^columns.add^& ( &^"column"^& &^,^& &^"MULTIPOLYGON (((40 40, 20 45, 45 30, 40 40)),((20 35, 10 30, 10 10, 30 5, 45 20, 20 35),(30 20, 20 15, 20 25, 30 20)))"^& &^) ^&;&
            &- Document document = new Document ( ) ;&
            &- mapper.addFields ( document , columns ) ;&
            &- assertEquals ( "Fields are not properly created" , 1 , document.getFields ( "field" ) .length ) ;&
            &- assertEquals ( "Fields are not properly created" , 1 , document.getFields ( ) .size ( ) ) ;&
            `+ `@@List@@` `@@< IndexableField > fields @@`= `@@mapper.indexableFields ( @@`new Columns ( ) `@@.add ( Column.apply ( "column" )@@ @@.withValue ( <@"@>MULTIPOLYGON (((40 40, 20 45, 45 30, 40 40)),(20 35, 10 30, 10 10, 30 5, 45 20, 20 35),(30 20, 20 15, 20 25, 30 20)))) ) ) @@`;`
            `+ `@@assertEquals@@` ( `@@"Fields are not properly created"@@` , `@@1@@` , `@@fields.size@@` ( ) ) ;`
            
            ----------------------------------
            
            &- &^Columns^& &^columns^& = new Columns ( ) ;&
            &- &^columns.add^& ( &^"column"^& &^,^& &^"MULTIPOLYGON (((40 40, 20 45, &^45^& 30, 40 40)),((20 35, 10 30, 10 10, 30 5, 45 20, 20 35),(30 20, 20 15, 20 25, 30 20)))"^& &^) ^&;&
            &- Document document = new Document ( ) ;&
            &- mapper.addFields ( document , columns ) ;&
            &- assertEquals ( "Fields are not properly created" , 1 , document.getFields ( "field" ) .length ) ;&
            &- assertEquals ( "Fields are not properly created" , 1 , document.getFields ( ) .size ( ) ) ;&
            `+ `@@List@@` `@@< IndexableField > fields @@`= `@@mapper.indexableFields ( @@`new Columns ( ) `@@.with ( "field" ) .withValue@@ @@( "MULTIPOLYGON (((40 40, 20 45, 45 30, 40 40)),((20 35, 10 30, 10 10, 30 5, 45 20, 20 35),(30 20, 20 15, 20 25, 30 20)))" ) ) @@`;`
            `+ `@@assertEquals@@` ( `@@"Fields are not properly created"@@` , `@@1@@` , `@@fields.size@@` ( ) ) ;`
            
            ----------------------------------
            
            &- &^Columns^& &^columns^& = new Columns ( ) ;&
            &- &^columns.add^& ( &^"column"^& &^,^& &^"MULTIPOLYGON (((40 40, 20 45, 45 30, 40 40)),((20 35, 10 30, 10 10, 30 5, 45 20, 20 35),(30 20, 20 15, 20 25, 30 20)))"^& &^) ^&;&
            &- Document document = new Document ( ) ;&
            &- mapper.addFields ( document , columns ) ;&
            &- assertEquals ( "Fields are not properly created" , 1 , document.getFields ( "field" ) .length ) ;&
            &- assertEquals ( "Fields are not properly created" , 1 , document.getFields ( ) .size ( ) ) ;&
            `+ `@@List@@` `@@< IndexableField > fields @@`= `@@mapper.indexableFields ( @@`new Columns ( ) `@@.add ( Column.apply ( "column" )@@ @@.withValue ( <@"@>MULTIPOLYGON (((40 40, 20 45, 45 30, 40 40)),((20 35, 10 30, 10 10, 30 5, 45 20, 20 35),(30 20, 20 15, 20 25, 30 20)))) ) ) @@`;`
            `+ `@@assertEquals@@` ( `@@"Fields are not properly created"@@` , `@@1@@` , `@@fields.size@@` ( ) ) ;`
            
            ----------------------------------
            
            &- &^Columns^& &^columns^& = new Columns ( ) ;&
            &- &^columns.add^& ( &^"column"^& &^,^& &^"MULTIPOLYGON (((40 40, 20 45, 45 30, 40 40)),((20 35, 10 30, 10 10, 30 5, 45 20, 20 35),(30 20, 20 15, 20 25, 30 20)))"^& &^) ^&;&
            &- Document document = new Document ( ) ;&
            &- mapper.addFields ( document , columns ) ;&
            &- assertEquals ( "Fields are not properly created" , 1 , document.getFields ( "field" ) .length ) ;&
            &-  assertEquals ( "Fields are not properly created" , 1 , document.getFields ( ) .size ( ) ) ;&
            `+ `@@List@@` `@@< IndexableField > fields @@`= `@@mapper.indexableFields ( @@`new Columns ( ) `@@.add ( Column.apply ( "column" )@@ @@.withValue ( <@"@>MULTIPOLYGON (((40 40, 20 45, 45 30, 40 40)),(20 35, 10 30, 10 10, 30 5, 45 20, 20 35),(30 20, 20 15, 20 25, 30 20)))) ) @@`;`
            `+ `@@assertEquals@@` ( `@@"Fields are not properly created"@@` , `@@1@@` , `@@fields.size@@` ( ) ) ;`
            
            ----------------------------------
            
            &- &^Columns^& &^columns^& = new Columns ( ) ;&
            &- &^columns.add^& ( &^"column"^& &^,^& &^"MULTIPOLYGON (((40 40, 20 45, 45 30, 40 40)),((20 35, 10 30, 10 10, 30 5, 45 20, 20 35),(30 20, 20 15, 20 25, 30 20)))"^& &^) ^&;&
            &- Document document = new Document ( ) ;&
            &- mapper.addFields ( document , columns ) ;&
            &- assertEquals ( "Fields are not properly created" , 1 , document.getFields ( "field" ) .length ) ;&
            &- assertEquals ( "Fields are not properly created" , 1 , document.getFields ( ) .size ( ) ) ;&
            `+ `@@List@@` `@@< IndexableField > fields @@`= `@@mapper.indexableFields ( @@`new Columns ( ) `@@.with ( "column" ) .withValue@@ @@( "MULTIPOLYGON (((40 40, 20 45, 45 30, 40 40)),((20 35, 10 30, 10 10, 30 5, 45 20, 20 35),(30 20, 20 15, 20 25, 30 20)))" ) ) @@`;`
            `+ `@@assertEquals@@` ( `@@"Fields are not properly created"@@` , `@@1@@` , `@@fields.size@@` ( ) ) ;`
            
        \end{lstlisting}
        \end{tcolorbox}
        \subcaption{Selected Top \textbf{Repair Candidates} Generated by \approachName{}}
        \label{fig:fail3_cr}
    \end{subfigure}
\end{figure}
\begin{figure}[H]\ContinuedFloat
    \begin{subfigure}{\linewidth}
        \centering
        \begin{tcolorbox}[codebox]
        \begin{lstlisting}[language=Java,gobble=10,numbers=left,style=codeHighlighting,showstringspaces=false]
            &-&^addFields^& ( &^new Document ( ) , ^&columns ) ;&
            `+`@@indexableFields@@` ( columns ) ;`
            
            ----------------------------------
            
            &-public &^void^& &^addFields^& &^(^& &^Document^& &^document^& &^,^& Columns columns ) {&
            `+public `@@List@@` `@@< IndexableField@@` `@@>@@` `@@indexableFields@@` `@@(@@` Columns columns ) {`
            
            ----------------------------------
            
            &-&^mapper^&.&^addFields^& ( &^document^& &^,^& columns ) ;&
            `+`@@fields@@`.`@@addAll@@` ( `@@mapper.indexableFields@@` `@@(@@` columns ) `@@) @@`;`
            
            ----------------------------------
            
            &-columns.add ( &^ColumnsMapper^&.&^column^& ( name &^,^& value , valueType ) ) ;&
            `+columns`@@ = columns@@`.add ( `@@Column@@`.`@@apply@@` ( name `@@)@@` `@@.withValue ( ColumnsMapper.compose ( @@`value , valueType ) ) `@@) @@`;`
        \end{lstlisting}
        \end{tcolorbox}
        \subcaption{Relevant \textbf{Repair Context} Selected by \approachName{} and Included in the Model's Input}
        \label{fig:fail3_ric}
    \end{subfigure}
\end{figure}


\subsubsection{Failure Example 4}

In the fourth failure example shown in Figure~\ref{fig:fail4}, the ground truth repair (Figure~\ref{fig:fail4_gt}) reveals that four assertions in the test case were repaired. Specifically, the leading zero values in the expected outputs of the \textit{Utils.formatDuration} method were removed.

Examining the top repair candidates in Figure~\ref{fig:fail4_cr}, we observe that none of the generated solutions move toward removing the leading zeros, as seen in the ground truth repair. Moreover, the model introduces incorrect or redundant changes in some candidates. For instance, in line 4, it changes \textit{"ms"} to \textit{"s"}, despite the string already containing \textit{"1sec"}. Similarly, in line 13, the time value is altered from \textit{100} to \textit{1000} milliseconds in both the input and expected output, which does not repair the test case. Additionally, in the second and fourth candidates (lines 13 and 29), the model changes only one of the four required assertions.

The relevant repair context provided to the model, shown in Figure~\ref{fig:fail4_ric}, indicates that the SUT changed its logic for formatting the duration. Previously, the method used \textit{String.format} to generate a string based on a predefined format stored in the \textit{DURATION\_FORMAT} variable. After the change, it relies on conditional logic to decide whether each time component should be included in the output. Despite the fact that this repair context should have been sufficient for repairing the test case, the model failed to fully grasp the change in logic, solely from the code differences.

In conclusion, this failure can be attributed to insufficient contextual understanding, similar to the example discussed in Section~\ref{sec:fe1}. To enhance repair performance in cases like this, a more robust strategy that incorporates repository-level repair context—-such as considering the unchanged parts of the \textit{formatDuration} method or analyzing usage examples of that method elsewhere in the SUT-—would likely yield better results.

\begin{figure}[H]
    \caption{\textbf{Failure Example 4} -- An example of missing additional repository-level unchanged contextual information from the model's input.\\\textbf{\benchmarkName{} ID:} \textit{j-easy/easy-batch:13}\\\textbf{GitHub Commit Reference:} \href{https://github.com/j-easy/easy-batch/commit/2882d33b}{\textcolor{blue}{\textit{\underline{https://github.com/j-easy/easy-batch/commit/2882d33b}}}}}
    \label{fig:fail4}
    \centering
    \begin{subfigure}{\linewidth}
        \centering
        \begin{tcolorbox}[codebox]
        \begin{lstlisting}[language=Java,gobble=10,numbers=left,style=codeHighlighting,showstringspaces=false,breaklines=true]
            @Test
            public void testFormatDuration() throws Exception {
            <@\quad@>&- assertThat(Utils.formatDuration(Duration.of(100, ChronoUnit.MILLIS))).isEqualTo(&^"0d 0hr 0min 0sec 100ms"^&);&
            <@\quad@>`+ assertThat(Utils.formatDuration(Duration.of(100, ChronoUnit.MILLIS))).isEqualTo(`@@"100ms"@@`);`
            <@\quad@>&- assertThat(Utils.formatDuration(Duration.of(1, ChronoUnit.SECONDS))).isEqualTo(&^"0d 0hr 0min 1sec 0ms"^&);&
            <@\quad@>`+ assertThat(Utils.formatDuration(Duration.of(1, ChronoUnit.SECONDS))).isEqualTo(`@@"1sec 0ms"@@`);`
            <@\quad@>&- assertThat(Utils.formatDuration(Duration.of(2, ChronoUnit.MINUTES))).isEqualTo(&^"0d 0hr 2min 0sec 0ms"^&);&
            <@\quad@>`+ assertThat(Utils.formatDuration(Duration.of(2, ChronoUnit.MINUTES))).isEqualTo(`@@"2min 0sec 0ms"@@`);`
            <@\quad@>&- assertThat(Utils.formatDuration(Duration.of(2, ChronoUnit.HOURS))).isEqualTo(&^"0d 2hr 0min 0sec 0ms"^&);&
            <@\quad@>`+ assertThat(Utils.formatDuration(Duration.of(2, ChronoUnit.HOURS))).isEqualTo(`@@"2hr 0min 0sec 0ms"@@`);`
            <@\quad@>assertThat(Utils.formatDuration(Duration.of(2, ChronoUnit.DAYS))).isEqualTo("2d 0hr 0min 0sec 0ms");
            <@\quad@>// in the extremely unlikely event that the job takes more than a year
            <@\quad@>assertThat(Utils.formatDuration(Duration.of(366, ChronoUnit.DAYS))).isEqualTo("366d 0hr 0min 0sec 0ms");
            }
        \end{lstlisting}
        \end{tcolorbox}
        \subcaption{\textbf{Ground Truth} Repair}
        \label{fig:fail4_gt}
    \end{subfigure}
\end{figure}
\begin{figure}[H]\ContinuedFloat
    \begin{subfigure}{\linewidth}
        \centering
        \begin{tcolorbox}[codebox]
        \begin{lstlisting}[language=Java,gobble=10,numbers=left,style=codeHighlighting,showstringspaces=false,basicstyle=\fontsize{6}{6}\ttfamily, numberstyle=\fontsize{6}{6}]
            &- assertThat ( Utils.formatDuration ( Duration.of ( 100 , ChronoUnit.&^MILLIS^& ) ) ) .isEqualTo& &( &^"0d 0hr 0min 0sec 100ms"^& ) ;&
            `+ assertThat ( Utils.formatDuration ( Duration.of ( 100 , ChronoUnit.`@@SECONDS@@` ) ) ) .isEqualTo` `( `@@"0d 0hr 0min 0sec 100s"@@` ) ;`
            &- assertThat ( Utils.formatDuration ( Duration.of ( 1 , ChronoUnit.SECONDS ) ) ) .isEqualTo& &( &^"0d 0hr 0min 1sec 0ms"^& ) ;&
            `+ assertThat ( Utils.formatDuration ( Duration.of ( 1 , ChronoUnit.SECONDS ) ) ) .isEqualTo` `( `@@"0d 0hr 0min 1sec 0s"@@` ) ;`
            &- assertThat ( Utils.formatDuration ( Duration.of ( 2 , ChronoUnit.MINUTES ) ) ) .isEqualTo& &( &^"0d 0hr 2min 0sec 0ms"^& ) ;&
            `+ assertThat ( Utils.formatDuration ( Duration.of ( 2 , ChronoUnit.MINUTES ) ) ) .isEqualTo` `( `@@"0d 0hr 2min 0sec 0s"@@` ) ;`
            &- assertThat ( Utils.formatDuration ( Duration.of ( 2 , ChronoUnit.HOURS ) ) ) .isEqualTo& &( &^"0d 2hr 0min 0sec 0ms"^& ) ;&
            `+ assertThat ( Utils.formatDuration ( Duration.of ( 2 , ChronoUnit.HOURS ) ) ) .isEqualTo ( `@@"0d 2hr 0min 0sec 0s"@@` ) ;`
            
            ----------------------------------
            
            &- assertThat ( Utils.formatDuration ( Duration.of ( &^100^& , ChronoUnit.MILLIS ) ) ) .isEqualTo& &( &^"0d 0hr 0min 0sec 100ms"^& ) ;&
            `+ assertThat ( Utils.formatDuration ( Duration.of ( `@@1000@@` , ChronoUnit.MILLIS ) ) ) .isEqualTo` `( `@@"0d 0hr 0min 0sec 1000ms"@@` ) ;`
            
            ----------------------------------
            
            &- assertThat ( Utils.formatDuration ( Duration.of ( 100 , ChronoUnit.MILLIS ) ) ) .isEqualTo& &( &^"0d 0hr 0min 0sec 100ms"^& ) ;&
            `+ assertThat ( Utils.formatDuration ( Duration.of ( 100 , ChronoUnit.MILLIS ) ) ) .isEqualTo` `( `@@"0 d 0hr 0min 0sec 100ms"@@` ) ;`
            &- assertThat ( Utils.formatDuration ( Duration.of ( 1 , ChronoUnit.SECONDS ) ) ) .isEqualTo& &( &^"0d 0hr 0min 1sec 0ms"^& ) ;&
            `+ assertThat ( Utils.formatDuration ( Duration.of ( 1 , ChronoUnit.SECONDS ) ) ) .isEqualTo` `( `@@"0 d 0hr 0min 1sec 0ms"@@` ) ;`
            &- assertThat ( Utils.formatDuration ( Duration.of ( 2 , ChronoUnit.MINUTES ) ) ) .isEqualTo& &( &^"0d 0hr 2min 0sec 0ms"^& ) ;&
            `+ assertThat ( Utils.formatDuration ( Duration.of ( 2 , ChronoUnit.MINUTES ) ) ) .isEqualTo` `( `@@"0 d 0hr 2min 0sec 0ms"@@` ) ;`
            &- assertThat ( Utils.formatDuration ( Duration.of ( 2 , ChronoUnit.HOURS ) ) ) .isEqualTo& &( &^"0d 2hr 0min 0sec 0ms"^& ) ;&
            `+ assertThat ( Utils.formatDuration ( Duration.of ( 2 , ChronoUnit.HOURS ) ) ) .isEqualTo` `( `@@"0 d 2hr 0min 0sec 0ms"@@` ) ;`
            
            ----------------------------------
            
            &- assertThat ( Utils.formatDuration ( Duration.of ( 100 , ChronoUnit.MILLIS ) ) ) .isEqualTo& &( &^"0d 0hr 0min 0sec 100ms"^& ) ;&
            `+ assertThat ( Utils.formatDuration ( Duration.of ( 100 , ChronoUnit.MILLIS ) ) ) .isEqualTo` `( `@@"0d 0hr 0min 0sec 0ms"@@` ) ;`
        \end{lstlisting}
        \end{tcolorbox}
        \subcaption{Selected Top \textbf{Repair Candidates} Generated by \approachName{}}
        \label{fig:fail4_cr}
    \end{subfigure}
\end{figure}
\begin{figure}[H]\ContinuedFloat
    \begin{subfigure}{\linewidth}
        \centering
        \begin{tcolorbox}[codebox]
        \begin{lstlisting}[language=Java,gobble=10,numbers=left,style=codeHighlighting,showstringspaces=false]
            @@+ long days = duration.toDays ( ) ;@@
            @@+ long hours = duration.toHours ( ) ;@@
            @@+ long minutes = duration.toMinutes ( ) ;@@
            @@+ long seconds = duration.getSeconds ( ) ;@@
            &- long &^millis^& = duration.toMillis ( ) ;&
            `+ long `@@milliseconds@@` = duration.toMillis ( ) ;`
            @@+ return ( days > 0 ? days + "d " : "" ) +@@
            <@\quad@>@@+ ( hours > 0 ? ( hours - TimeUnit.DAYS.toHours ( days ) ) + "hr " : "" ) +@@
            &- &^return^& &^String.format^& ( &^DURATION_FORMAT^& &^,^& &^MILLISECONDS^&.&^toDays^& ( &^millis^& ) &^,^& &^MILLISECONDS.toHours^& &^(^& &^millis^& &^)^& &^%^& &^24^& &^,^
            <@\quad@>`+ `@@(@@` `@@minutes@@` `@@> 0 ? @@`( `@@minutes@@` `@@-@@` `@@TimeUnit@@`.`@@HOURS.toMinutes@@` ( `@@hours@@` ) `@@)@@` `@@+@@` `@@"min "@@` `@@:@@` `@@""@@` `@@)@@` `@@+@@
            <@\quad@>&- &^MILLISECONDS.toMinutes ^&( &^millis^& &^)^& &^%^& &^60^& &^,^& &^MILLISECONDS^&.toSeconds ( &^millis^& ) &^%^& &^60^& &^,^& &^millis^& &^%^& &^1000^& ) &^;^
            <@\quad@>`+ ( `@@seconds@@` `@@>@@` `@@0@@` `@@?@@` `@@(@@` `@@seconds - TimeUnit@@`.`@@MINUTES.@@`toSeconds ( `@@minutes@@` ) `@@)@@` `@@+@@` `@@"sec "@@` `@@:@@` `@@""@@` ) `@@+@@
            <@\quad@>@@+ ( milliseconds - TimeUnit.SECONDS.toMillis ( seconds ) ) + "ms" ;@@
        \end{lstlisting}
        \end{tcolorbox}
        \subcaption{Relevant \textbf{Repair Context} Selected by \approachName{} and Included in the Model's Input}
        \label{fig:fail4_ric}
    \end{subfigure}
\end{figure}

\subsection{Successful Examples}
\subsubsection{Successful Example 1}

\approachName{} generated an exact-match repair candidate in the first successful example shown in Figure~\ref{fig:success1}. As shown in Figure~\ref{fig:success1_gt}, the repair involves multiple changes, including replacing the \textit{add} method with \textit{addComposed} and removing the usage of the \textit{Column.builder} and \textit{buildWithComposed} methods.

Among the nine hunks included in the model's input by \approachName{}, the most relevant is highlighted in Figure~\ref{fig:success1_ric}. This particular hunk represents a change in the SUT that provides the necessary pattern for repairing the test case. Although the hunk utilizes a different method (\textit{addDecomposed}) and is not directly invoked during construction, the model effectively grasps the essential concept without merely replicating it.

The commit associated with this example comprised a total of 47 hunks. Due to input size constraints, \approachName{} could include only 9 of these in the input. Nevertheless, the hunk selection and prioritization process ensured that the critical hunk required for the repair was selected first, accompanied by eight other relevant hunks. This successful exact-match example demonstrates the efficacy of our hunk selection strategy and the model's fine-tuning, which enables it to learn patterns rather than blindly replicating SUT changes.

\begin{figure}[H]
    \caption{\textbf{Successful Example 1} -- An exact-match repair where \approachName{} prioritizes the critical hunk and effectively applies its change pattern.\\\textbf{\benchmarkName{} ID:} \textit{stratio/cassandra-lucene-index:485}\\\textbf{GitHub Commit Reference:} \href{https://github.com/Stratio/cassandra-lucene-index/commit/fdd34e53}{\textcolor{blue}{\textit{\underline{https://github.com/Stratio/cassandra-lucene-index/commit/fdd34e53}}}}}
    \label{fig:success1}
    \centering
    \begin{subfigure}{\linewidth}
        \centering
        \begin{tcolorbox}[codebox]
        \begin{lstlisting}[language=Java,gobble=10,numbers=left,style=codeHighlighting,showstringspaces=false,breaklines=true]
            @Test
            public void testValidateColumns() {
            <@\quad@>Schema schema = SchemaBuilders.schema().mapper("field1", stringMapper()).build();
            <@\quad@>&- Columns columns = new Columns ( ) .&^add^& ( &^Column.builder ( ^&"field1" &^)^& &^.buildWithComposed ( ^&"value" ,& &UTF8Type.instance ) &^) ^&;&
            <@\quad@>`+ Columns columns = new Columns ( ) .`@@addComposed@@` ( "field1" `@@,@@` "value" , UTF8Type.instance ) ;`
            <@\quad@>schema.validate(columns);
            <@\quad@>schema.close();
            }
        \end{lstlisting}
        \end{tcolorbox}
        \subcaption{\textbf{Ground Truth} Repair and \textbf{the Exact Match Repair Candidate} Generated by \approachName{}}
        \label{fig:success1_gt}
    \end{subfigure}
\end{figure}
\begin{figure}[H]\ContinuedFloat
    \begin{subfigure}{\linewidth}
        \centering
        \begin{tcolorbox}[codebox]
        \begin{lstlisting}[language=Java,gobble=10,numbers=left,style=codeHighlighting,showstringspaces=false]
            &- columns.&^add^& ( &^Column.builder ( ^&name &^)^& &^.buildWithDecomposed ( ^&value , valueType ) &^) ^&;&
            `+ columns.`@@addDecomposed@@` ( name `@@,@@` value , valueType ) ;`
        \end{lstlisting}
        \end{tcolorbox}
        \subcaption{Relevant \textbf{Repair Context} Selected by \approachName{} and Included in the Model's Input}
        \label{fig:success1_ric}
    \end{subfigure}
\end{figure}


\subsubsection{Successful Example 2}

Another exact-match example is presented in Figure~\ref{fig:success2}. As depicted in Figure~\ref{fig:success2_gt}, one of the assertion statements (lines 7 and 8) has been changed. Initially, the assertion checked if the singleton instance \textit{UnauthorizedAction.INSTANCE} was exactly equal to the \textit{action} variable. After the repair, the assertion was updated to verify only that the type of the \textit{action} variable was \textit{UnauthorizedAction}. This change occurred because the SUT update removed the singleton implementation of the instance.

\approachName{} included 10 hunks in the repair context for this input, with the highest-ranked hunk shown in Figure~\ref{fig:success2_ric}. This particular hunk illustrates that the return value of the singleton instance was replaced by a new instance of the class. The remaining nine hunks in the repair context either represented the same change for other classes or were irrelevant to the repair.

In this case, the commit involved 44 hunks. Once again, \approachName{} successfully identified and prioritized the most relevant hunk as the top selection in the input. Moreover, there were no explicit examples or clues in the input regarding the use of \textit{instanceof}, yet the model accurately predicted it solely based on the provided hunk. This demonstrates the effectiveness of the model’s learned knowledge, acquired through pre-training and fine-tuning. 

\begin{figure}[H]
    \caption{\textbf{Successful Example 2} -- An exact-match repair example where \approachName{} correctly prioritizes the most critical hunk, demonstrating the model’s ability to leverage its learned knowledge for an effective repair.\\\textbf{\benchmarkName{} ID:} \textit{pac4j/pac4j:397}\\\textbf{GitHub Commit Reference:} \href{https://github.com/pac4j/pac4j/commit/5f7dfe653}{\textcolor{blue}{\textit{\underline{https://github.com/pac4j/pac4j/commit/5f7dfe653}}}}}
    \label{fig:success2}
    \centering
    \begin{subfigure}{\linewidth}
        \centering
        \begin{tcolorbox}[codebox]
        \begin{lstlisting}[language=Java,gobble=10,numbers=left,style=codeHighlighting,showstringspaces=false,breaklines=true]
            @Test
            public void testBuildUnauthenticated403WithHeader() {
            <@\quad@>HttpActionHelper.setAlwaysUse401ForUnauthenticated(false);
            <@\quad@>final WebContext context = MockWebContext.create();
            <@\quad@>context.setResponseHeader(HttpConstants.AUTHENTICATE_HEADER, VALUE);
            <@\quad@>final var action = HttpActionHelper.buildUnauthenticatedAction(context);
            <@\quad@>&- &^assertEquals^& ( &^UnauthorizedAction.INSTANCE^& &^,^& &^action^& ) ;&
            <@\quad@>`+ `@@assertTrue@@` ( `@@action@@` `@@instanceof@@` `@@UnauthorizedAction@@` ) ;`
            <@\quad@>assertEquals(VALUE, context.getResponseHeader(HttpConstants.AUTHENTICATE_HEADER).get());
            }
        \end{lstlisting}
        \end{tcolorbox}
        \subcaption{\textbf{Ground Truth} Repair and \textbf{the Exact Match Repair Candidate} Generated by \approachName{}}
        \label{fig:success2_gt}
    \end{subfigure}
\end{figure}
\begin{figure}[H]\ContinuedFloat
    \begin{subfigure}{\linewidth}
        \centering
        \begin{tcolorbox}[codebox]
        \begin{lstlisting}[language=Java,gobble=10,numbers=left,style=codeHighlighting,showstringspaces=false]
            &-return UnauthorizedAction&^.INSTANCE^& ;&
            `+return `@@new @@`UnauthorizedAction `@@( ) @@`;`
        \end{lstlisting}
        \end{tcolorbox}
        \subcaption{Relevant \textbf{Repair Context} Selected by \approachName{} and Included in the Model's Input}
        \label{fig:success2_ric}
    \end{subfigure}
\end{figure}


\subsubsection{Successful Example 3}
\label{sec:se3}

The ground truth repair for this example (Figure~\ref{fig:success3}) occurs in the final assert statement. As illustrated in Figure~\ref{fig:success3_gt}, the last assertion originally checks if the \textit{httpRequestMatchers} list is empty. The repair involves invoking the \textit{toSortedList} method on this attribute, which is necessary because the type of \textit{httpRequestMatchers} was changed in the update to the SUT.

The relevant repair context, provided in the input (Figure~\ref{fig:success3_ric}), indicates the change in how the SUT handles \textit{httpRequestMatchers}. Specifically, instead of treating it as a list, the \textit{stream} method is now invoked on it. However, the changed type is not explicitly stated. In response to this change, \approachName{} generates a plausible repair, as shown in Figure~\ref{fig:success3_cr}. The plausible repair adopts a more generalized strategy: instead of converting the variable to a list to check for emptiness, it directly calls the \textit{size} method and compares the result to zero. This approach is broadly applicable, as the \textit{size} method is typically implemented for most collection types, and it maintains semantic equivalence with the ground truth repair without altering the intent of the test case.

This example demonstrates \approachName{}'s capability to explore diverse, yet correct, repair candidates. It also underscores the importance of metrics, such as plausible repair accuracy (introduced in our study), which evaluate model performance when the prediction does not exactly match the ground truth.

\begin{figure}[H]
    \caption{\textbf{Successful Example 3} -- An example of a plausible repair generated by \approachName{}, which is an equivalent repair to the ground truth that preserves the original test's intent.\\\textbf{\benchmarkName{} ID:} \textit{mock-server/mockserver:2081}\\\textbf{GitHub Commit Reference:} \href{https://github.com/mock-server/mockserver/commit/8ddaa46ea}{\textcolor{blue}{\textit{\underline{https://github.com/mock-server/mockserver/commit/8ddaa46ea}}}}}
    \label{fig:success3}
    \centering
    \begin{subfigure}{\linewidth}
        \centering
        \begin{tcolorbox}[codebox]
        \begin{lstlisting}[language=Java,gobble=10,numbers=left,style=codeHighlighting,showstringspaces=false,breaklines=true]
            @Test
            public void shouldNotRemoveAfterTimesCompleteOrExpired() {
            <@\quad@>// when
            <@\quad@>requestMatchers.add(new Expectation(new HttpRequest().withPath("somepath"), Times.exactly(0), TimeToLive.unlimited(), 0).thenRespond(response().withBody("someBody")), API);
            <@\quad@>requestMatchers.add(new Expectation(request().withPath("someOtherPath"), Times.unlimited(), TimeToLive.exactly(TimeUnit.MICROSECONDS, 0L), 0).thenRespond(response().withBody("someOtherBody")), API);
            <@\quad@>requestMatchers.add(new Expectation(request().withPath("someOtherPath"), Times.exactly(0), TimeToLive.exactly(TimeUnit.MICROSECONDS, 0L), 0).thenRespond(response().withBody("someOtherBody")), API);
            <@\quad@>// then
            <@\quad@>assertThat(requestMatchers.firstMatchingExpectation(new HttpRequest().withPath("somepath")), nullValue());
            <@\quad@>assertThat(requestMatchers.firstMatchingExpectation(new HttpRequest().withPath("someOtherPath")), nullValue());
            <@\quad@>&- assertThat(requestMatchers.httpRequestMatchers, empty());&
            <@\quad@>`+ assertThat(requestMatchers.httpRequestMatchers`@@.toSortedList()@@`, empty());`
            }
        \end{lstlisting}
        \end{tcolorbox}
        \subcaption{\textbf{Ground Truth} Repair}
        \label{fig:success3_gt}
    \end{subfigure}
\end{figure}
\begin{figure}[H]\ContinuedFloat
    \begin{subfigure}{\linewidth}
        \centering
        \begin{tcolorbox}[codebox]
        \begin{lstlisting}[language=Java,gobble=10,numbers=left,style=codeHighlighting,showstringspaces=false]
            &- assertThat ( requestMatchers.httpRequestMatchers , &^empty^& ( ) ) ;&
            `+ assertThat ( requestMatchers.httpRequestMatchers`@@.size@@` `@@( ) @@`, `@@is@@` ( `@@0 @@`) ) ;`
        \end{lstlisting}
        \end{tcolorbox}
        \subcaption{Top \textbf{Plausible Repair Candidate} Generated by \approachName{}}
        \label{fig:success3_cr}
    \end{subfigure}
\end{figure}
\begin{figure}[H]\ContinuedFloat
    \begin{subfigure}{\linewidth}
        \centering
        \begin{tcolorbox}[codebox]
        \begin{lstlisting}[language=Java,gobble=10,numbers=left,style=codeHighlighting,showstringspaces=false]
            &- return &^new^& &^ArrayList^& &^<^& &^>^& ( &^httpRequestMatchers^& ) ;&
            `+ return `@@httpRequestMatchers.stream@@` `@@(@@` `@@)@@` `@@.collect@@` ( `@@Collectors.toList@@` `@@( @@`) `@@) @@`;`
            
            ----------------------------------
            
            &- &^new^& &^ArrayList < > ^&( &^httpRequestMatchers ^&) .forEach ( httpRequestMatcher -> removeHttpRequestMatcher (& &httpRequestMatcher , cause , false ) ) ;&
            `+ `@@httpRequestMatchers.stream@@` ( ) .forEach ( httpRequestMatcher -> removeHttpRequestMatcher ( httpRequestMatcher` `, cause , false ) ) ;`
            
            ----------------------------------
            
            &- private &^List^& < HttpRequestMatcher > getHttpRequestMatchersCopy ( ) {&
            &- &^httpRequestMatchersCopy.compareAndSet ( null , httpRequestMatchers.toSortedList ( ) ) ;^&&
            &- return &^httpRequestMatchersCopy.get^& ( ) ;&
            `+ private `@@Stream@@` < HttpRequestMatcher > getHttpRequestMatchersCopy ( ) {`
            `+ return `@@httpRequestMatchers.stream@@` ( ) ;`
        \end{lstlisting}
        \end{tcolorbox}
        \subcaption{Relevant \textbf{Repair Context} Selected by \approachName{} and Included in the Model's Input}
        \label{fig:success3_ric}
    \end{subfigure}
\end{figure}


\subsubsection{Successful Example 4}

In this successful case (Figure~\ref{fig:success4}), the ground truth repair involves two changes to the point where the test case expects an exception to be thrown. As illustrated in Figure~\ref{fig:success4_gt}, the first change replaces the exception type \textit{FailedPreconditionRuntimeException} with \textit{InternalRuntimeException}. The second change updates the API used by the testing framework to check for the thrown exception: the new version of the test case uses \textit{Assert.assertThrows} instead of \textit{mThrown.expect}, which was used in the earlier version. It is important to note that this second change is related to the testing framework itself and is unrelated to any changes in the SUT.

The relevant repair context, depicted in Figure~\ref{fig:success4_ric}, provides clear guidance for the model. In this instance, \approachName{} generates a plausible repair candidate, shown in Figure~\ref{fig:success4_cr}. This candidate repair only addresses the change to the exception type. Similar to the example discussed in Section~\ref{sec:se3}, this repair is functionally equivalent to the ground truth, preserving the testing intent. Furthermore, the additional change present in the ground truth is a test refactoring and is thus unnecessary for the repair. This type of plausible repair is advantageous because it remains simple, effectively isolating the required code repair from any unrelated test refactoring.

\begin{figure}[H]
    \caption{\textbf{Successful Example 4} -- An example of a plausible repair generated by \approachName{}, which is an equivalent repair to the ground truth that isolates test repair from refactoring.\\\textbf{\benchmarkName{} ID:} \textit{alluxio/alluxio:1164}\\\textbf{GitHub Commit Reference:} \href{https://github.com/Alluxio/alluxio/commit/6982d6c759}{\textcolor{blue}{\textit{\underline{https://github.com/Alluxio/alluxio/commit/6982d6c759}}}}}
    \label{fig:success4}
    \centering
    \begin{subfigure}{\linewidth}
        \centering
        \begin{tcolorbox}[codebox]
        \begin{lstlisting}[language=Java,gobble=10,numbers=left,style=codeHighlighting,showstringspaces=false,breaklines=true]
            @Test
            public void close() throws Exception {
            <@\quad@>ByteBuffer buf = BufferUtils.getIncreasingByteBuffer(TEST_BLOCK_SIZE);
            <@\quad@>Assert.assertEquals(TEST_BLOCK_SIZE, mWriter.append(buf));
            <@\quad@>mWriter.close();
            <@\quad@>&- &^mThrown^&.&^expect^&(&^FailedPreconditionRuntimeException^&.class);&
            <@\quad@>^- mWriter.append(buf);^
            <@\quad@>`+ `@@Assert@@`.`@@assertThrows@@`(`@@InternalRuntimeException@@`.class`@@, (@@`)`@@ -> mWriter.append(buf))@@`;`
            }
        \end{lstlisting}
        \end{tcolorbox}
        \subcaption{\textbf{Ground Truth} Repair}
        \label{fig:success4_gt}
    \end{subfigure}
\end{figure}
\begin{figure}[H]\ContinuedFloat
    \begin{subfigure}{\linewidth}
        \centering
        \begin{tcolorbox}[codebox]
        \begin{lstlisting}[language=Java,gobble=10,numbers=left,style=codeHighlighting,showstringspaces=false]
            &- mThrown.expect ( &^FailedPreconditionRuntimeException^&.class ) ;&
            `+ mThrown.expect ( `@@InternalRuntimeException@@`.class ) ;`
        \end{lstlisting}
        \end{tcolorbox}
        \subcaption{Top \textbf{Plausible Repair Candidate} Generated by \approachName{}}
        \label{fig:success4_cr}
    \end{subfigure}
\end{figure}
\begin{figure}[H]\ContinuedFloat
    \begin{subfigure}{\linewidth}
        \centering
        \begin{tcolorbox}[codebox]
        \begin{lstlisting}[language=Java,gobble=10,numbers=left,style=codeHighlighting,showstringspaces=false] 
            &- return new &^FailedPreconditionRuntimeException^& ( t ) ;&
            `+ return new `@@InternalRuntimeException@@` ( t ) ;`
        \end{lstlisting}
        \end{tcolorbox}
        \subcaption{Relevant \textbf{Repair Context} Selected by \approachName{} and Included in the Model's Input}
        \label{fig:success4_ric}
    \end{subfigure}
\end{figure}

\clearpage
\section{A Comparative Analysis of Test Case Repairs Generated by \approachName{} and \ceprot{}}
This section supplements our paper's discussion by presenting five detailed examples where \approachName{} outperformed \ceprot{}, the baseline approach analyzed in our study. These examples illustrate the limitations of \ceprot{} and highlight the strengths of \approachName{}. Explanations for each example are provided in the corresponding figure captions.

\begin{figure}[H]
    \caption{\textbf{Comparison 1} -- Test repair where \approachName{} generated an exact match repair while \ceprot{} made no changes to the broken test, despite having sufficient repair context.\\\textbf{\ceprot{} \textit{test\_db} ID:} 4244531\\\textbf{GitHub Commit Reference:} \href{https://github.com/apache/wicket/commit/7c693db546}{\textcolor{blue}{\textit{\underline{https://github.com/apache/wicket/commit/7c693db546}}}}}
    \label{fig:cepCompare1}
    \centering
    \begin{subfigure}{\linewidth}
        \centering
        \begin{tcolorbox}[codebox]
        \begin{lstlisting}[language=Java,gobble=10,numbers=left,style=codeHighlighting,showstringspaces=false,breaklines=true]
            @Test
            `+ public void writeStream()`@@ throws IOException@@``
            {
            <@\quad@>WriteCallback callback = new WriteCallback()
            <@\quad@>{
            <@\quad@><@\quad@>@Override
            <@\quad@><@\quad@>public void writeData(Attributes attributes)
            <@\quad@><@\quad@>{
            <@\quad@><@\quad@>}
            <@\quad@>};
            <@\quad@>ByteArrayResponse response = new ByteArrayResponse();
            <@\quad@>Attributes attributes = new Attributes(new MockWebRequest(new Url()), response);
            <@\quad@>byte[] srcData = new byte[5000];
            <@\quad@>for (int i = 0; i < srcData.length; i++)
            <@\quad@>{
            <@\quad@><@\quad@>srcData[i] = (byte)i;
            <@\quad@>}
            <@\quad@>InputStream in = new ByteArrayInputStream(srcData);
            <@\quad@>callback.writeStream(attributes, in);
            <@\quad@>assertTrue("Content not equal", Arrays.equals(response.getBytes(), srcData));
            <@\quad@>}
            }
        \end{lstlisting}
        \end{tcolorbox}
        \subcaption{\textbf{Ground Truth} Repair and \textbf{the Exact Match Repair Candidate} Generated by \approachName{}}
        \label{fig:cepCompare1_gt}
    \end{subfigure}
\end{figure}
\begin{figure}[H]\ContinuedFloat
    \begin{subfigure}{\linewidth}
        \centering
        \begin{tcolorbox}[codebox]
        \begin{lstlisting}[language=Java,gobble=10,numbers=left,style=codeHighlighting,showstringspaces=false]
            &- protected final void writeStream ( Attributes attributes , InputStream stream )&
            `+ protected final void writeStream ( Attributes attributes , InputStream stream )`@@ throws IOException@@``           
        \end{lstlisting}
        \end{tcolorbox}
        \subcaption{Relevant \textbf{Repair Context} Selected by \approachName{} and Included in the Model's Input}
        \label{fig:cepCompare1_targetContext}
    \end{subfigure}
\end{figure}
\begin{figure}[H]\ContinuedFloat
    \begin{subfigure}{\linewidth}
        \centering
        \begin{tcolorbox}[codebox]
        \begin{lstlisting}[language=Java,gobble=10,numbers=left,style=codeHighlighting,showstringspaces=false]
            // No change to broken test
        \end{lstlisting}
        \end{tcolorbox}
        \subcaption{Incorrect Repair Generated by \ceprot{}. \ceprot{} Makes No Change to the Broken Test.}
        \label{fig:cepCompare1_ceprotRep}
    \end{subfigure}
\end{figure}
\begin{figure}[H]\ContinuedFloat
    \begin{subfigure}{\linewidth}
        \centering
        \begin{tcolorbox}[codebox]
        \begin{lstlisting}[language=Java,gobble=10,numbers=left,style=codeHighlighting,showstringspaces=false]
            &-protected final void writeStream(Attributes attributes, InputStream stream) { final Response response =& &attributes.getResponse(); &^try { ^&Streams.copy(stream, response.getOutputStream()); } &^catch(IOException e) {^ ^throw new WicketRuntimeException(e); } } ^&&
            `+protected final void writeStream(Attributes attributes, InputStream stream)`@@throws@@` `@@IOException @@`{ final Response` `response = attributes.getResponse(); Streams.copy(stream, response.getOutputStream()); } `

        \end{lstlisting}
        \end{tcolorbox}
        \subcaption{\ceprot{} Repair Context. Contains Sufficient Context to Perform the Required Repair.}
        \label{fig:cepCompare1_ceprotContext}
    \end{subfigure}
\end{figure}
\begin{figure}[H]
    \caption{\textbf{Comparison 2} -- Test repair where \approachName{} generated an exact match repair while \ceprot{} removed much of the broken test without generating a replacement, despite having sufficient repair context.\\\textbf{\ceprot{} \textit{test\_db} ID:} 262742\\\textbf{GitHub Commit Reference:} \href{https://github.com/Alluxio/alluxio/commit/b90ffa1af1}{\textcolor{blue}{\textit{\underline{https://github.com/Alluxio/alluxio/commit/b90ffa1af1}}}}}

    \label{fig:cepCompare2}
    \centering
    \begin{subfigure}{\linewidth}
        \centering
        \begin{tcolorbox}[codebox]
        \begin{lstlisting}[language=Java,gobble=10,numbers=left,style=codeHighlighting,showstringspaces=false,breaklines=true]
            @Test
            public void deleteTest() throws Exception {
            <@\quad@>CreateDirectoryOptions recMkdir = CreateDirectoryOptions.defaults().setRecursive(true);
            <@\quad@>DeleteOptions recDelete =
            <@\quad@><@\quad@>&- &^new^& DeleteOptions.&^Builder^&().setRecursive(true)&^.build()^&;&
            <@\quad@><@\quad@>`+ `@@DeleteOptions@@` `@@recDelete = @@`DeleteOptions.`@@defaults@@`().setRecursive(true);`
            <@\quad@>for (int i = 0; i < 10; i ++) {
            <@\quad@><@\quad@>String dirPath = "/i" + i;
            <@\quad@><@\quad@>mTfs.createDirectory(new TachyonURI(dirPath), recMkdir);
            <@\quad@><@\quad@>for (int j = 0; j < 10; j ++) {
            <@\quad@><@\quad@><@\quad@>CreateFileOptions option = CreateFileOptions.defaults().setBlockSizeBytes((i + j + 1) * 64);
            <@\quad@><@\quad@><@\quad@>String filePath = dirPath + "/j" + j;
            <@\quad@><@\quad@><@\quad@>mTfs.createFile(new TachyonURI(filePath), option).close();
            <@\quad@><@\quad@><@\quad@>if (j >= 5) {
            <@\quad@><@\quad@><@\quad@><@\quad@>mTfs.delete(new TachyonURI(filePath), recDelete);
            <@\quad@><@\quad@><@\quad@>}
            <@\quad@><@\quad@>}
            <@\quad@><@\quad@>if (i >= 5) {
            <@\quad@><@\quad@><@\quad@>mTfs.delete(new TachyonURI(dirPath), recDelete);
            <@\quad@><@\quad@>}
            <@\quad@>}
            <@\quad@>mLocalTachyonCluster.stopTFS();
            <@\quad@>deleteTestUtil();
            <@\quad@>deleteFsMasterJournalLogs();
            <@\quad@>deleteTestUtil();
            }
        \end{lstlisting}
        \end{tcolorbox}
        \subcaption{\textbf{Ground Truth} Repair and \textbf{the Exact Match Repair Candidate} Generated by \approachName{}}
        \label{fig:cepCompare2_gt}
    \end{subfigure}
\end{figure}
\begin{figure}[H]\ContinuedFloat
    \begin{subfigure}{\linewidth}
        \centering
        \begin{tcolorbox}[codebox]
        \begin{lstlisting}[language=Java,gobble=10,numbers=left,style=codeHighlighting,showstringspaces=false]
            &- DeleteOptions options = &^new ^&DeleteOptions.&^Builder^& ( ) .setRecursive ( true ) &^.build ( ) ^&;&
            `+ DeleteOptions options = DeleteOptions.`@@defaults@@` ( ) .setRecursive ( true ) ;`
            
            --------------------------------------------------
            
            &- DeleteOptions options = &^new ^&DeleteOptions.&^Builder^& ( ) .setRecursive ( recursive ) &^.build ( ) ^&;&
            `+ DeleteOptions options = DeleteOptions.`@@defaults@@` ( ) .setRecursive ( recursive ) ;`          
        \end{lstlisting}
        \end{tcolorbox}
        \subcaption{Relevant \textbf{Repair Context} Selected by \approachName{} and Included in the Model's Input}
        \label{fig:cepCompare2_targetContext}
    \end{subfigure}
\end{figure}
\begin{figure}[H]\ContinuedFloat
    \begin{subfigure}{\linewidth}
        \centering
        \begin{tcolorbox}[codebox]
        \begin{lstlisting}[language=Java,gobble=10,numbers=left,style=codeHighlighting,showstringspaces=false]
            &- @Test public void deleteTest()throws Exception { CreateDirectoryOptions recMkdir =& &CreateDirectoryOptions.defaults().setRecursive(true); DeleteOptions recDelete = new& &DeleteOptions.Builder().setRecursive(true).build(); for(int i = 0; i < 10; i ++ ) { String dirPath = "/i" + i;& &mTfs.createDirectory(new TachyonURI(dirPath), recMkdir); for(int j = 0; j < 10; j ++ ) { CreateFileOptions& &option = CreateFileOptions.defaults().setBlockSizeBytes((i + j + 1) * 64); String filePath = dirPath + "/j" + j;& &mTfs.createFile(new &^TachyonURI(filePath), option).close(); if(j >= 5) { mTfs.delete(new TachyonURI(filePath),^ ^recDelete); } } if(i >= 5) { mTfs.delete(new TachyonURI(dirPath), recDelete); } } mLocalTachyonCluster.stopTFS();^ ^deleteTestUtil(); deleteFsMasterJournalLogs(); deleteTestUtil(); } ^&&
            `+ @Test public void deleteTest()throws Exception { CreateDirectoryOptions recMkdir =` `CreateDirectoryOptions.defaults().setRecursive(true); DeleteOptions recDelete = new` `DeleteOptions.Builder().setRecursive(true).build(); for(int i = 0; i < 10; i ++ ) { String dirPath = "/i" + i;` `mTfs.createDirectory(new TachyonURI(dirPath), recMkdir); for(int j = 0; j < 10; j ++ ) { CreateFileOptions` `option = CreateFileOptions.defaults().setBlockSizeBytes((i + j + 1) * 64); String filePath = dirPath + "/j" + j;` `mTfs.createFile(new `@@T@@``
        \end{lstlisting}
        \end{tcolorbox}
        \subcaption{Incorrect Repair Generated by \ceprot{}. \ceprot{} Deletes Much of the Test.}
        \label{fig:cepCompare2_ceprotRep}
    \end{subfigure}
\end{figure}
\begin{figure}[H]\ContinuedFloat
    \begin{subfigure}{\linewidth}
        \centering
        \begin{tcolorbox}[codebox]
        \begin{lstlisting}[language=Java,gobble=10,numbers=left,style=codeHighlighting,showstringspaces=false]
            &- @Override public boolean delete(Path cPath, boolean recursive)throws IOException { LOG.info("delete({}, {})", cPath,& &recursive); if(mStatistics != null) { mStatistics.incrementWriteOps(1); } TachyonURI path = new& &TachyonURI(Utils.getPathWithoutScheme(cPath)); DeleteOptions options = &^new^ ^^&DeleteOptions.&^Builder^&().setRecursive(recursive)&^.build()^&; try { mTFS.delete(path, options); return true; }& &catch(InvalidPathException e) { LOG.info("delete failed: {}", e.getMessage()); return false; }& &catch(FileDoesNotExistException e) { LOG.info("delete failed: {}", e.getMessage()); return false; }& &catch(TachyonException e) { throw new IOException(e); } } &
            `+ @Override public boolean delete(Path cPath, boolean recursive)throws IOException { LOG.info("delete({}, {})", cPath,` `recursive); if(mStatistics != null) { mStatistics.incrementWriteOps(1); } TachyonURI path = new` `TachyonURI(Utils.getPathWithoutScheme(cPath)); DeleteOptions options =` `DeleteOptions.`@@defaults@@`().setRecursive(recursive); try { mTFS.delete(path, options); return true; }` `catch(InvalidPathException e) { LOG.info("delete failed: {}", e.getMessage()); return false; }` `catch(FileDoesNotExistException e) { LOG.info("delete failed: {}", e.getMessage()); return false; }` `catch(TachyonException e) { throw new IOException(e); } } `

        \end{lstlisting}
        \end{tcolorbox}
        \subcaption{\ceprot{} Repair Context. Contains Sufficient Context to Perform the Required Repair.}
        \label{fig:cepCompare2_ceprotContext}
    \end{subfigure}
\end{figure}
\begin{figure}[H]
    \caption{\textbf{Comparison 3} -- Test repair where \approachName{} generated an exact match repair while \ceprot{} removed much of the broken test without generating a replacement, possibly because of its insufficient repair context.\\\textbf{\ceprot{} \textit{test\_db} ID:} 2229339\\\textbf{GitHub Commit Reference:} \href{https://github.com/hazelcast/hazelcast/commit/83d814eefb}{\textcolor{blue}{\textit{\underline{https://github.com/hazelcast/hazelcast/commit/83d814eefb}}}}}
    \label{fig:cepCompare3}
    \centering
    \begin{subfigure}{\linewidth}
        \centering
        \begin{tcolorbox}[codebox]
        \begin{lstlisting}[language=Java,gobble=10,numbers=left,style=codeHighlighting,showstringspaces=false,breaklines=true]
            @Test
            public void addAllAsync_manyTimesRoundTheRing() throws Exception {
            <@\quad@>RingbufferConfig c = config.getRingbufferConfig(ringbuffer.getName());
            <@\quad@>Random random = new Random();
            <@\quad@>for (int iteration = 0; iteration < 1000; iteration++) {
            <@\quad@><@\quad@>List<String> items = randomList(max(1, <@\quad@><@\quad@>random.nextInt(c.getCapacity())));
            <@\quad@><@\quad@>long previousTailSeq = ringbuffer.tailSequence();
            <@\quad@><@\quad@>&- long result = ringbuffer.addAllAsync(items, OVERWRITE).get();&
            <@\quad@><@\quad@>`+ long result = ringbuffer.addAllAsync(items, OVERWRITE).`@@toCompletableFuture().@@`get();`
            <@\quad@><@\quad@>assertEquals(previousTailSeq + items.size(), <@\quad@><@\quad@>ringbuffer.tailSequence());
            <@\quad@><@\quad@>if (ringbuffer.tailSequence() < c.getCapacity()) {
            <@\quad@><@\quad@><@\quad@>assertEquals(0, ringbuffer.headSequence());
            <@\quad@><@\quad@>} else {
            <@\quad@><@\quad@><@\quad@>assertEquals(ringbuffer.tailSequence() - c.getCapacity() + 1, ringbuffer.headSequence());
            <@\quad@><@\quad@>}
            <@\quad@><@\quad@>assertEquals(ringbuffer.tailSequence(), result);
            <@\quad@><@\quad@>long startSequence = previousTailSeq + 1;
            <@\quad@><@\quad@>for (int k = 0; k < items.size(); k++) {
            <@\quad@><@\quad@><@\quad@>assertEquals(items.get(k), ringbuffer.readOne(startSequence + k));
            <@\quad@><@\quad@>}
            <@\quad@>}
            }
        \end{lstlisting}
        \end{tcolorbox}
        \subcaption{\textbf{Ground Truth} Repair and \textbf{the Exact Match Repair Candidate} Generated by \approachName{}}
        \label{fig:cepCompare3_gt}
    \end{subfigure}
\end{figure}
\begin{figure}[H]\ContinuedFloat
    \begin{subfigure}{\linewidth}
        \centering
        \begin{tcolorbox}[codebox]
        \begin{lstlisting}[language=Java,gobble=10,numbers=left,style=codeHighlighting,showstringspaces=false]
            &- return ringbuffer.addAsync ( message , OverflowPolicy.OVERWRITE ) .get ( ) ;&
            `+ return ringbuffer.addAsync ( message , OverflowPolicy.OVERWRITE ) .`@@toCompletableFuture ( ) .@@`get ( ) ;`     
        \end{lstlisting}
        \end{tcolorbox}
        \subcaption{Relevant \textbf{Repair Context} Selected by \approachName{} and Included in the Model's Input}
        \label{fig:cepCompare3_targetContext}
    \end{subfigure}
\end{figure}
\begin{figure}[H]\ContinuedFloat
    \begin{subfigure}{\linewidth}
        \centering
        \begin{tcolorbox}[codebox]
        \begin{lstlisting}[language=Java,gobble=10,numbers=left,style=codeHighlighting,showstringspaces=false]
            &- @Test public void addAllAsync_manyTimesRoundTheRing()throws Exception { RingbufferConfig c =& &config.getRingbufferConfig(ringbuffer.getName()); Random random = new Random(); for(int iteration = 0; iteration& &< 1000; iteration ++ ) { List < String > items = randomList(max(1, random.nextInt(c.getCapacity()))); long& &previousTailSeq = ringbuffer.tailSequence(); long result = ringbuffer.addAllAsync(items, OVERWRITE).get();& &assertEquals(previousTailSeq + items.size(), ringbuffer.tailSequence()); if(ringbuffer.tailSequence() <& &c.getCapacity()) { assertEquals(0, ringbuffer.&^headSequence()); } else { assertEquals(ringbuffer.tailSequence()^ ^- c.getCapacity() + 1, ringbuffer.headSequence()); } assertEquals(ringbuffer.tailSequence(), result); long^ ^startSequence = previousTailSeq + 1; for(int k = 0; k < items.size(); k ++ ) { assertEquals(items.get(k),^ ^ringbuffer.readOne(startSequence + k)); } } } ^&&
            `+ @Test public void addAllAsync_manyTimesRoundTheRing()throws Exception { RingbufferConfig c =` `config.getRingbufferConfig(ringbuffer.getName()); Random random = new Random(); for(int iteration = 0; iteration` `< 1000; iteration ++ ) { List < String > items = randomList(max(1, random.nextInt(c.getCapacity()))); long` `previousTailSeq = ringbuffer.tailSequence(); long result = ringbuffer.addAllAsync(items, OVERWRITE).get();` `assertEquals(previousTailSeq + items.size(), ringbuffer.tailSequence()); if(ringbuffer.tailSequence() <` `c.getCapacity()) { assertEquals(0, ringbuffer.`@@head@@``
        \end{lstlisting}
        \end{tcolorbox}
        \subcaption{Incorrect Repair Generated by \ceprot{}. \ceprot{} Deletes Much of the Test.}
        \label{fig:cepCompare3_ceprotRep}
    \end{subfigure}
\end{figure}
\begin{figure}[H]\ContinuedFloat
    \begin{subfigure}{\linewidth}
        \centering
        \begin{tcolorbox}[codebox]
        \begin{lstlisting}[language=Java,gobble=10,numbers=left,style=codeHighlighting,showstringspaces=false]
            &- &^ICompletableFuture^& < Long > addAllAsync(@Nonnull Collection < ? extends E > collection, @Nonnull OverflowPolicy& &overflowPolicy); &
            `+ `@@CompletionStage@@` < Long > addAllAsync(@Nonnull Collection < ? extends E > collection, @Nonnull OverflowPolicy` `overflowPolicy); `
        \end{lstlisting}
        \end{tcolorbox}
        \subcaption{\ceprot{} Repair Context. Contains Insufficient Context to Perform the Required Repair.}
        \label{fig:cepCompare3_ceprotContext}
    \end{subfigure}
\end{figure}
\begin{figure}[H]
    \caption{\textbf{Comparison 4} -- Test repair where \approachName{} generated an exact match repair while \ceprot{} produced a repair that, although compilable, was incorrect, despite having sufficient repair context.\\\textbf{\ceprot{} \textit{test\_db} ID:} 3675646\\\textbf{GitHub Commit Reference:} \href{ https://github.com/ThreeTen/threetenbp/commit/7138c860e2}{\textcolor{blue}{\textit{\underline{ https://github.com/ThreeTen/threetenbp/commit/7138c860e24}}}}}
    \label{fig:cepCompare4}
    \centering
    \begin{subfigure}{\linewidth}
        \centering
        \begin{tcolorbox}[codebox]
        \begin{lstlisting}[language=Java,gobble=10,numbers=left,style=codeHighlighting,showstringspaces=false,breaklines=true]
            @Test(expectedExceptions=NullPointerException.class)
            public void test_appendValueReduced_null() throws Exception {
            <@\quad@>&- builder.appendValueReduced(null, 2, 2000);&
            <@\quad@>`+ builder.appendValueReduced(null, 2, `@@2, @@`2000);`
            }

        \end{lstlisting}
        \end{tcolorbox}
        \subcaption{\textbf{Ground Truth} Repair and \textbf{the Exact Match Repair Candidate} Generated by \approachName{}}
        \label{fig:cepCompare4_gt}
    \end{subfigure}
\end{figure}
\begin{figure}[H]\ContinuedFloat
    \begin{subfigure}{\linewidth}
        \centering
        \begin{tcolorbox}[codebox]
        \begin{lstlisting}[language=Java,gobble=10,numbers=left,style=codeHighlighting,showstringspaces=false]
            &- appendValueReduced ( field , 2 , 2000 ) ;&
            `+ appendValueReduced ( field , 2 , `@@2 , @@`2000 ) ;`
        \end{lstlisting}
        \end{tcolorbox}
        \subcaption{Relevant \textbf{Repair Context} Selected by \approachName{} and Included in the Model's Input}
        \label{fig:cepCompare4_targetContext}
    \end{subfigure}
\end{figure}
\begin{figure}[H]\ContinuedFloat
    \begin{subfigure}{\linewidth}
        \centering
        \begin{tcolorbox}[codebox]
        \begin{lstlisting}[language=Java,gobble=10,numbers=left,style=codeHighlighting,showstringspaces=false]
            &- @Test(expectedExceptions = NullPointerException.class)public void test_appendValueReduced_null()throws Exception {& &builder.appendValueReduced(null, 2, 2000); }&
            `+ @Test(expectedExceptions = NullPointerException.class)public void test_appendValueReduced_null()throws Exception {` `builder.appendValueReduced(null, 2, 2000`@@, 2@@`); } `
        \end{lstlisting}
        \end{tcolorbox}
        \subcaption{Incorrect Repair Generated by \ceprot{}. \ceprot{} Generates an Incorrect Repair.}
        \label{fig:cepCompare4_ceprotRep}
    \end{subfigure}
\end{figure}
\begin{figure}[H]\ContinuedFloat
    \begin{subfigure}{\linewidth}
        \centering
        \begin{tcolorbox}[codebox]
        \begin{lstlisting}[language=Java,gobble=10,numbers=left,style=codeHighlighting,showstringspaces=false]
            &- public DateTimeFormatterBuilder appendValueReduced(TemporalField field, int width, int baseValue) {& &Objects.requireNonNull(field, "field"); ReducedPrinterParser pp = new ReducedPrinterParser(field, width,& &baseValue); appendFixedWidth(width, pp); return this; }&
            `+ public DateTimeFormatterBuilder appendValueReduced(TemporalField field, int width, int `@@maxWidth, int @@`baseValue) {` `Objects.requireNonNull(field, "field"); ReducedPrinterParser pp = new ReducedPrinterParser(field, width,` @@maxWidth, @@`baseValue); `@@if(width == maxWidth) { @@`appendFixedWidth(width, pp); `@@} else { appendInternal(pp); }@@ `return this; }`
        \end{lstlisting}
        \end{tcolorbox}
        \subcaption{\ceprot{} Repair Context. Contains Sufficient Context to Perform the Required Repair.}
        \label{fig:cepCompare4_ceprotContext}
    \end{subfigure}
\end{figure}
\begin{figure}[H]
    \caption{\textbf{Comparison 5} -- Test repair where \approachName{} generated an exact match repair while \ceprot{} only modified whitespace, possibly due to irrelevant repair context.\\\textbf{\ceprot{} \textit{test\_db} ID:} 4640902\\\textbf{GitHub Commit Reference:} \href{ https://github.com/BaseXdb/basex/commit/5570aea8cb}{\textcolor{blue}{\textit{\underline{https://github.com/BaseXdb/basex/commit/5570aea8cb}}}}}
    \label{fig:cepCompare5}
    \centering
    \begin{subfigure}{\linewidth}
        \centering
        \begin{tcolorbox}[codebox]
        \begin{lstlisting}[language=Java,gobble=10,numbers=left,style=codeHighlighting,showstringspaces=false,breaklines=true]
            @Test public void serialize() {
            <@\quad@>for(final String[] test : TOCSV) {
            <@\quad@><@\quad@>final String query = test[1].isEmpty() ? _CSV_SERIALIZE.args(test[0]) :
            <@\quad@><@\quad@>_CSV_SERIALIZE.args(test[0], " {" + test[1] + "}");
            <@\quad@><@\quad@>if(test.length == 2) {
            <@\quad@><@\quad@><@\quad@>&- error(query, Err.BXCS_CONFSEP, Err.&^ELMOPTION^&);&
            <@\quad@><@\quad@><@\quad@>`+ error(query, Err.BXCS_CONFSEP, Err.`@@INVALIDOPT@@`);`
            <@\quad@><@\quad@>} else {
            <@\quad@><@\quad@><@\quad@>query(query, test[2]);
            <@\quad@><@\quad@>}
            <@\quad@>}
            }
        \end{lstlisting}
        \end{tcolorbox}
        \subcaption{\textbf{Ground Truth} Repair and \textbf{the Exact Match Repair Candidate} Generated by \approachName{}}
        \label{fig:cepCompare5_gt}
    \end{subfigure}
\end{figure}
\begin{figure}[H]\ContinuedFloat
    \begin{subfigure}{\linewidth}
        \centering
        \begin{tcolorbox}[codebox]
        \begin{lstlisting}[language=Java,gobble=10,numbers=left,style=codeHighlighting,showstringspaces=false]
            &- if ( mode == null ) &^ELMOPTION^&.thrw ( info , md ) ;&
            `+ if ( mode == null ) `@@INVALIDOPT@@`.thrw ( info , `@@Util.info ( Text.INVALID_VALUE_X_X , "mode" , @@`md ) `@@) @@`;`
            
            --------------------------------------------------
            
            &- &^ELMOPTION^& ( FORG , 6 , "&^Option '^&%&^' is invalid^&." ) ,&
            `+ `@@INVALIDOPT@@` ( FORG , 6 , "%." ) ,`
            @@+ / ** FORG0006. * /@@
            @@+ INVALIDOPTX ( FORG , 6 , Text.UNKNOWN_OPTION_X ) ,@@

        \end{lstlisting}
        \end{tcolorbox}
        \subcaption{Relevant \textbf{Repair Context} Selected by \approachName{} and Included in the Model's Input}
        \label{fig:cepCompare5_targetContext}
    \end{subfigure}
\end{figure}
\begin{figure}[H]\ContinuedFloat
    \begin{subfigure}{\linewidth}
        \centering
        \begin{tcolorbox}[codebox]
        \begin{lstlisting}[language=Java,gobble=10,numbers=left,style=codeHighlighting,showstringspaces=false]
            &- @Test public void serialize() { for(final String[]test : TOCSV) { final String query = test[1].isEmpty()&^ ^&?& &_CSV_SERIALIZE.args(test[0]) : _CSV_SERIALIZE.args(test[0], " {" + test[1] + "}"); if(test.length == 2) {& &error(query, Err.BXCS_CONFSEP, Err.ELMOPTION); } else { query(query, test[2]); } } } &
            `+ @Test public void serialize() { for(final String[]test : TOCSV) { final String query = test[1].isEmpty()?` `_CSV_SERIALIZE.args(test[0]) : _CSV_SERIALIZE.args(test[0], " {" + test[1] + "}"); if(test.length == 2) {` `error(query, Err.BXCS_CONFSEP, Err.ELMOPTION); } else { query(query, test[2]); } } }`
        \end{lstlisting}
        \end{tcolorbox}
        \subcaption{Incorrect Repair Generated by \ceprot{}. \ceprot{} Generates a Repair that Only Contains Whitespace Changes.}
        \label{fig:cepCompare5_ceprotRep}
    \end{subfigure}
\end{figure}
\begin{figure}[H]\ContinuedFloat
    \begin{subfigure}{\linewidth}
        \centering
        \begin{tcolorbox}[codebox]
        \begin{lstlisting}[language=Java,gobble=10,numbers=left,style=codeHighlighting,showstringspaces=false]
            &- public final void serialize(final Item item)throws IOException { openResult(); if(item instanceof ANode) { final& &Type type = item.type; if(type == NodeType.ATT)SERATTR.&^thrwSerial^&(item); if(type ==& &NodeType.NSP)SERNS.&^thrwSerial^&(item); serialize((ANode)item); } else if(item instanceof FItem) {& &SERFUNC.&^thrwSerial^&(item.description()); } else { finishElement(); atomic(item); } closeResult(); } &
            `+ public final void serialize(final Item item)throws IOException { openResult(); if(item instanceof ANode) { final` `Type type = item.type; if(type == NodeType.ATT)SERATTR.`@@thrwIO@@`(item); if(type == NodeType.NSP)SERNS.`@@thrwIO@@`(item);` `serialize((ANode)item); } else if(item instanceof FItem) { SERFUNC.`@@thrwIO@@`(item.description()); } else {` `finishElement(); atomic(item); } closeResult(); } `
        \end{lstlisting}
        \end{tcolorbox}
        \subcaption{\ceprot{} Repair Context. Contains Insufficient Context to Perform the Required Repair.}
        \label{fig:cepCompare5_ceprotContext}
    \end{subfigure}
\end{figure}

\end{appendices}

\twocolumn

\end{document}